\documentclass{article}
\usepackage[utf8]{inputenc}
\usepackage{hyperref}
\usepackage{color}
\hypersetup{colorlinks=false, linktoc=all,linkcolor=blue}
\usepackage{caption}
\usepackage{subcaption}
\usepackage{verbatim}
\usepackage{mwe}
\usepackage[ruled,vlined]{algorithm2e}
\usepackage[english]{babel}
\usepackage{amsmath}
\usepackage{verbatim}
\DeclareMathOperator\erf{erf}
\usepackage[autostyle]{csquotes}
\usepackage{amsfonts}
\usepackage{amssymb}
\usepackage{float}
\makeatletter
 \usepackage{url}

\makeatother

\usepackage{geometry}

\DeclareMathOperator\erfc{erfc}

\usepackage{graphicx}
\usepackage{multicol}
\usepackage[version=4]{mhchem}
\usepackage[table,xcdraw]{xcolor}
\usepackage{float}

\usepackage{authblk}

\title{Fitness-Based Growth of Directed Networks with Hierarchy}

\author[1,2]{Niall Rodgers \thanks{nxr081@student.bham.ac.uk}}
\author[3]{Peter Ti\v{n}o}
\author[1]{Samuel Johnson \thanks{s.johnson.4@bham.ac.uk}}

\affil[1]{School of Mathematics, University of Birmingham, Birmingham, United Kingdom}
\affil[2]{Centre for Doctoral Training Topological Design, University of Birmingham, Birmingham, United Kingdom}
\affil[3]{School of Computer Science, University of Birmingham, Birmingham, United Kingdom}
\date{\today}
\begin{document}

\maketitle


\begin{abstract}

Growing attention has been brought to the fact that many real directed networks exhibit hierarchy and directionality as measured through techniques like Trophic Analysis and non-normality. 
We propose a simple growing network model where the probability of connecting to a node is defined by a preferential attachment mechanism based on degree and the difference in fitness between nodes. 
In particular, we show  how mechanisms such as degree-based preferential attachment and node fitness interactions can lead to the emergence of the spectrum of hierarchy and directionality observed in real networks.  In this work, we study various features of this model relating to network hierarchy, as measured by Trophic Analysis. This includes (I) how preferential attachment can lead to network hierarchy, (II) how scale-free degree distributions and network hierarchy can coexist, (III) the correlation between node fitness and trophic level, (IV) how the fitness parameters can predict trophic incoherence and how the trophic level difference distribution compares to the fitness difference distribution, (V) the relationship between trophic level and degree imbalance and the unique role of nodes at the ends of the fitness hierarchy and (VI) how fitness interactions and degree-based preferential attachment can interplay to generate networks of varying coherence and degree distribution. We also provide an example of the intuition this work enables in the analysis of a real historical network. This work provides insight into simple mechanisms which can give rise to hierarchy in directed networks and quantifies the usefulness and limitations of using Trophic Analysis as an analysis tool for real networks.
    
\end{abstract}

\section{Introduction}

Complex networks are evolving systems and much of the research in network science has been dedicated to finding models which explain how networks can be formed in such a way which reproduces a certain property observed in real-world systems. Real directed networks have properties which are unique to the directed case such as non-normality, global directionality and hierarchical organisation of the nodes. It has been shown that networks which are non-normal \cite{Duan2022NetworkSystems,Asllani2018StructureNetworks,Nartallo-Kaluarachchi2024BrokenNetworks} or have a global directionality as measured by trophic incoherence \cite{MacKay2020HowNetwork,Rodgers2023StrongNetworks,Rodgers2023InfluenceNetworks,Johnson2020DigraphsSystems,Nartallo-Kaluarachchi2024BrokenNetworks} are very common in real-world settings. We propose a generative model based on the interactions between node fitness and degree-based preferential attachment which explains this behaviour in real-world systems. We also show how the properties of the fitness model can be linked to the structural features related to hierarchy and directionality as measured by the technique of Trophic Analysis \cite{MacKay2020HowNetwork} which justifies its usefulness as an analytical tool. Trophic Analysis is a method for studying directed networks which allows the local position in the hierarchy, Trophic Level, and the global directionality, Trophic Incoherence, to be measured. It has been related to the spectral properties  \cite{Johnson2017LooplessnessCoherence,Hazan2023ProductionAnalysis,MacKay2020HowNetwork} and strong connectivity in real directed networks \cite{Rodgers2023StrongNetworks}. As well as notions of network influence \cite{Rodgers2023InfluenceNetworks}, directed Hopfield networks \cite{Rodgers2022NetworkNetworks}, historical networks \cite{Shuaib2022TrophicInformation} as well as  trade, social systems, and economic networks \cite{Sornette2023Non-normalBubbles,Dawes2022System-levelLevel, Dawes2022SDGHierarchies} with additional equivalent formulations \cite{MacKay2020HowNetwork} also used \cite{Fujiwara2021MoneyJapan}.   

We study a growing network model where the probability of connecting to a node is a function of the fitness difference in combination with preferential attachment based on degree. This idea builds on many models present in the literature which use node fitness as a way to build networks. 

The earliest models of complex network evolution based on fitness and preferential attachment, which used connection probabilities based on the fitness of a node without interaction, allowed the previously observed scale-free network property to be better understood \cite{Bianconi2001CompetitionNetworks}. For example, the Bianconi–Barab\'asi model which combined node fitness with degree-based attachment \cite{Bianconi2001CompetitionNetworks}, where nodes with higher fitness are more likely to accrue more connections. This provided a fitness-based extension to the earlier models highlighting the ubiquity of scale-free networks \cite{Barabasi1999EmergenceNetworks}. With later models also studying how the fitness distribution can translate into properties of the the degree distribution \cite{GhadgeLogNormal}. In this work we take a similar approach but tackle the problem of network hierarchy and show how fitness models allow the emergence of network hierarchy to be understood.

 In previous work, networks of varying non-normality and trophic coherence have either been created artificially using degree imbalances \cite{Duan2022NetworkSystems}, using node arrival time and a reciprocity parameter to create hierarchical structure \cite{Asllani2018StructureNetworks, Nartallo-Kaluarachchi2024BrokenNetworks} or using static models \cite{DeBacco2018ANetworks}, like the variants of the generalised preferential preying model \cite{Rodgers2022NetworkNetworks,Rodgers2023StrongNetworks,Rodgers2023InfluenceNetworks,Klaise2016FromProcesses} which is based on the initial trophic level of the nodes of a seed network. However, these methods are not realistic in that in a real systems networks grow over time, the possible structures are limited \cite{Asllani2018StructureNetworks} or individuals may not have access to the network properties and choose nodes to connect to based only on fitness of the nodes they observe. Hence we demonstrate how the emergence of wide range of hierarchical structures can be found in fitness-based growth models.

The range of possible phenomena that can be explained by fitness models increases when  one allows for interactions between the fitness of the connecting nodes.
In order to induce a hierarchy within the network we study models where the probability to attach depends on the fitness differences between nodes with an offset. Fitness-based models without an offset, which can be seen as a specific case of our model when this parameter is set to zero, have been widely used to study a variety of phenomena. This includes models which create scale-free networks with homophilly \cite{DeAlmeida2013Scale-freeNetwork,Santos2020CriticalNetwork, Piva2021NetworksApplications}, used to study inequality \cite{Lee2019HomophilyNetworks,Nettasinghe2022Scale-freeNetworks} and spreading processes in social settings \cite{Gargiulo2017TheControversies}. Fitness interactions have also been used to study the interactions between countries in trade networks \cite{Hoppe2015ANetwork}. Models to study citation networks can also be viewed as models based on fitness difference interactions as the time of publication of a paper can be used as fitness and the decay in the citation probability with time viewed as the interaction between fitness parameters as well as any other metrics based on the success of the paper \cite{Wang2013QuantifyingImpact,Golosovsky2018MechanismsModels, Hajra2005AgingNetworks}. In some systems however, it makes sense to try to connect to another node who has a different fitness to 
the parent node. For example, in ecology an animal may
as a first approximation be able to eat a smaller animal rather than connecting to a prey of the same size \cite{Loeuille2005EvolutionaryWebs}. This leads to models describing ecosystem formation based on the niche widths of species \cite{Loeuille2005EvolutionaryWebs}, which is closely related to one of the cases our model captures.

A model of network formation based on fitness differences with an offset makes sense in many settings in which hierarchy has been observed. The most obvious case is ecology, where a species consumes a species which is just below it in the trophic hierarchy however this may also make sense in other areas. In a network of trades between sports teams you may want to buy players from a team which is below in the hierarchy and sell to teams above. In trade networks there is a fitness hierarchy, known in economics as upstreamness, depending on how far along the chain of the processing a good is. For example, a company which makes batteries may want to buy raw materials directly but a company making consumer electronics wants to buy the batteries and does not buy the raw materials directly imposing a hierarchy of transactions. Similar affects could also be seen in neural or communications networks where connections want to be made into the next layer of the system and to respect this ordering. Related phenomena may also arise in certain social settings, where the reciprocity of interactions depends on the social status of the individuals \cite{Ball2013FriendshipStatus}.

\section{Background}

\subsection{Hierarchy in Directed Networks} 

Hierarchy in directed networks can be quantified in many ways \cite{Mones2012HierarchyNetworks,Corominas-Murtra2013OnNetworks} in terms of either flow hierarchy or a nested hierarchy. In this work, we speak of hierarchy in terms of flow hierarchy. That is the nodes can be ordered on a one dimensional axis via a method like trophic analysis \cite{MacKay2020HowNetwork}, SpringRank \cite{DeBacco2018ANetworks}, an ordered stochastic block-model \cite{Peixoto2022OrderedNetworks} or another method so that edges point in the direction of the flow hierarchy as much as possible given the constraints of the topology and ranking algorithm. This flow hierarchy can have many effects on network structure and dynamics \cite{Johnson2020DigraphsSystems,Rodgers2023InfluenceNetworks,Rodgers2023StrongNetworks,Johnson2017LooplessnessCoherence,Johnson2014TrophicStability, Zamani2017GlassyOrganizations} and understanding how it is arises it vital for understanding complex systems. The hierarchy and directionality we use specifically describes a layered flow hierarchy, where edges point upwards a specified distance on the one dimensional axis, which differs from some ideas of flow hierarchy where it is not important how far up the direction the edges point.

\subsubsection{Trophic Analysis} \label{section:Trophic_Analysis}

Trophic Analysis is a method which allows the study of linear hierarchy and global directionality in directed networks \cite{MacKay2020HowNetwork}. It was originally conceived as a tool for networks science in \cite{Johnson2014TrophicStability} where it was proposed as a solution to ``May's paradox'' \cite{May1972WillStable} in ecosystems. However this original definition was constrained so that a network required nodes of zero in-degree and that these nodes were always at the bottom of the hierarchy. An improved version of Trophic Analysis which removed this constraint was proposed by MacKay {et al.} \cite{MacKay2020HowNetwork}, which is the definition we use in this work. Trophic Analysis is composed of two parts, the local measure of where individual nodes sit in the hierarchy known as {\it trophic level}, and the global quantity, {\it trophic incoherence}, which measures the global directionality of the network and how much the edges point in the same direction with respect to the hierarchy.

In this work we studied unweighted directed complex networks represented by the an $N \times N$ adjacency matrix with the following convention, 
\begin{equation} \label{Adj_eq}
A_{ij}=
    \begin{cases}
     1  \text{ \quad if there exists an edge } i \to j \\ 
     0  \text{ \quad  otherwise }
    \end{cases}.
\end{equation}

Trophic Analysis can simply extend to the weighted case \cite{MacKay2020HowNetwork} but here we present the unweighted case for simplicity. Trophic Analysis can be formulated as an optimisation problem where we want to select the optimal set of trophic levels for the nodes which minimises the trophic incoherence. The trophic incoherence, $F$, is defined as  
\begin{equation} \label{eq:F}
    F = \frac{\sum_{ij}A_{ij}(h_{j} -h_{i}-1)^2}{\sum_{ij}A_{ij}}.
\end{equation} 
Equations \ref{eq:F} states that we wish to select trophic levels, $h_i$ such that if there is an edge from node $i$ to $j$ then node $j$ has trophic level, $h_j$, which is exactly one more than the trophic level of node $i$, with any deviations from this incurring penalties in the square of the deviation. Due to the structure of $F$ it can be minimised by simply taking the derivative with respect to the trophic levels. This leads to the following linear equation for the vector of trophic levels 
\begin{equation} 
    \Lambda h = v,
    \label{eq_h}
\end{equation}
where $v$ is the degree imbalance vector defined as $v_i = k^{\text{in}}_i - k^{\text{out}}_i$ and $\Lambda$ is defined as 
\begin{equation}
    \Lambda = diag(u) - A - A^{T},
\end{equation}
with $u$ being the vector of the sum of the in and out degrees of each node, $u_i = k^{\text{in}}_i + k^{\text{out}}_i$. As equation \ref{eq_h} is linear it can be solved relatively easily with a few caveats. The matrix $\Lambda$ cannot be inverted as it always has a zero eigenvalue. This issue can be resolved by fixing
the level for one of the nodes, \cite{MacKay2020HowNetwork}, which can be done as equation \ref{eq_h} is invariant up to the addition of a constant vector to $h$, so the minimum value of $h$ can be rescaled to any value. This relates to the fact $F$ only depends on the difference between the levels of nodes. Where we refer to trophic incoherence, $F$, throughout this work we refer to equation \ref{eq:F} evaluated using the value of $h$ which minimises it. We choose to work in the convention where the minimum trophic level is set to zero. Another approach which can be taken is to solve for $h$ iteratively using techniques of sparse linear algebra. 

When the trophic level vector $h$ which minimises $F$ is substituted into \ref{eq:F} it acts of a measure of how hierarchical the network is and it quantifies the global directionality of the edges. This minimum value of $F$ is what we refer to as the incoherence of a network for the remainder of this work. This value is bound between $0$ and $1$ \cite{MacKay2020HowNetwork}. With $F=0$  being attained when the network has a perfect hierarchy like a directed path where the levels are integers with steps of exactly $+1$ between nodes. While $F=1$ when the network is balanced (i.e. the in-degree matches the out-degree for each node), so every node takes level $h_i=0$ and there is no hierarchy. Real networks exhibit a wide spectrum of trophic incoherence. Some, like food-webs which have a clear ordering of the edges, have low values of $F$; while networks which have many reciprocal connections or cycles, like social networks or those with little structure like Erd\H{o}s-R\'enyi random graphs, have a higher $F$. Trophic Incoherence is 
linked to other network structure features such as the spectral radius \cite{Johnson2017LooplessnessCoherence,MacKay2020HowNetwork}, non-normality \cite{Johnson2020DigraphsSystems} or the emergence of a strongly connected component \cite{Rodgers2023StrongNetworks}.

The method of Trophic Analysis is closely related to several other ranking methods, SpringRank \cite{DeBacco2018ANetworks} and a method for analysing `circularity' \cite{Kichikawa2019CommunityNetwork} based on Helmholtz-Hodge decomposition. The method of \cite{Kichikawa2019CommunityNetwork} has been shown to be equivalent to Trophic Analysis \cite{MacKay2020HowNetwork} however with a different set of terminology. SpringRank \cite{DeBacco2018ANetworks} views the ranking problem as minimising the energy of network of directed springs and devises a very similar function to minimise as $F$, just with a different normalisation. However, some versions of SpringRank introduce a regularisation term to remove the invariance of the ranks to the addition of a constant vector and give the ranking equation a unique solution.

In \cite{Johnson2020DigraphsSystems,MacKay2020HowNetwork} it was proposed that there is a relationship between the non-normality of the adjacency matrix and trophic incoherence. This was shown by measuring the normality in \cite{MacKay2020HowNetwork,Nartallo-Kaluarachchi2024BrokenNetworks} and the pseudospectral radius, a property related to non-normality, in \cite{Rodgers2023InfluenceNetworks}. A matrix is normal if \begin{equation}
    AA^T = A^TA
\end{equation}
and a non-normal matrix is one where the adjacency matrix does not commute with its transpose \cite{Trefethen2005SpectraOperators}. It can be quantified the extent to which a matrix does not commute with it's transpose.
For a example, a network with very few reciprocal edges is much more non-normal than a network with many reciprocal connections with the extreme case being an undirected graph where every edge is both bidirectional and $A$ is symmetric. This also shows why trophic incoherence and normality can be related, as the fraction of reciprocal edges and cycles also impacts the coherence and hierarchical structure of the network \cite{Nartallo-Kaluarachchi2024BrokenNetworks} (with some work on non-normal networks referring to these networks as flow hierarchical \cite{OBrien2021HierarchicalNetworks}).

The fact that a matrix is non-normal can play a role in the stability of the dynamics and the sensitivity of the spectra to perturbations \cite{Asllani2018StructureNetworks,Trefethen2005SpectraOperators}. Matrix non-normality impacts many fields \cite{Trefethen2005SpectraOperators} including condensed matter physics, acoustics, the behaviour of numerical methods and fluid mechanics \cite{Trefethen2005SpectraOperators,Okuma2020HermitianPseudospectra,Gebhardt1994ChaosStability,Symon2018Non-normalityAnalysis,Sujith2016Non-normalityInstabilities}. Its  importance as a framework to understand directed networks has also been highlighted in network science \cite{Duan2022NetworkSystems,Asllani2018StructureNetworks,Sornette2023Non-normalBubbles,OBrien2021HierarchicalNetworks,Muolo2023PersistenceNetworks,Rodgers2023InfluenceNetworks,Baggio2020EfficientNon-normality, Nartallo-Kaluarachchi2024BrokenNetworks} where it has been linked to the behaviour of dynamics and the stability of complex systems. In addition, 
non-normality has been shown to be a very common phenomenon \cite{Asllani2018StructureNetworks,Johnson2020DigraphsSystems,MacKay2020HowNetwork,Duan2022NetworkSystems,Nartallo-Kaluarachchi2024BrokenNetworks} present in many systems. In this work we propose a fitness-based model which, by explaining the wide spectrum of trophic incoherence found in nature, also gives an explanation as to the ubiquity of non-normal networks. However, we also note that the relationship between non-normality and trophic incoherence proposed in \cite{Johnson2020DigraphsSystems,MacKay2020HowNetwork} is for a particular graph ensemble, although it fits well with real network data. As such there are examples of specific adjacency matrices where the relationship between trophic incoherence and matrix normality does not hold in the same way \cite{Nartallo-Kaluarachchi2024BrokenNetworks}, particularly in the case of weighted networks. In \cite{Nartallo-Kaluarachchi2024BrokenNetworks} small example network structures (some weighted) where this is the case are demonstrated, however, a correlation between trophic and non-normality based measured of directionality is still observed in the real-data used in \cite{Nartallo-Kaluarachchi2024BrokenNetworks}.

\subsection{Preferential Attachment and Fitness Models}

Preferential attachment is one of the main paradigms of network science and the scale-free property of networks has been of great interest to science since it was discovered, and it has been analysed and interpreted in various ways \cite{Zuev2016HamiltonianAttachment}. The discovery that preferential attachment based on degree can lead to scale-free networks was one of the great achievements of early network science \cite{Barabasi1999EmergenceNetworks}, and much of the field has been built on this work. Soon afterwards, however, it was noted that nodes may not attach based on only the degree a node but also based on the fitness of a node \cite{Bianconi2001CompetitionNetworks}. We base our study of degree-based preferential attachment on the Directed Scale-Free Model (DSF) \cite{Imae2018OnNetworks} which provides analytical expression for the power-law exponent in certain regimes in the directed case and highlights the impact of varying the ratio of incoming and outgoing edges to the newly added nodes.

Preferential attachment models can however have some caveats associated with them.  The initial conditions of the network can play a role and affect the equilibration time of the model \cite{Berset2013TheAttachment} or the scale-free effect may also be hidden by finite size affects in real systems \cite{Serafino2021TrueEffects}. Additionally, node age can play a role in the dynamics of networks grown with degree-based preferential attachment as the oldest nodes are most likely to gain the highest degree as the system evolves \cite{Sun2020Time-invariantModels}. This a natural place where fitness models can play a role as in many systems, such as citations networks. This is not realistic as the number of citations should decrease with the age of the work and the growth of the average number of citations should not depend on the time frame sampled \cite{Sun2020Time-invariantModels}. As such there are various methods to modify preferential attachment to account for this such as adding a time varying fitness, an ageing fitness parameter or a factor based on the age difference of the nodes \cite{Sun2020Time-invariantModels,Peng2022PreferentialAttractiveness,Medo2011TemporalNetworks,Golosovsky2018MechanismsModels}.  Additionally, given a real network the parameters that generated it may not be known and there has been work on how the fitness and preferential attachment parameters can be inferred from network growth data \cite{Pham2015PAFit:Networks,Pham2016JointNetworks}.

It has been shown \cite{Servedio2004VertexNetworks} that fitness interaction based preferential attachment and fitness distributions can lead to scale-free networks in the case of symmetric interactions and undirected networks, which helps to explain the ubiquity of the scale-free phenomenon
\cite{Caldarelli2002Scale-FreeFitness}. Network models using fitness interactions have been further studied in \cite{Bedognea2006ComplexFitness} with work on relating model properties to the degree distributions \cite{Smolyarenko2013NetworkFitness} and percolation in network models based on fitness interactions \cite{Hoppe2014PercolationResilience}. However, this work differs from the model we propose as we use a fitness offset to study the hierarchical nature of directed networks. Our model, in the case on only interactions based on degree, is most closely related to \cite{Imae2018OnNetworks}. Fitness based preferential attachment models can also be used to model specific phenomena by coupling the evolution of the network to dynamics taking place in the network \cite{Poncela2008ComplexAttachment}. 

Fitness based models without degree-based preferential attachment are also commonly used in ecology, motivated by the concept of a niche axis \cite{Loeuille2005EvolutionaryWebs}. 
However, it is also possible to make an attachment model which attaches based on the ratio of in and out degree \cite{Sevim2008NetworkOutdegree} designed to model food webs where you want to attach to nodes with high prey and low predator numbers. Niche models have limitations and it has 
been proposed that a one dimensional niche axis does not capture ecosystem complexity
\cite{Allesina2008AStructure}. Moreover, 
existing niche models fail to reproduce the trophic coherence of real food webs \cite{Johnson2014TrophicStability}. In this work, we use a one-dimensional fitness as we wish to study linear hierarchy, but fitness distributed in a different topological space could be an avenue of future work. In the study of social networks there is also a class of model called latent space models where nodes connect based on the distance between nodes in this space \cite{Hoff2002LatentAnalysis}, which may be higher dimensional \cite{Gaisbauer2023GroundingModels}. Our fitness space could be interpreted as latent space model with several modifications, such as the fact we use a growing network model with fixed edge number.

There has also been work on using degree-based preferential attachment based models and edge reciprocity to create networks of varying non-normality \cite{Asllani2018StructureNetworks}. In \cite{Asllani2018StructureNetworks} a model is introduced which uses Price's model \cite{DeSollaPrice1965NetworksPapers} as a base, to which they add an extra step to control the non-normality. In this model of directed network formation, when new nodes are introduced they connect into the existing network using degree-based preferential attachment in such a way which would create a directed acyclic graph as nodes are ordered by arrival time. However, there is an additional parameter which tunes the probability for the edge added to be reciprocal. This can then tune the amount of hierarchy in the system by adding edges which break the ordering. In \cite{Nartallo-Kaluarachchi2024BrokenNetworks}, Trophic Incoherence and non-normality were used to study the entropy production rate of various dynamics on networks and it was shown that this parameter could be linked to Trophic Incoherence found in the generated networks \cite{Nartallo-Kaluarachchi2024BrokenNetworks}. An extension of this model, with a modification to the placement probability of reciprocal edges based on the the inverse of the out-degree, was used to study the emergence of ``leader'' nodes (source/sink nodes depending on convention) \cite{OBrien2021HierarchicalNetworks} and their importance to dynamics in networks which are non-normal and hierarchical, where the position in the hierarchy is measured by steps from the ``leader nodes''.   

 Non-normality is also created in networks with preferential attachment in \cite{Sornette2023Non-normalBubbles} where the impact on of non-normality and hierarchy on the behaviour of financial bubbles and meme stocks was studied. In this model, they start with several top nodes and then introduce nodes via preferential attachment in a similar way to \cite{Asllani2018StructureNetworks,OBrien2021HierarchicalNetworks}, but with the condition that probability for an edge to be reciprocal is a function of the trophic level of a node. This is explained by the observation that in a Reddit discussion forum users with more popular comments are less likely to reciprocally comment with a user replying to them \cite{Sornette2023Non-normalBubbles}. However, this is not the case in all work on non-normal networks, since these can also be created by starting from a ring-like structure similar to that found in small-world network models and using edge weights to induce non-normality which does not involve preferential attachment, as in \cite{Asllani2018TopologicalSystems}. Other work  growing network models with varying reciprocity includes \cite{Cirkovic2024ModelingReciprocity}, where a network growth model was studied where the likelihood to reciprocate edges varies with different groups of users in social networks, something which is observed in real-network data \cite{Cirkovic2024ModelingReciprocity}.

\section{Fitness Difference Model}

\begin{figure}[H]
      		\centering
            \includegraphics[width=0.9\linewidth]{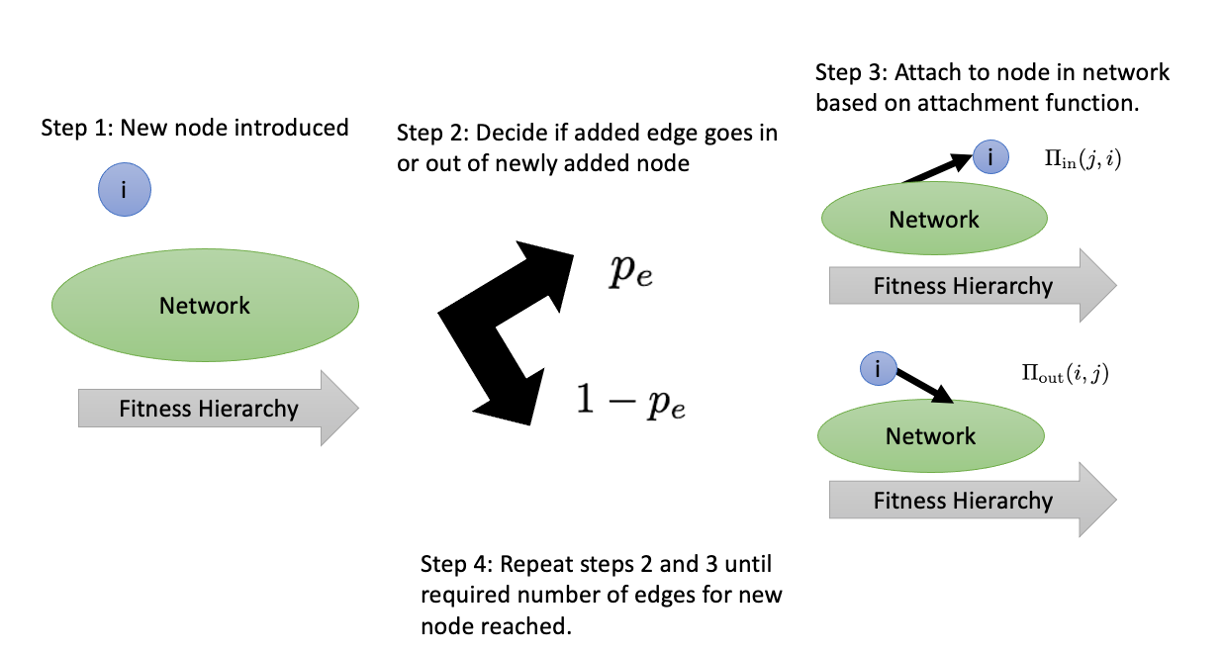}
        	\caption{ Schematic Diagram of the Generative model used.}
        	
        \label{fig:model_diagram}
        \end{figure}

In this work, we define a model of network formation where at each time step a new node is introduced and $m$ edges are added to it, which connect to the other nodes dependant on the fitness interaction and degree of the nodes. We define a probability, $p_e$, which is the probability that 
the edge currently being added points to the new node (incoming edge).
The edge points out of the new node (outgoing edge) with probability
$1-p_e$.
We use the value $p_e = 0.5$ unless otherwise stated so that the fraction of in and out edges are roughly similar most of the time and the distributions of in- and out- degrees are the same. This model is similar to the Directed Scale Free (DSF) model introduced in \cite{Imae2018OnNetworks} apart from 
the fact that we allow in the number of in- and out- edges to randomly fluctuate  and that we add the fitness interaction, as well as degree-based preferential attachment. Our model and the DSF model \cite{Imae2018OnNetworks} differ from previous work on models which lead to directed scale free networks \cite{Bollobas2003DirectedGraphs} in that the number of edges is known and fixed beforehand. We do not consider weighted networks in this work, but our model could be extended to the case of weighted networks by adding a mechanism to update the edge weights \cite{Pi2021AMechanism}, which could be made a function of fitness interactions.

A general way to formulate connection probability based on fitness is that edges connect out of the new node $i$ with probability \begin{equation} \label{eq:out_probablities}
    \Pi_{\text{out}}(i,j) = \frac{m(k^{\text{in}}_j) f(\beta_j,\beta_i)g_{\text{in}}(\beta_j)}{ \sum_q m(k^{\text{in}}_q) f(\beta_q,\beta_i)g_{\text{in}}(\beta_q)},
\end{equation} 
where $\beta_j$ is the fitness parameter of node $j$ which could be drawn from a fitness distribution or calculated from a property of the network; $m(\cdot) $ is a function which represents how degree is taken into account when selecting a node to attach; and $f(\beta_j,\beta_i)$ is a function which represents the interaction of the fitness of the two nodes which may attach and $g_{\text{in}}(\beta_j)$ is function representing the contribution of the fitness of node $j$ to the probability of attachment. 
With a similar equation and set of generic functions for edges incoming to the new node, 
\begin{equation} \label{eq:in_probablities}
    \Pi_{\text{in}}(j,i) = \frac{m(k^{\text{out}}_j) f(\beta_i,\beta_j)g_{\text{out}}(\beta_j)}{ \sum_q m(k^{\text{out}}_q) f(\beta_i,\beta_q)g_{\text{out}}(\beta_q)}. 
\end{equation}

The probabilities could also be defined as the probability of connections into/out of the network so in/out labels could be swapped \cite{Imae2018OnNetworks}. Care must be taken with the convention choice here.

We simplify to focus on the combination of degree-based preferential attachment, parameterised by $\alpha$, 
and functions $S(\cdot)$ of fitness differences 
(which may be offset to create hierarchy):
\begin{equation} \label{eq:edges_prob_1}
    \Pi_{\text{out}}(i,j) = \frac{(k^{\text{in}}_j)^\alpha S(\beta_j-\beta_i)}{\sum_q (k^{\text{in}}_q)^\alpha S(\beta_q-\beta_i)},
\end{equation}

\begin{equation} \label{eq:edges_prob_2}
    \Pi_{\text{in}}(j,i) = \frac{(k^{\text{out}}_j)^\alpha S(\beta_i-\beta_j)}{\sum_q (k^{\text{out}}_q)^\alpha S(\beta_i-\beta_q)}.
\end{equation}

The function $S(\beta_i-\beta_j)$ can always be multiplied by a constant as this will not change the behaviour of the model, as it will cancel in equations \ref{eq:edges_prob_1} and \ref{eq:edges_prob_2}. This allows $S(\beta_i-\beta_j)$ to be associated to a probability distribution if the normalisation constant is appropriately chosen. This is not necessary but this is something we will do later in this work in order to characterise the functions using standard features of probability distributions and to relate to the fitness differences observed in the generated networks.

Many fitness based ideas can be expressed as specific cases of this general model if the functions are selected to reproduce the desired behaviour. By setting the fitness function to one for all inputs we create a variant of the generalised BA preferential attachment to directed networks found in \cite{Imae2018OnNetworks}. Directed variants of homophilly models, inspired by the scale-free homophillic model \cite{DeAlmeida2013Scale-freeNetwork,Santos2020CriticalNetwork}, can be created by setting the offset parameter to zero and having the function $S$ representing the preferences of interaction between discrete groups labelled by fitness, \begin{equation}
    S(\beta_i-\beta_j ) = 1 - B_{ij},
\end{equation} 
where $B_{ij}$ takes a value between 0 and 1 depending on the interactions between groups. Models of this type have been used in the study of network inequality looking at the differing experiences of groups within the network \cite{Lee2019HomophilyNetworks,Nettasinghe2022Scale-freeNetworks}. It is also possible to produce variants of  homophilly models based on an exponentially decaying function to connect with fitness difference \cite{Gargiulo2017TheControversies}, which have been used to study the spread of opinion on network constructed with homophilly. This can be done be choosing 
    \begin{equation}
    S(\beta_j-\beta_i ) = \exp{(-c\lvert \beta_j - \beta_i \rvert}),
\end{equation}
where $\beta_j$ is node fitness and then $c$ is a non-negative control parameter. Our model can also link to the concepts used in the study of citation networks. By choosing the fitness of a node to represent time of publication we can create a model very similar in concept to those used in citation network analysis \cite{Hajra2005AgingNetworks} by choosing a function which decays with the time between the publication of the papers \begin{equation}
    S(\beta_j-\beta_i ) = {( t_j - t_i})^{-d}.
\end{equation} 
Here we need to place extra restrictions on the model so that $t_j$ is always greater than $t_i$ for two connected nodes and we preserve causality.

An offset could quite naturally be included in this fitness function as it can be difficult to cite a paper which is published very close to that date when your own is published. However, most of the models used in the literature based on the interaction between the fitness of nodes do not use an offset parameter, with exception being ecology. In ecology, it has been assumed that organisms have a niche profile \cite{Cai2020MutualisticInteractions}. This mean that they consume species at a certain fitness (approximated by body size) below their own level. 
For example, this could be when $S(\beta_j-\beta_i)$ is a Gaussian in the fitness differences. This is \begin{equation}
    S(\beta_j-\beta_i) \propto \exp{\left[ -\frac{1}{2}\left(\frac{\beta_j-\beta_i - \mu_f}{\sigma_f}\right)^2 \right]},
\end{equation} 
which is very similar to the one used to model the niche width in ecology in the influential paper where it was introduced \cite{Loeuille2005EvolutionaryWebs}.

In this work, we assume the fitness of each node is uniformly distributed between 0 and 10 but there is no reason why this restriction has to be in place and any range will work if the parameters are modified accordingly. Node fitness could be distributed according to an exponential, Gaussian or any other physically relevant distribution if that better reflects the system being represented. A non-uniform distribution could be used to modify the degree distribution, to modify the size of the community structures, to create variations in the edge density across the fitness scale or to represent the time distribution of events. It has been shown that using log-normally distributed fitness and a simpler attachment function based only on fitness can lead to power-law degree distributions \cite{GhadgeLogNormal}, which is an example of the range of possibility that could be induced by varied fitness distributions. As such we expect a variety of complex effects could be observed by mixing various attachment functions with non-trivial fitness distributions.

\subsection{Fitness Functions Used}\label{all:equations}

Many possible functions could be chosen to represent the interaction between the fitness parameters of the nodes. In general, we need functions which are defined for positive and negative fitness differences. However, in order to induce hierarchy and for simplicity we choose functions that can be normalised to standard unimodal distributions and for mathematical convenience have a well defined mean and standard deviation, apart from in some specific cases which we use as examples. Any normalisation will cancel in the computation of the edge addition probabilities, equations \ref{eq:edges_prob_1} and \ref{eq:edges_prob_2}. However, we give the form normalised to a probability density function associated with the functions where it is useful to show that the input fitness function can reflect the empirically measured distribution of fitness differences; and so that we can use the standard analytical expressions for mean, $\mu$, and standard deviation, $\sigma$, of the distributions. 

The simplest function we use is a function which represents a Gaussian distribution
\begin{equation} \label{eq:gaussian}
    S(\beta_j-\beta_i) = \frac{1}{\sqrt{2\pi \sigma_f^2}}\exp{\left[ -\frac{1}{2}\left(\frac{\beta_j-\beta_i - \mu_f}{\sigma_f}\right)^2 \right]},
\end{equation}
where the mean $\mu = \mu_f$ and standard deviation $\sigma=\sigma_f$.

We also introduce the Laplace distribution, which has a more pronounced peak than a Gaussian but decays more slowly
(wider, longer tails) 
\begin{equation}\label{eq:Laplace}
    S(\beta_j-\beta_i) = \frac{1}{2b}\exp{\left[ -\left\lvert\frac{\beta_j-\beta_i - \mu_f}{b} \right\rvert \right]},
\end{equation}
where the mean $\mu = \mu_f$ and standard deviation $\sigma=\sqrt{2}b$.

The Laplace and Gaussian distributions can also be encompassed by the generalised normal distribution, which allows the exponent to be varied and distributions of different shapes to be created. When the exponent $\nu$ is very large the distribution tends to a top-hat shape,
\begin{equation}\label{eq:gen_normal1}
    S(\beta_j-\beta_i) = \frac{\nu}{2b\Gamma(\frac{1}{\nu})}\exp{\left[ -\left\lvert\frac{\beta_j-\beta_i - \mu_f}{b} \right\rvert^\nu \right]},
\end{equation}
where the mean $\mu = \mu_f$ and standard deviation $\sigma= b \sqrt{\frac{\Gamma(\frac{3}{\nu})}{\Gamma(\frac{1}{\nu})}}$. With $\Gamma(x)$ being the standard gamma function of $x$.

We also include distributions which can be skewed such as the Gumbel distribution that can used to model extreme values such as  the maximum of a sample (the form used here),
\begin{equation} \label{eq:gumbel}
    S(\beta_j-\beta_i) = \frac{1}{b}\exp{\left[ -\left( \frac{\beta_j-\beta_i - \mu_f}{b}  + \exp{\left(-\frac{\beta_j-\beta_i - \mu_f}{b} \right)}  \right) \right]},
\end{equation}
where the mean $\mu = \mu_f + b\gamma_{\text{E}} $ with $\gamma_{\text{E}}$ being the Euler–Mascheroni constant and the standard deviation $\sigma=\frac{\pi b}{\sqrt{6}}$.

We further include the exponentially modified Gaussian distribution which reflects the distribution of the sum of normally and exponentially distributed random variables and can be skewed. This distribution is given by 

\begin{equation}\label{eq:exp_gaussian}
    S(\beta_j-\beta_i) = \frac{\lambda}{2}\exp{\left[ -\frac{\lambda}{2}\left( -2(\beta_j-\beta_i ) + \lambda \sigma_{e}^2 + 2 \mu_{f}\right) \right]} \erfc{\left[ \frac{-(\beta_j-\beta_i ) + \lambda \sigma_{e}^2 +  \mu_{f}}{\sqrt{2}\sigma_{e}}\right ]},
\end{equation}
where the mean $\mu = \mu_f + \frac{1}{\lambda}$ and standard deviation $\sigma=\sqrt{\sigma_e^2 + \frac{1}{\lambda^2}} $.

We also introduce several fitness functions, equations \ref{eq:tanh}, \ref{eq:absolute} and \ref{eq:uniform}, which do not have a single peak and do not lead to the same kind of hierarchical structure which we use when studying the distribution of trophic levels with fitness, section \ref{Sec:trophic_level_distrubtions}. Firstly, we introduce the hyperbolic tangent function of the fitness differences which plateaus to a constant value for large positive fitness differences, 
\begin{equation}\label{eq:tanh}
    S(\beta_j-\beta_i) = {C_1\left(\tanh{\left(\frac{\beta_j-\beta_i - a}{T_1}\right)} +1\right)},
\end{equation}
where $C_1$ is an appropriate normalisation constant found by integrating over the fitness range and $T_1$ is a parameter which sets how sharp the transition in the hyperbolic tangent function is with fitness difference.

We additionally include a fitness function where edges try to maximise the absolute value of the fitness difference
\begin{equation}\label{eq:absolute}
    S(\beta_j-\beta_i) = C_2\left\lvert\frac{\beta_j-\beta_i}{T_2}\right\rvert,
\end{equation}
where $C_2$ is an appropriate normalisation constant found by integrating over the fitness range and $T_2$ is a scaling parameter to set the growth of the penalty.

Finally, we have a fitness interaction which is uniform which does not create any structure based on fitness.
\begin{equation}\label{eq:uniform}
    S(\beta_j-\beta_i) = \text{constant},
\end{equation}
where $constant$ can be chosen as the inverse of the span of the fitness range if fitness normalisation is required.

\section{Results}

\subsection{Degree Distributions}

\begin{figure}[H]
    \centering
\begin{subfigure}[t]{0.4\textwidth}    
    \includegraphics[width=\textwidth]{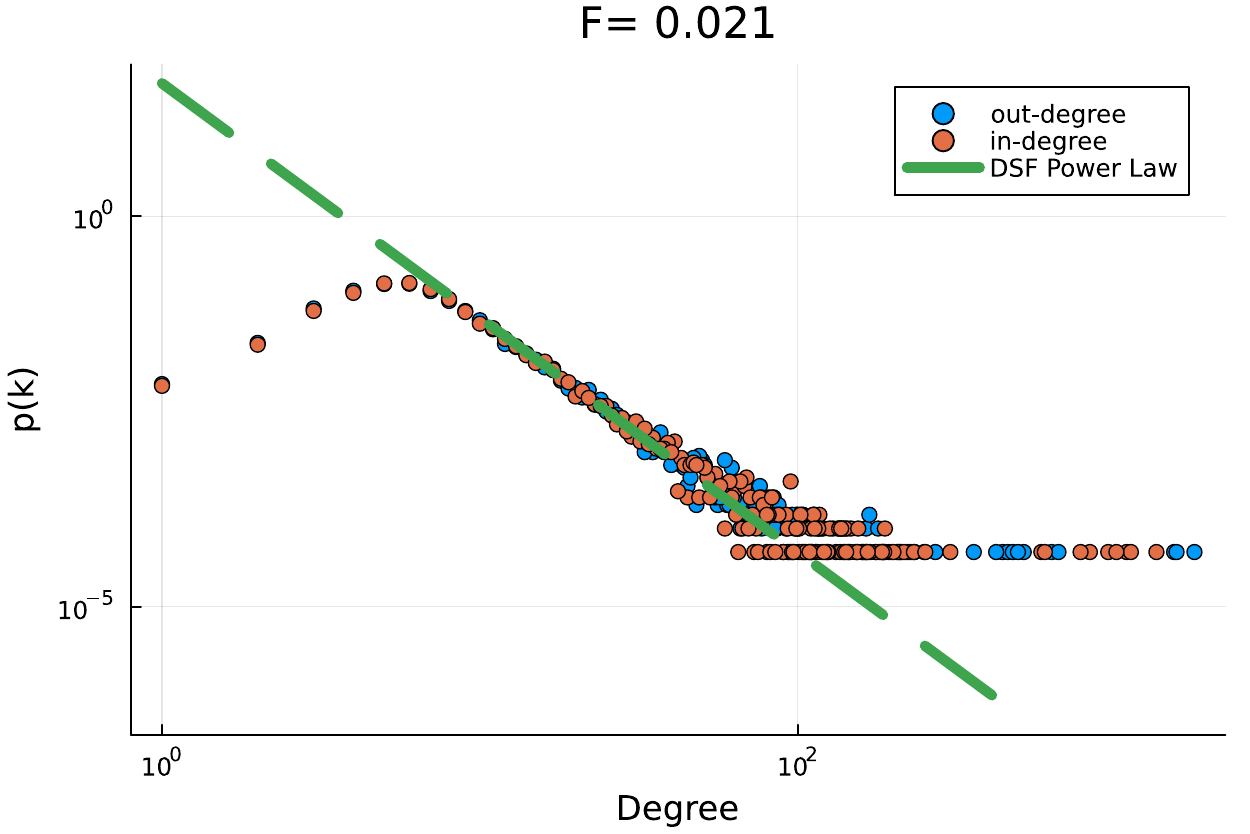}
    \caption{ Degree Probability Distribution Functions. Gaussian Function. Equation \ref{eq:gaussian} $\mu_f= 1$, $\sigma_f= 0.1$. }  	
    \label{fig:pdf_pref_gaussain_low_T}
    \end{subfigure} 
    \hfill
\begin{subfigure}[t]{0.45\textwidth}
    \centering
        \includegraphics[width=\textwidth]{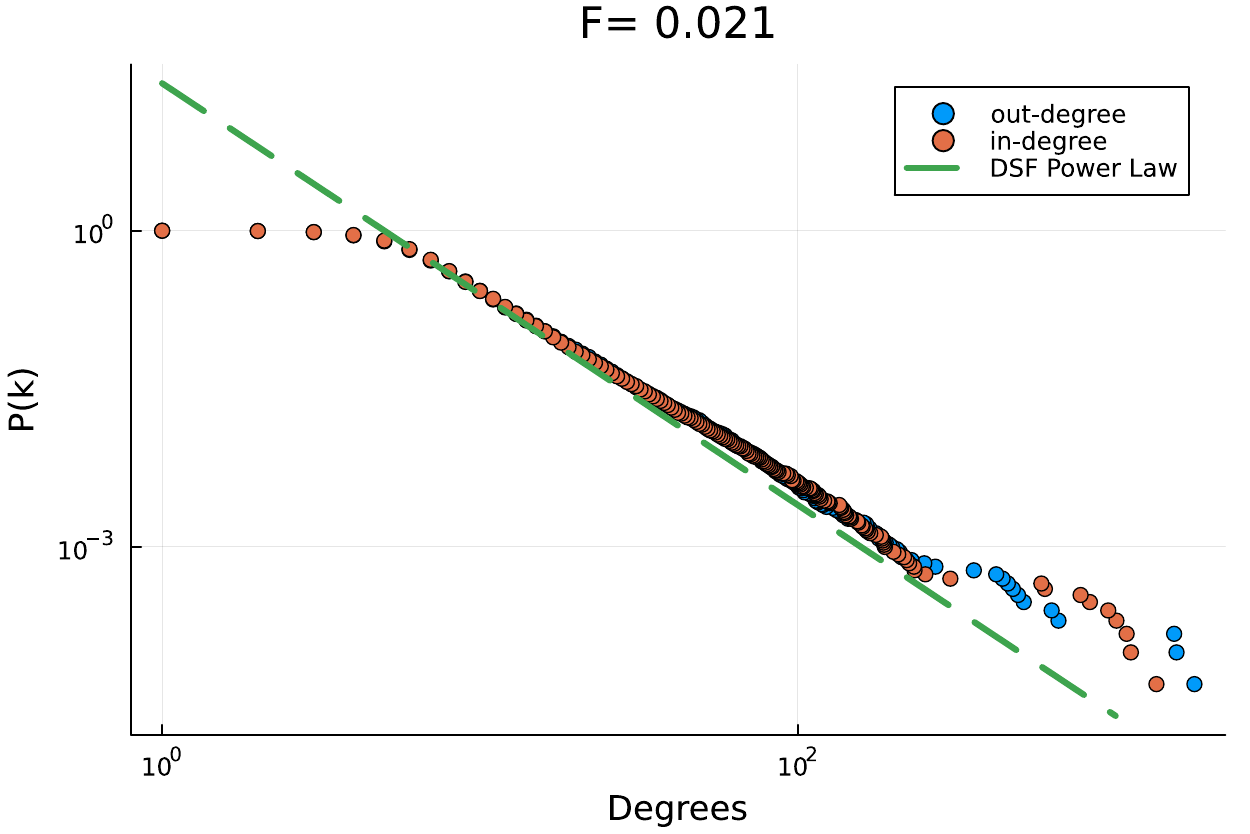}
        \caption{ Degree Cumulative Distributions. Gaussian Function. Equation \ref{eq:gaussian} $\mu_f= 1$, $\sigma_f= 0.1$.}
        \label{fig:cdf_pref_gaussain_low_T}
        \end{subfigure} 
         \hfill
\begin{subfigure}[t]{0.45\textwidth}    
    \includegraphics[width=\textwidth]{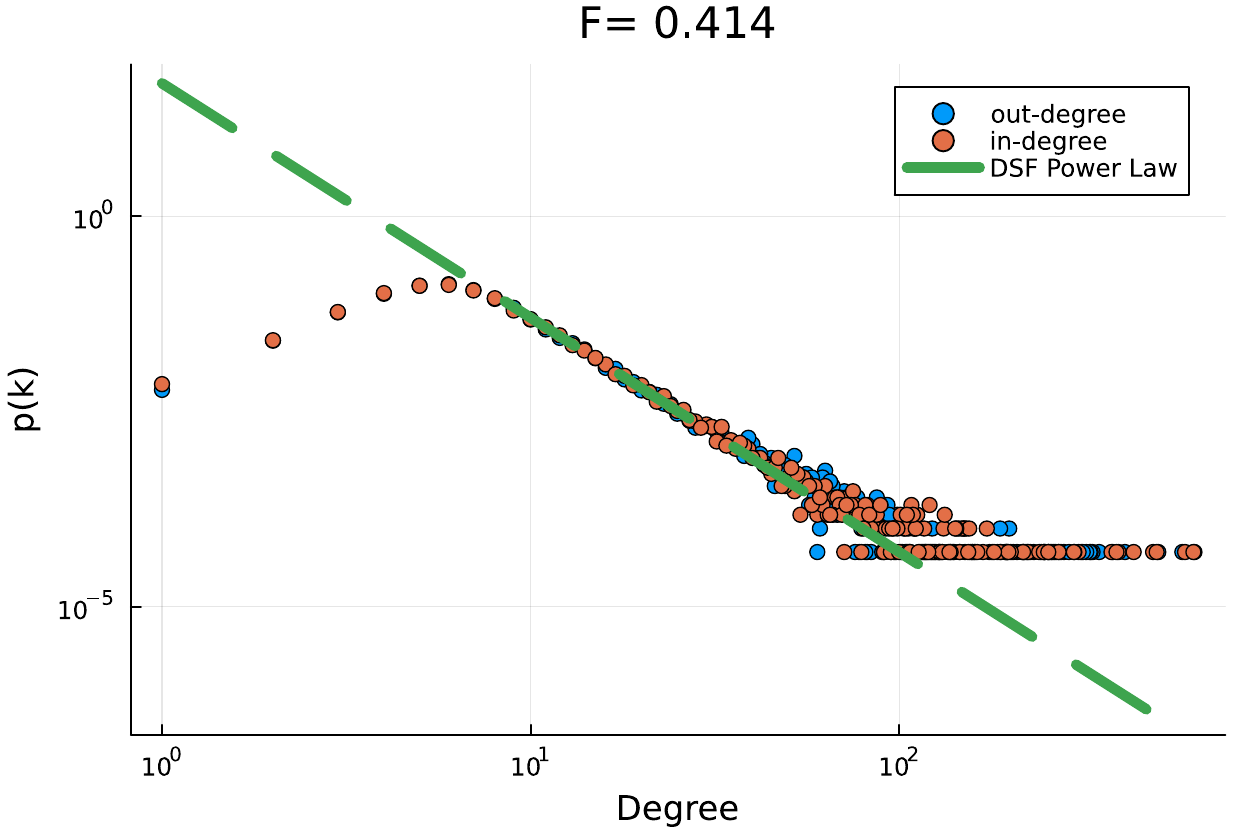}
    \caption{ Degree Probability Distributions. Gaussian Function. Equation \ref{eq:gaussian} $\mu_f= 1$, $\sigma_f= 1$.}  	
    \label{fig:pdf_pref_gaussain_med_T}
    \end{subfigure} 
    \hfill     
\begin{subfigure}[t]{0.45\textwidth}
    \centering
        \includegraphics[width=\textwidth]{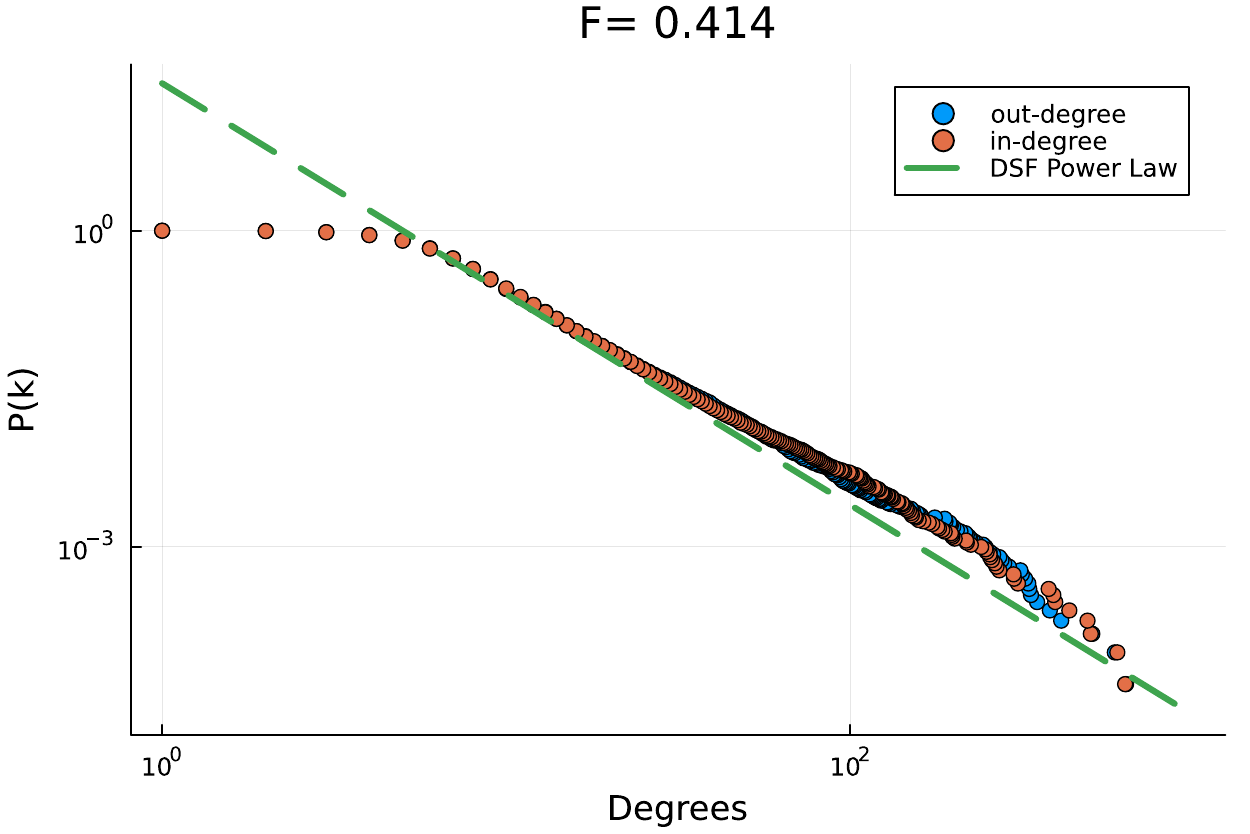}
        \caption{ Degree Cumulative Distributions. Gaussian Function. Equation \ref{eq:gaussian} $\mu_f= 1$, $\sigma_f= 1$. }
        \label{fig:cdf_pref_gaussain_med_T}
        \end{subfigure} 
\begin{subfigure}[t]{0.45\textwidth}    
    \includegraphics[width=\textwidth]{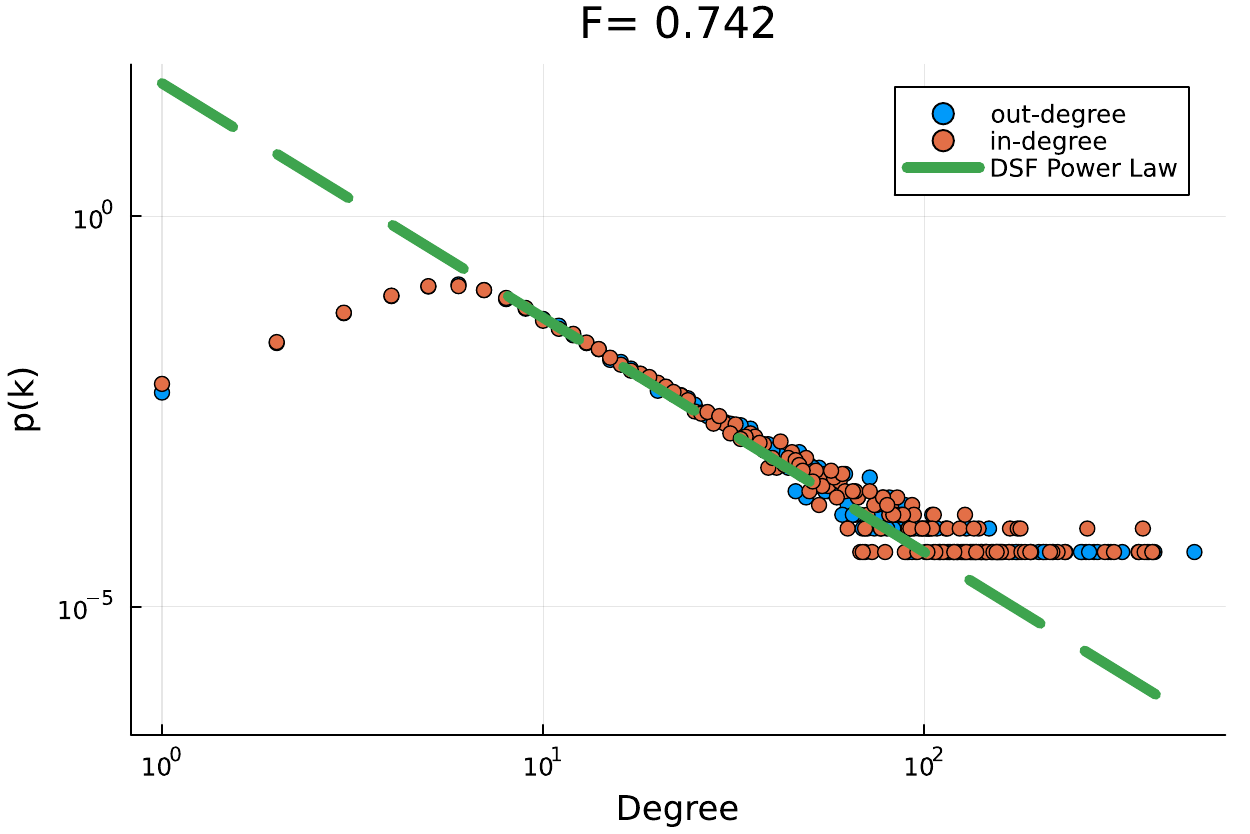}
    \caption{ Degree Probability Distributions. Preferential Attachment only. $S(\beta_j-\beta_i) = 1$ }  	
    \label{fig:pdf_pref_only}
    \end{subfigure} 
    \hfill     
\begin{subfigure}[t]{0.45\textwidth}
    \centering
        \includegraphics[width=\textwidth]{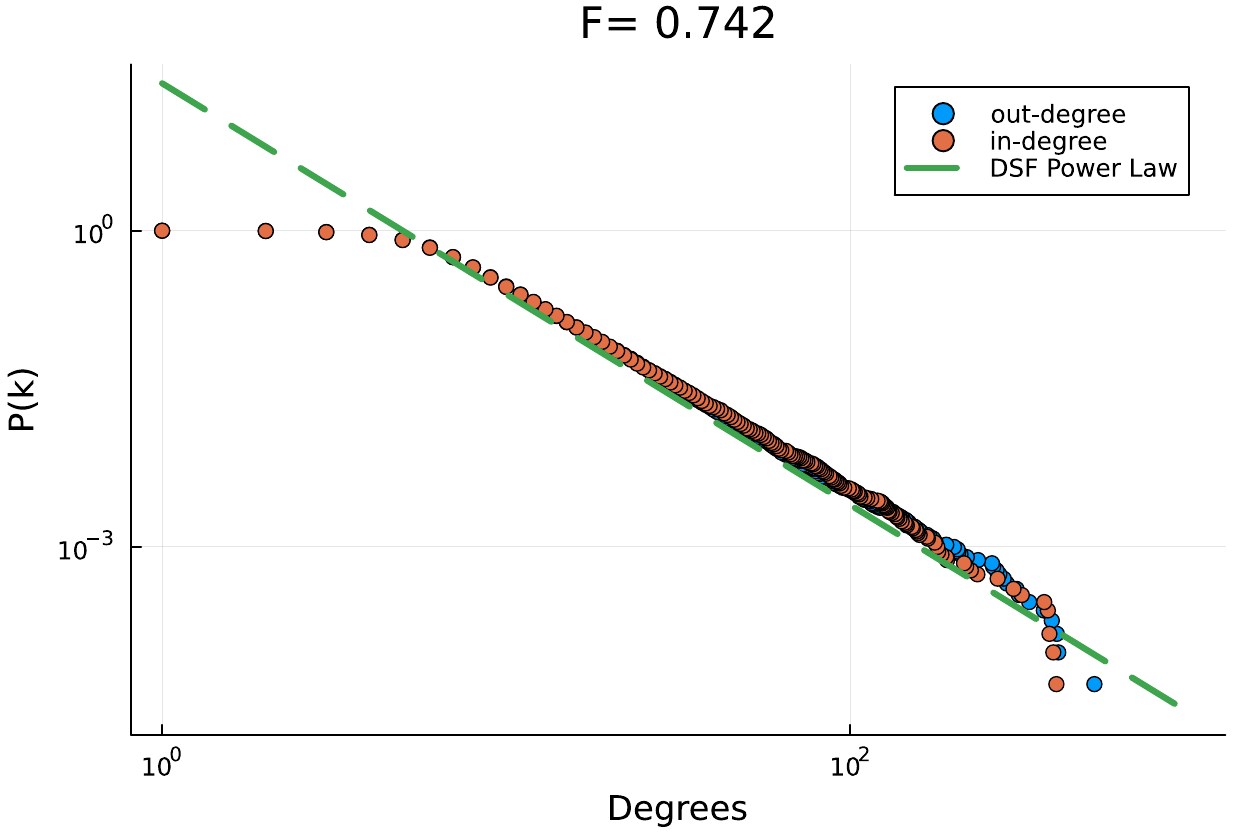}
        \caption{ Degree Cumulative Distributions. Preferential Attachment Only. $S(\beta_j-\beta_i) = 1$.}
        \label{fig:cdf_pref_only}
        \end{subfigure}   
\caption{Degree distributions of networks with Fitness Functions and Preferential Attachment. $N=20000$, $m =10$, $\alpha=1$. Seed graph is a path of length $m=10$. Power-law for derived from DSF model \cite{Imae2018OnNetworks} in the case of equal probability of an edge being incoming or outgoing to a new node, equations \ref{eq:DSF_in_power_law} and \ref{eq:DSF_out_power_law}.}
\label{fig:degre_distrubtions}
\end{figure}

Firstly, we study how the addition of the hierarchy affects the degree distribution of the networks compared to the preferential attachment without the imposition of hierarchy. This is shown in figure \ref{fig:degre_distrubtions}. We plot the probability and cumulative degree distribution for networks with $\alpha=1$ and a Gaussian fitness function of various standard deviations and compare to degree-based preferential attachment alone where the fitness function is constant. We choose $\alpha= 1$ so we can plot our results against the analytical power law for the Directed Scale-Free (DSF) model \cite{Imae2018OnNetworks} which our model is very similar to. This model generalises the BA model to generate directed scale-free networks with fixed number of nodes and edges as well as control of the number of in and out edges added with each new node. The degree exponents for this model are a function of the ratio of in and out edges added. The mean-field degree distribution in this model can be calculated by taking the continuous approximation so that the evolution of the node degree \cite{Imae2018OnNetworks} can be written as a set of differential equations, given by 

\begin{equation}
    \frac{dk^{\text{in}}_j}{dt} = m_{\text{in}}\frac{k^{\text{in}}_j} {\sum_q k^{\text{in}}_q},
\end{equation}

\begin{equation}
    \frac{dk^{\text{out}}_j}{dt} = m_{\text{out}}\frac{k^{\text{out}}_j}{\sum_q k^{\text{out}}_q}.
\end{equation}

Which leads to degree probability distributions which can be approximated as \cite{Imae2018OnNetworks} \begin{equation} \label{eq:DSF_in_power_law}
\begin{aligned}
 p(k^{\text{in}}) &\sim (1 + \frac{m_{\text{in}}}{m_{\text{out}}})m_{\text{in}}^{(1 + \frac{m_{\text{in}}}{m_{\text{out}}})}(k^{\text{in}})^{-(2 + \frac{m_{\text{in}}}{m_{\text{out}}})}, \\
p(k^{\text{in}}) &\sim  (k^{\text{in}})^{-\gamma^{\text{in}}}, 
\end{aligned}
\end{equation}
where the in-degree exponent $\gamma^{\text{in}}= 2 + \frac{m_{\text{in}}}{m_{\text{out}}}$ and 
\begin{equation} \label{eq:DSF_out_power_law}
\begin{aligned}
 p(k^{\text{out}}) &\sim (1 + \frac{m_{\text{out}}}{m_{\text{in}}})m_{\text{out}}^{(1 + \frac{m_{\text{out}}}{m_{\text{in}}})}(k^{\text{out}})^{-(2 + \frac{m_{\text{out}}}{m_{\text{in}}})} \\
p(k^{\text{out}}) &\sim   (k^{\text{out}})^{-\gamma^{\text{out}}} 
\end{aligned}
\end{equation}
where the out-degree exponent $\gamma^{\text{out}}= 2 + \frac{m_{\text{out}}}{m_{\text{in}}}$, following the notation convention of \cite{Imae2018OnNetworks}.

In the case where the in and out degree of the added nodes are the same the exponent reduces to the same value as in an undirected BA network, $\gamma^{\text{out}}= \gamma^{\text{in}}= 3$, and the power law is the same for both in and out degree. This is the regime that we work in and the power-law that we plot as we set the probability of an incoming or outgoing edge to be equal.

The main difference between our model and the DSF model, apart from the imposition of a fitness function, \cite{Imae2018OnNetworks} is that we do not fix the number of in and out edges added to the new nodes and instead add incoming or outgoing with a prescribed probability. So in the equation for the power-law we replace the number of in and out edges added with the expected value of their ratio. This is simply an approximation which works well enough when we add $m=10$ edges to each node but may fail when the number of edges added is lower or the  edge type probability is imbalanced, so there are relatively larger fluctuations about this ratio.

Despite the addition of the hierarchy, at standard deviations small enough to induce hierarchy of the fitness functions, the power-law from the DSF model is quite well reproduced, figure \ref{fig:degre_distrubtions}, in a similar way to when we only have degree-based preferential attachment and no fitness, figures \ref{fig:pdf_pref_only} and \ref{fig:cdf_pref_only}. 
This makes sense as imposing the fitness hierarchy does not change which nodes grow in popularity due to the preferential attachment, it merely restricts the range of nodes which can be connected to it by splitting the network into different fitness regions. 
This may lead to finite size affects being more pronounced, particularly in the case of very strict fitness functions as the effective network seen by a node is reduced in size even if the same power-law behaviour happens in that subset. An analytical argument of why the hierarchical fitness functions have little impact of the power-law behaviour is given in the next section, section \ref{Analytical_arg_degree_dist}. There is some difference in the behaviour of the tails of the power-laws between the networks of stronger hierarchy. This may be due to fitness edge affects which we explore further in section \ref{Degree_imbalnce_section}. When the hierarchy is very strict nodes at the very top of the hierarchy have no other nodes above them to which a connection is preferred, so will have very low out-degree and while connections from below are not penalised by the fitness function leading to high degree imbalance. A very similar effect occurs at the bottom of the fitness hierarchy with the sharp edge of the fitness distribution limiting the range of connections available.  

The numerical tests in figure \ref{fig:degre_distrubtions} demonstrates that fitness hierarchy can coexist with a scale-free degree distribution, which makes sense as these phenomena are both known to be features of real networks.

\subsubsection{Mean-Field Analysis of Degree Distribution} \label{Analytical_arg_degree_dist}

We now conduct mean-field analysis to justify why the impact of the fitness function on the formation of the power-law degree distribution is small by showing the mean-field calculation approximately reduces to the same equations as found in  the DSF model \cite{Imae2018OnNetworks}. In the numerical case, we used finite sized networks and a finite span of fitness values where it worked well, however in order to conduct the mean-field analysis we need to make several assumptions regarding large networks sizes and range of fitness values which we detail below. This difference means that the effects highlighted by the mean-field may exist outside the regime of parameters where it can be conveniently calculated. 

By taking a continuous approximation for the degree evolution of our model, as network size tends to infinity, we can write a mean-field expression for evolution of the in-degree of a node with fitness $\beta_j$ as 

\begin{equation} \label{eq:mean_field}
    \frac{dk^{\text{in}}_{j, \beta_j}}{dt} = m_{\text{in}}\frac{k^{\text{in}}_j \int_{\beta_{\text{min}}}^{\beta_{\text{max}}} d\beta' \ S(\beta_j - \beta' -a) \rho(\beta')}{\sum_q k^{\text{in}}_q \int_{\beta_{\text{min}}}^{\beta_{\text{max}}} d\beta'' \ S(\beta_q - \beta'' -a) \rho(\beta'')},
\end{equation}
where $\rho(\beta)$ is the fitness density distribution. In this case, we write the fitness interaction function as $S(\beta_j - \beta' -a)$ to explicitly highlight the fact that we are using a non-zero fitness offset, given by the parameter $a$, which can be accounted for in the change of variable required. We also select a fitness function which can be normalised to a probability distribution so that it can be quantified via the standard deviation. A similar expression and for which the following arguments hold identically can be constructed for the out-degree.  

The expression given, equation \ref{eq:mean_field},  is very similar to the expression which arises in the DSF model \cite{Imae2018OnNetworks} apart form the addition of the integral involving the fitness. This may explain why in some cases the DSF power-law exponent approximately holds when a hierarchical ordering function is imposed. In the case of uniform fitness, $\rho(\beta)$ is a constant for all $\beta$. So this cancels leading to the expression, \begin{equation}
    \frac{dk^{\text{in}}_{j, \beta_j}}{dt} = m_{\text{in}}\frac{k^{\text{in}}_j \int_{\beta_{\text{min}}}^{\beta_{\text{max}}} d\beta' \ S(\beta_j - \beta' -a)}{\sum_q k^{\text{in}}_q \int_{\beta_{\text{min}}}^{\beta_{\text{max}}} d\beta'' \ S(\beta_q - \beta'' -a)},
\end{equation}
which differs from the DSF equation only by the integrals of the form \begin{equation}
    I(\beta_j) = \int_{\beta_{\text{min}}}^{\beta_{\text{max}}} d\beta' \ S(\beta_j - \beta' -a).
    \end{equation}
This integral can be analysed by making the change of variables $x= \beta_j - \beta' -a$, leading to \begin{equation}
    I = -\int_{x_1}^{x_2} dx \ S(x),
\end{equation}
where $x_1= \beta_j - \beta_{\text{min}} -a$ and $x_2= \beta_j - \beta_{\text{max}} -a$. $x_1$ and $x_2$ can be rewritten as multiples of the standard deviation of the normalised function $S(x)$ and the limits reversed to absorb the sign. This leads to \begin{equation}
    I = \int_{\theta_2\sigma_f}^{\theta_1\sigma_f} dx \ S(x),
\end{equation}
where \begin{equation}
    \theta_1 = \frac{ \beta_j - \beta_{\text{min}} -a}{\sigma_f}
\end{equation} 
and 
\begin{equation}
    \theta_2 = \frac{  \beta_j - \beta_{\text{max}} -a }{\sigma_f}.
\end{equation} 
Due to the steup, $\theta_2$ is always negative as $\beta_j \leq \beta_\text{max}$.

We wish to argue that the integral, $I$, is approximately a constant for the majority of values of $\beta_j$ and hence will approximately cancel in the mean-field equations leading to the same set of equations as in the DSF model with fitness \cite{Imae2018OnNetworks}. This can be done by considering the assumptions of the continuum approximation and the properties of the function, $S(x)$. One of the assumptions of mean-field theory is that the network size, $N>>0$, as $t>>0$ as such we can also assume that the network has many layers and there is a large number of trophic levels induced by a large range of uniformly distributed fitness values and that $\frac{\beta_{\text{max}}}{\sigma_f} >> 1$.
We also assume that the mean-field theory only covers nodes in the bulk of the fitness range, nodes which are far from either the maximum or the minimum of the fitness. This means we can assume that $ \theta_1 >> 1$ and  $\lvert \theta_2 \rvert >> 1$. As the fitness is uniformly distributed, the fraction of nodes not in this regime is the number of nodes within $2l$ standard deviations of the fitness endpoints at either $\beta_{\text{max}} $ or $\beta_{\text{min}}$. This is given by \begin{equation} \label{eq_2std}  
    2l\frac{\sigma_f}{\beta_{\text{max}} - \beta_{\text{min}}} = 2l\frac{\frac{\sigma_f}{\beta_{\text{max}}}}{1 - \frac{\beta_{\text{min}}}{\beta_{\text{max}}} },
\end{equation} where $l$ is a finite positive parameter setting how close to the end points are nodes to be neglected. When the fitness span and number of levels becomes very large, $\frac{\beta_{\text{max}}}{\sigma_f} >> 1$, this quantity becomes very small and the contribution from these nodes to the degree distribution can be neglected. An alternative view of eq. \eqref{eq_2std} is that its LHS contains a ratio of the fitness spread (expressed as the standard deviation $\sigma_f$ of the normalised fitness interaction function $S(x)$) and the $\beta$-value spread $\beta_{\text{max}} - \beta_{\text{min}}$. If the spread of $\beta$ values is much larger than the fitness spread, the term (and hence the fraction of nodes not in the regime needed for our approximation) becomes negligible.

If the function $S(x)$ is uni-modal with finite mean and the function decays quickly enough for values away from the mean then as long as the fitness satisfies the constraints that $ \theta_1  >> 1$ and  $\lvert \theta_2 \rvert >> 1$, the integral will give approximately the same value for any fitness $\beta_j$. This is due to the fact that the tails of the distribution contribute very little to the value of the integral as the fitness function goes to zero for large differences and we obtain the main contribution from the peak of $S(x)$. Meaning that the integral can be cancelled in the mean field equations and we return to the DSF case. This assumption works well if $S(x)$ is a Gaussian function for example, as the contributions to the integral become small after a few standard deviations. Some of examples of simple unnormalised unimodal functions, $\tilde{S}(x)$, are given in table \ref{tab:example_functions}, where it can be seen that the integral value quickly approaches a constant when the values of the limits move far away from zero.

\begin{table}[h]
\centering
    \begin{tabular}{ || c || c || c || } 
    \hline ~\\[-2ex]
    \boldmath$ \tilde{S}(x)$ &   \boldmath$\int_{\theta_2\sigma_f}^{\theta_1\sigma_f} dx \ \tilde{S}(x) $ & \boldmath$\int_{-\infty}^{+\infty} dx \ \tilde{S}(x) $ \\ [7pt] \hline  ~\\[-2ex]
    $e^{-\lvert x \rvert}$ &  $2 - e^{(- \theta_1\sigma_f)}  - e^{(- \lvert \theta_2 \rvert \sigma_f)}  $ & $ 2$\\ [7pt] \hline ~\\[-2ex]
    $e^{-x^2}$ &   $\frac{\sqrt{\pi}}{2}( \erf{(\theta_1\sigma_f}) + \erf{(\lvert \theta_2 \rvert \sigma_f))}$ & $\sqrt{\pi}$ \\ [7pt]\hline ~\\[-2ex]
    $\frac{1}{1 + x^2}$ & $\tan^{-1}(\theta_1\sigma_f) + \tan^{-1}(\lvert \theta_2 \rvert \sigma_f) $ & $\pi $\\ [7pt]
    \hline
    \end{tabular}
        \caption{Examples of functions that when chosen as $\tilde{S}(x)$ satisfy the required property of converging to an approximate constant when integrated with the upper limit much greater than zero and the lower limit much less than zero.}
        \label{tab:example_functions}
\end{table}

However, this can also be more rigorously quantified for a more general class of functions. If we assume that function $S(x)$ is normalised such that it represents a probability density function which gives the probability to observe a fitness difference, $x$, we can treat the analysis of the tails of the integral, $I$, as the probability of observing large fitness differences. In figure \ref{fig:level_distrubtions} we show that for a variety of fitness functions the  empirically measured fitness difference distribution is well represented by the probability density, when  $\alpha=0$. $S(x)$ can be normalised by any constant as it cancels in the edge addition probability as previously stated. We can then use Chebyshev's inequality, which states that for a random variable $X$ with mean $\mu_1$ and standard deviation $\sigma_1$, the probability of making an observation $\kappa$ times the standard deviation away from the mean is \begin{equation}
    \text{Pr}(\lvert X - \mu_1\rvert \geq \kappa \sigma_1) \leq \frac{1}{\kappa^2}.
\end{equation}

This means fitness differences which are many standard deviations away from the mean are very unlikely to be observed. If we assume the that integral, $I$, represents the integral over the probability density function of observing certain fitness differences then modification of the integral limits $\theta_1$ and $\theta_2$ when they are very large in absolute value is unlikely to impact the value of the integral and hence $I$ can be said to be approximately constant for any fitness, $\beta_j$, far from the edge of the fitness range. An identical calculation can be done for the out-degree simply with the function arguments and labelling of the degrees swapped. This argument is simply an approximation to explain what we observe in the numerical data in particular we do not expect this argument to hold when the distribution of fitness is non-uniform, for model parameters resulting in $S(x)$ not having the properties detailed above or for all possible values of $\alpha$. Even though this arguments is rough and approximate, as in the previous section, it makes sense that imposing a hierarchy in this way should not affect the degree distribution, aside form imposing finite size affects. Indeed, we do not impose any restriction on which nodes grow in degree - instead of a single large network the system is now organised through a linearly ordered fitness space where edges only connect within a specific region but degree-based preferential attachment can occur as normal in that range.

\subsubsection{Degree Imbalance and Trophic Level} \label{Degree_imbalnce_section}

\begin{figure}[H]
    \centering
\begin{subfigure}[t]{0.48\textwidth}
    \centering
        \includegraphics[width=\textwidth]{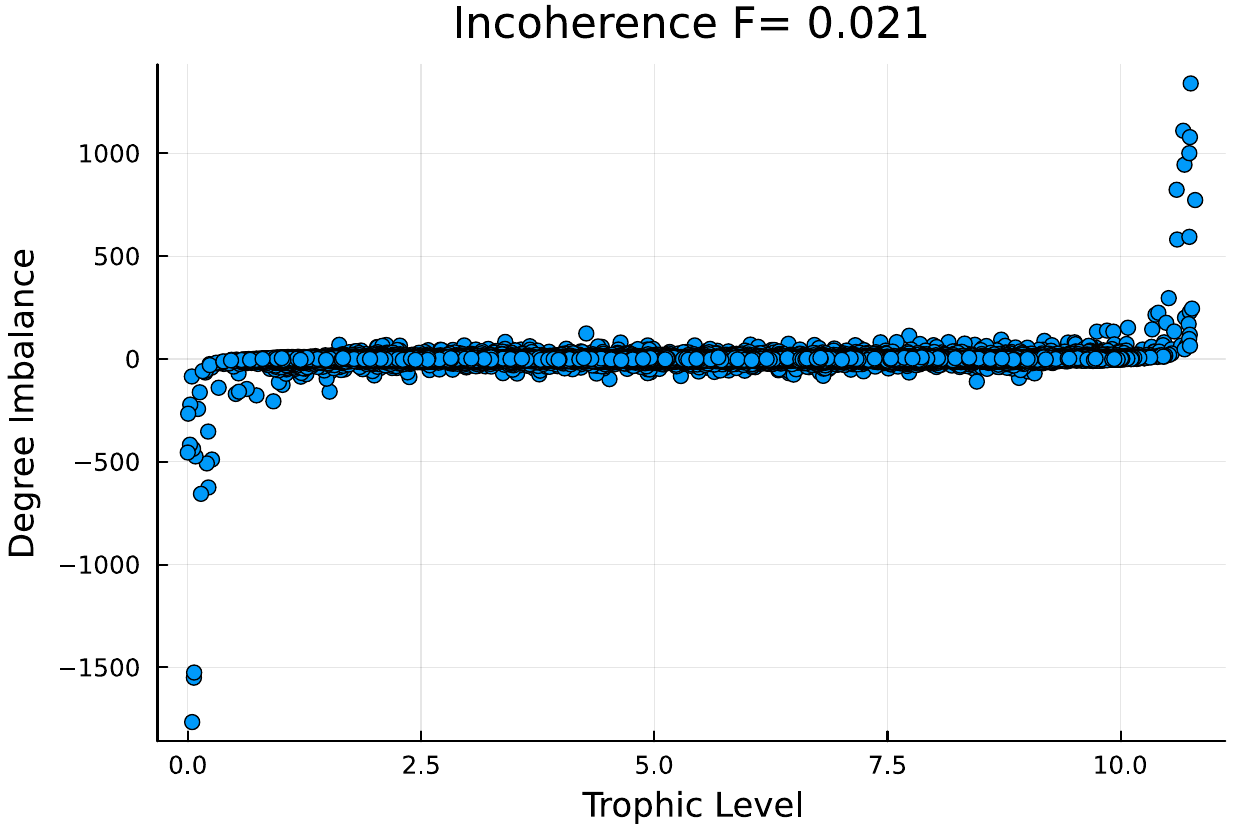}
        \caption{ Degree Imbalance (in-degree minus out-degree) with Trophic Level. Gaussian Function. Equation \ref{eq:gaussian} $\mu_f= 1$, $\sigma_f= 0.1$.}
        \label{fig:imb_pref_gaussain_low_T}
        \end{subfigure} 
         \hfill
\begin{subfigure}[t]{0.48\textwidth}
    \centering
        \includegraphics[width=\textwidth]{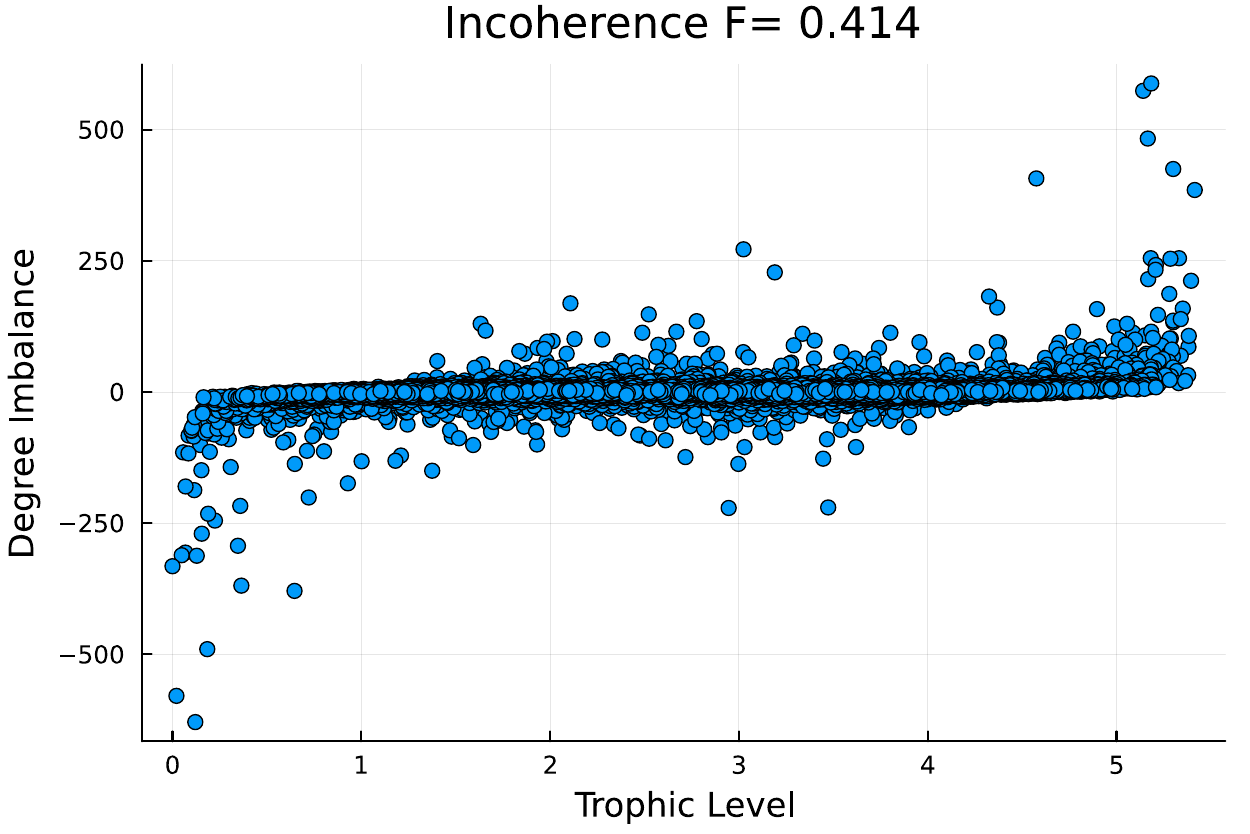}
        \caption{ Degree Imbalance (in-degree minus out-degree) with Trophic Level. Gaussian Function. Equation \ref{eq:gaussian} $\mu_f= 1$, $\sigma_f= 1$.}
        \label{fig:imb_pref_gaussain_med_T}
        \end{subfigure} 
\hfill     
\begin{subfigure}[t]{0.48\textwidth}
    \centering
        \includegraphics[width=\textwidth]{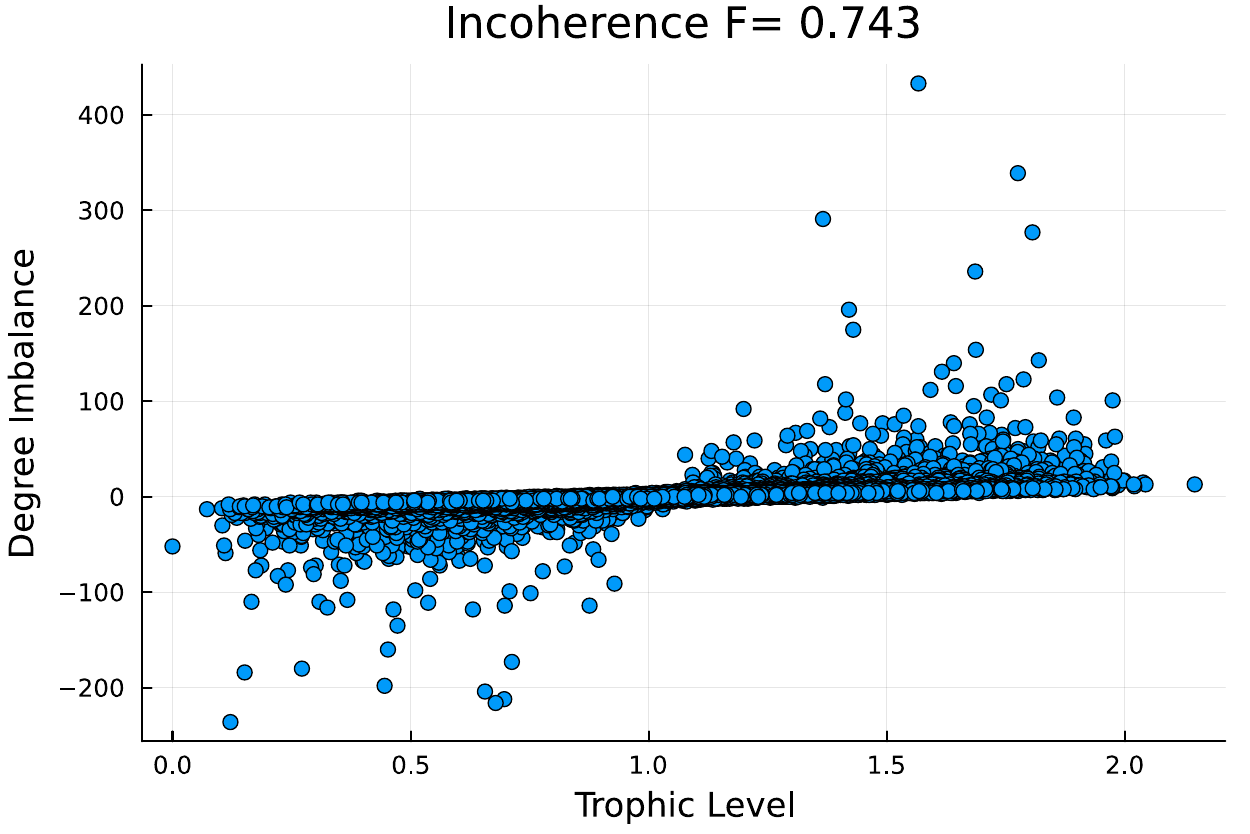}
        \caption{ Degree Imbalance (in-degree minus out-degree) with Trophic Level. Gaussian Function. Equation \ref{eq:gaussian} $\mu_f= 1$, $\sigma_f= 1000$. }
        \label{fig:imb_pref_gaussain_high_T}
        \end{subfigure}   
\caption{Degree Imbalance  with Trophic Level of evolving networks with Fitness Functions and Preferential Attachment. $N=20000$, $m =10$. Seed graph is a path of length $m=10$. Degree-based preferential attachment exponent set to one.}
\label{fig:degre_distrubtions_in_out}
\end{figure}

It has been observed previously that in some situations trophic level correlates with degree imbalance \cite{Rodgers2023StrongNetworks,Rodgers2023InfluenceNetworks} hence we investigated the relationship between trophic level and degree imbalance when the networks are constructed with degree-based preferential attachment and fitness interactions. When the network is highly coherent (i.e. the hierarchy is very strong), figure \ref{fig:imb_pref_gaussain_low_T}, the trophic level is not strongly related to the degree imbalance apart from at the ends of the hierarchy. This is due to the uniform fitness and that position in the hierarchy is strongly enforced by fitness. This shows that this kind of hierarchy is distinct from hierarchy induced by degree imbalance. Nodes at the bottom of the hierarchy have no nodes below them which favourably connect into them so they have a higher out-degree than their in-degree while the nodes at the top of the hierarchy have nowhere to connect out to meaning they have high degree imbalance. This highlights a key feature of systems which display linear hierarchy - if the system is finite there will be ``ends'' which may have different behaviour compared to a node which lies in the middle of the hierarchy. This also puts a caveat on the results of the degree distribution mean-field and numerical tests as the analysis fails to take into account this inhomogeneity in the system. When the network becomes more incoherent, figure \ref{fig:imb_pref_gaussain_med_T}, the trophic level starts to correlate more with the imbalance but we still observe the same behaviour in the tails with less sharp peaks. As the network becomes maximally incoherent for this variant of the model, standard deviation is much larger than the mean, shown in figure \ref{fig:imb_pref_gaussain_high_T}, we see less sharp changes at the end of the distribution as the hierarchy is less strict and a there is a large regime where the degree imbalance and trophic level are correlated. This is a similar result to what would be found if the fitness function was set to a constant and the hierarchical ordering, as measured by trophic level, was not determined by fitness and a result of the degree imbalance in the system only. In the high standard deviation regime, figure \ref{fig:imb_pref_gaussain_high_T}, we also see the spread of the trophic levels shrink and they do not span the full fitness regime as when the hierarchy was strongly enforced by fitness, figure \ref{fig:imb_pref_gaussain_low_T}.

\subsection{ Degree-based Preferential Attachment and Trophic Incoherence}

Many networks can grow by simple degree-based preferential attachment and we investigate how this mechanism relates to the varied trophic incoherence we observe in these systems. This may also explain why we observe so many coherent and non-normal networks in nature \cite{Duan2022NetworkSystems,Johnson2020DigraphsSystems, MacKay2020HowNetwork,Asllani2018StructureNetworks, Nartallo-Kaluarachchi2024BrokenNetworks}. We do this by investigating the impact of the preferential attachment exponent and the fraction of edges which go in and out of the newly added nodes on the trophic coherence. 

\subsubsection{Attachment Exponent and Trophic Incoherence}

Firstly, we  look at the relationship between the Trophic Incoherence and the preferential attachment exponent $\alpha$ to investigate if preferential attachment alone can act as an explanation for the appearance of coherent networks in nature. We take the fitness function $S(\beta_j -\beta_i) = 1$ and then use attachment functions of the form \begin{align}
    \Pi_{\text{out}}(i,j) \propto  (k^{\text{in}}_j)^\alpha, \\
    \Pi_{\text{in}}(j,i) \propto (k^{\text{out}}_j)^\alpha.
\end{align}
We also study the functions 
\begin{align}
    \Pi_{\text{out}}(i,j) \propto  (k^{\text{in}}_j + \delta)^\alpha, \\
    \Pi_{\text{in}}(j,i) \propto (k^{\text{out}}_j + \delta)^\alpha,
\end{align}
where $\delta$ is a small positive constant to allow the study of negative alpha.

\begin{figure}[H]
    \centering
\begin{subfigure}[t]{0.48\textwidth}        		
\centering
            \includegraphics[width=\textwidth]{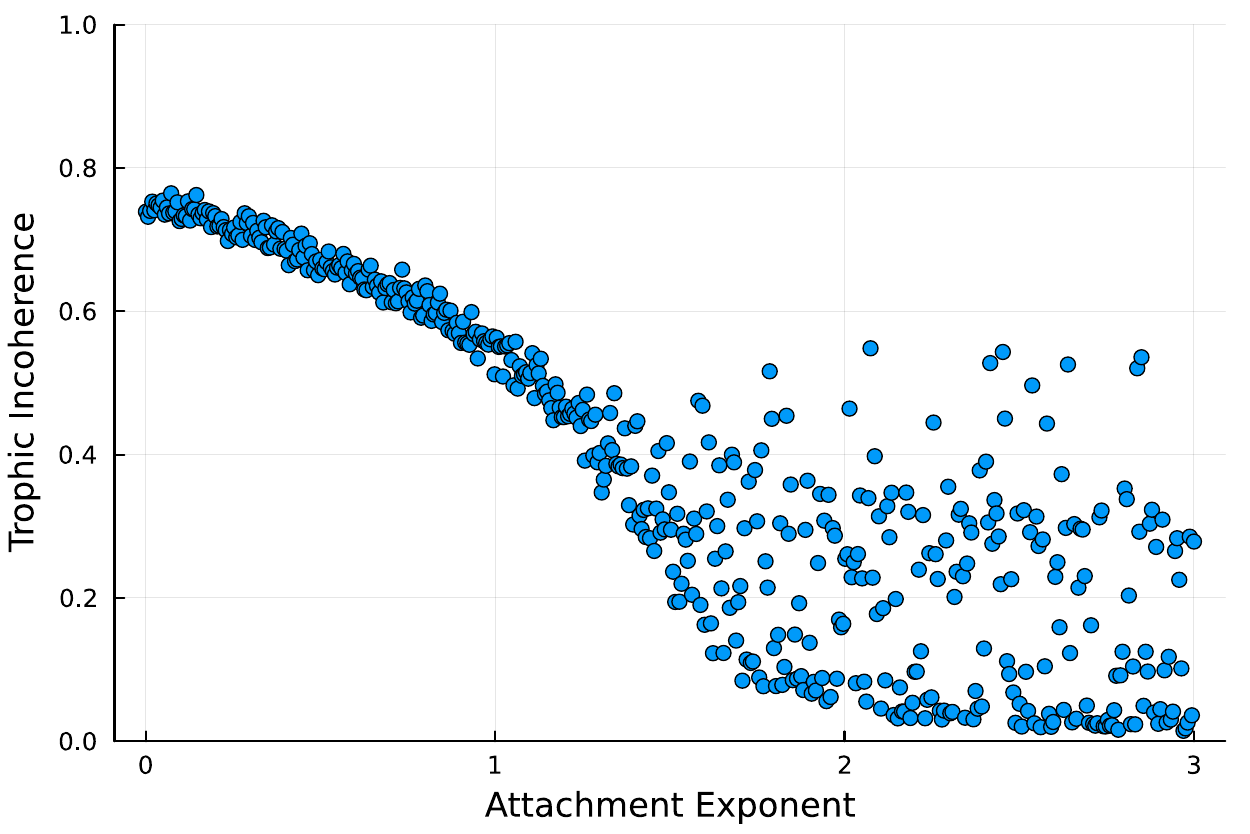}
        	\caption{Trophic Incoherence and Preferential Attachment for positive exponent, $\alpha$, which is evenly spaced between 0 and 3.}
        	
        \label{fig:pref_attach_positive_exp}\end{subfigure} 
        \hfill
\begin{subfigure}[t]{0.48\textwidth}   
\centering
            \includegraphics[width=\textwidth]{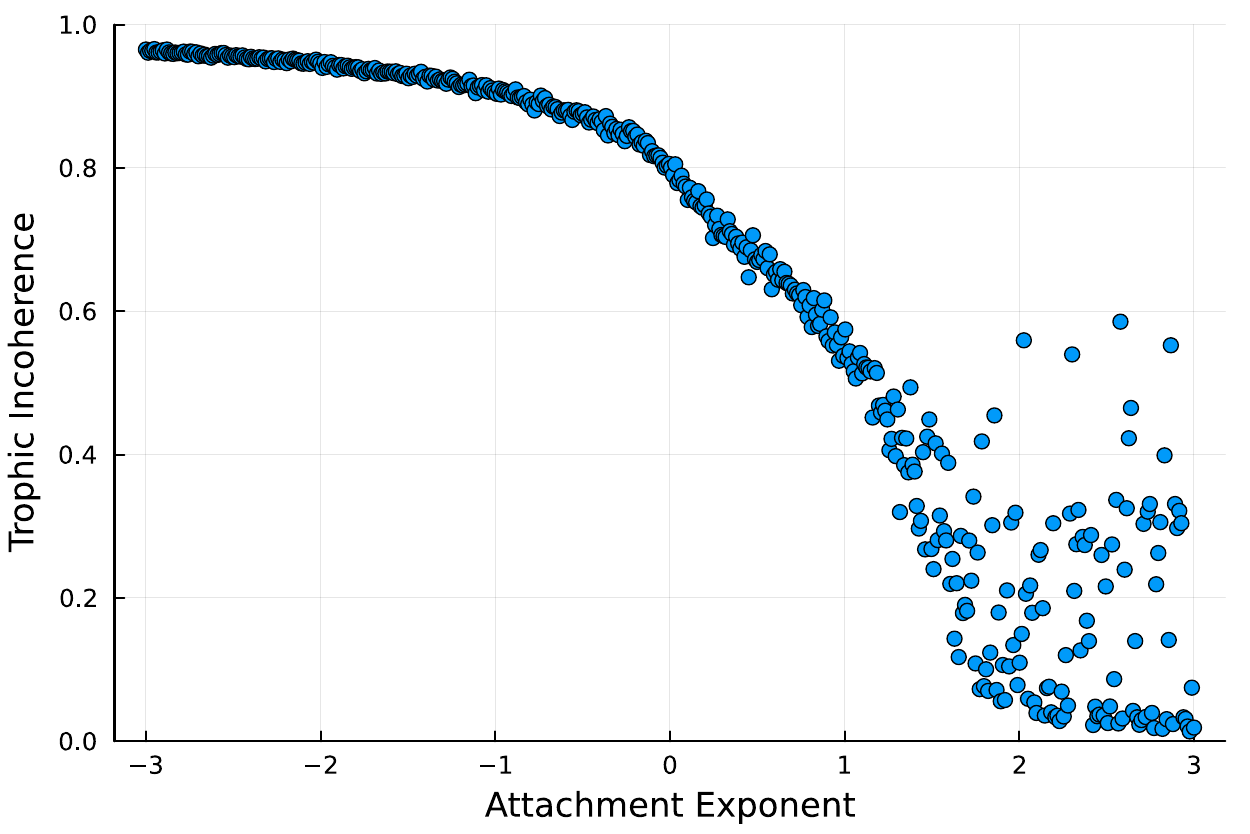}
        	\caption{Trophic Incoherence and degree-based attachment with a small positive constant added to the degree, $\delta = 10^{-6}$, to allow negative exponents. $\alpha$ is evenly spaced between -3 and 3.}
        	
        \label{fig:pref_attach_all_exp}
        \end{subfigure}  
\caption{Variation of Trophic Incoherence with preferential attachment exponent in standard preferential attachment and with a negative exponent. 500 networks were generated for each case with 1000 nodes and 5 edges added with each additional node. The seed network is a path of length 5. Edges have a probability 0.5 of being added as an incoming or outgoing edge of the newly added node. }
\label{fig:pref_attachement_incoherence}
\end{figure}

Preferential attachment with varying exponent is shown in figure \ref{fig:pref_attach_positive_exp}. When $\alpha =1$, this leads to networks which are more coherent than an ER random graph but not as coherent as a network which has a strong hierarchical ordering. In this case, much of the coherence comes from the large degree imbalance caused by the scale-free uncorrelated in and out degree distributions. This is expected as trophic level \cite{Rodgers2023InfluenceNetworks} and SpringRank \cite{Peixoto2022OrderedNetworks} have both been shown to be related to degree imbalance in some cases. As well as the fact that in the extreme case a star graph with all edges pointing either in or outwards is completely coherent with $F=0$.

When $\alpha=0$, we get the same kind of incoherence you see in random graphs as there is no preference for where the edges are added and no strong degree imbalance created. When we send $\alpha$ to be larger than one we reach the regime of super-linear preferential attachment where a few nodes connect to all other nodes. This leads to lots of variability and lack of control of the incoherence as the network structure is dominated by a few nodes. Extremely high degree imbalance leads to very coherent networks as would be expected tending towards the extreme of a star graph. 

This figure shows that preferential attachment can partially explain the emergence of coherent networks in nature. Preferential attachment with intermediate values of $\alpha$ leads to networks which are more coherent than random graphs explaining some of the networks we see in nature. Super-linear preferential attachment can lead to very coherent networks however it does this while enforcing degree distributions which are dominated by only a few nodes which is not the case in many real-world systems. It also comes with the additional constraint that the coherence is difficult to control in this regime.

In the case where we use the small constant to send $\alpha$ less than zero, figure \ref{fig:pref_attach_all_exp}, when $\alpha$ is negative the networks become more incoherent than uniform random attachment as edges are added to the nodes with the smallest degree which leads to a network with more balanced in and out degrees leading to higher incoherence. This is again expected as we tend closer to the extreme case where the in degree equals the out degree for each vertex and we have a network where $F=1$ \cite{MacKay2020HowNetwork}.

In \cite{Mones2013HierarchyNetworks} it was also found that the behaviour of hierarchy, as measured by Reaching Centrality, could also be affected by the changing the degree distribution of scale-free networks with the size of the out-components of nodes, and Global Reaching Centrality being affected as the exponent varied.

\subsubsection{Ratio of In and Out Edges Added and Trophic Incoherence}

\begin{figure}[H]
      		\centering
            \includegraphics[width=0.8\linewidth]{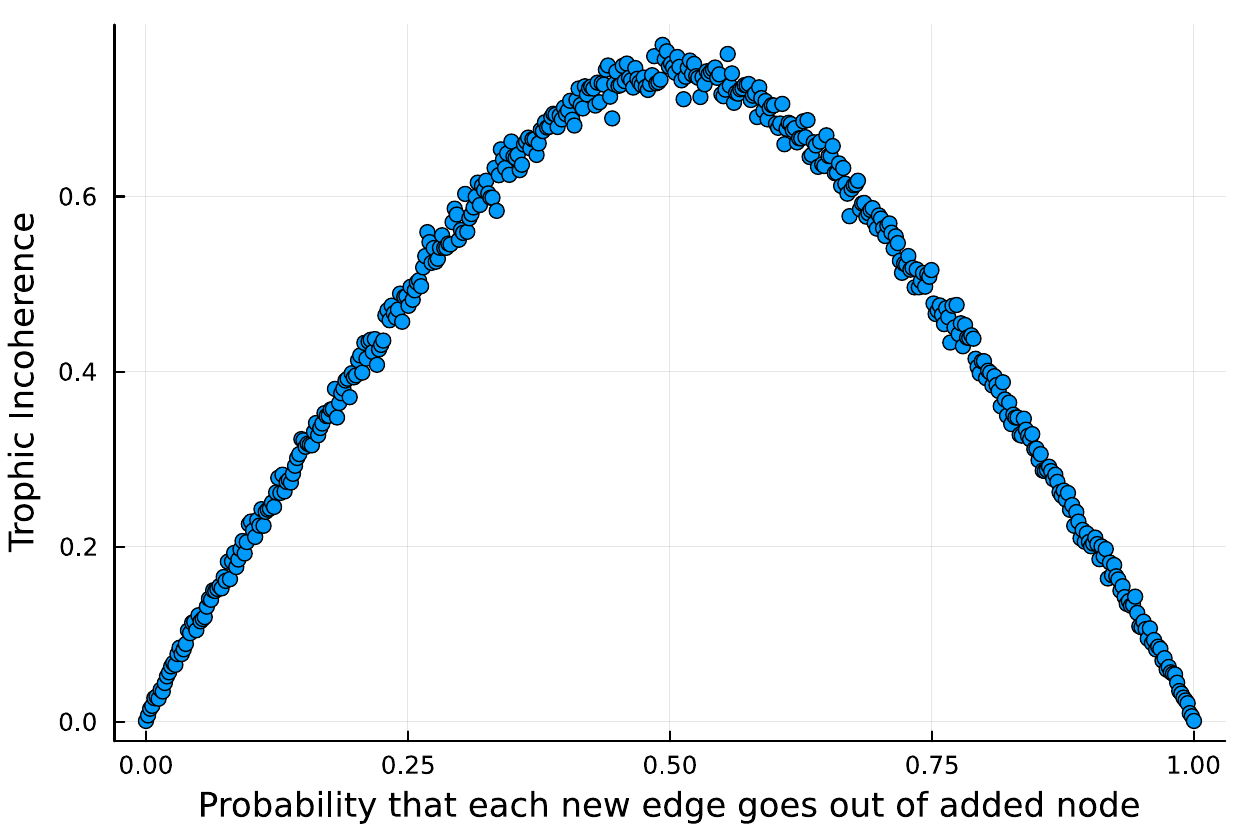}
        	\caption{Variation of Trophic Incoherence with probability of new edges being attached into and out of new nodes. 500 networks were generated with 1000 nodes and 10 edges added with each additional node. The seed network is a path of length 10. Degree-based preferential attachment parameter, $\alpha=1$.}
        	
        \label{fig:in_out_ratio_incoherence}
        \end{figure}

Trophic Incoherence and hence all the effects that depend on it \cite{Rodgers2023InfluenceNetworks,Rodgers2023StrongNetworks,MacKay2020HowNetwork,Klaise2016FromProcesses,Johnson2017LooplessnessCoherence,Johnson2014TrophicStability} can also be modified by varying the fraction of edges which attach into or out of the nodes as they are introduced, this is shown in figure \ref{fig:in_out_ratio_incoherence}. This figure shows that when edges strictly go only into or out of the newly added nodes this leads to networks which are very coherent at the ordering of the node addition implicitly leads to a directionality as nodes only connect to nodes which are younger or older (by node age). This explains why citation networks are very coherent as you can only cite papers which exist as the time of writing, creating a time ordering in the network.  Figure \ref{fig:in_out_ratio_incoherence} also shows how this process is symmetric about the point where the in/out probabilities are equal and this is where the networks are maximally coherent for the model as edges are equally likely to connect in any direction in terms of node age. In the cases where the fraction of edges added is between these extreme points we see that the network has coherence determined by the size of the majority and minority direction split. This behaviour can be interpreted by thinking about the age of nodes as a fitness parameter where new nodes connect a certain fraction of their edges from new nodes to older or from older nodes to the most recently added. 

The combination of the effects highlighted may partially explain the range of trophic incoherence we see in real networks \cite{MacKay2020HowNetwork,Rodgers2023StrongNetworks}. Preferential attachment can lead to more coherence than observed in random graphs due to the stronger degree imbalance which may be present in some systems. While if new nodes join the network and have an unequal probability to create incoming and outgoing edges this can lead to networks which are very coherent. For example, in citation networks where the age of the node acts as an ordering, limiting the range of connectivity structures available. The result in figure \ref{fig:in_out_ratio_incoherence} agrees with the results found in the models of network non-normality \cite{Asllani2018StructureNetworks,OBrien2021HierarchicalNetworks,Sornette2023Non-normalBubbles, Nartallo-Kaluarachchi2024BrokenNetworks} where varying the probability of reciprocal edges and hence varying  the probability that edges go against the ordering induced by the node arrival time modifies the non-normality and trophic incoherence.

\subsection{Correlation of Fitness and Trophic Level}

\begin{figure}[H]
    \centering
    \begin{subfigure}[t]{0.48\textwidth}       		\centering
            \includegraphics[width=\textwidth]{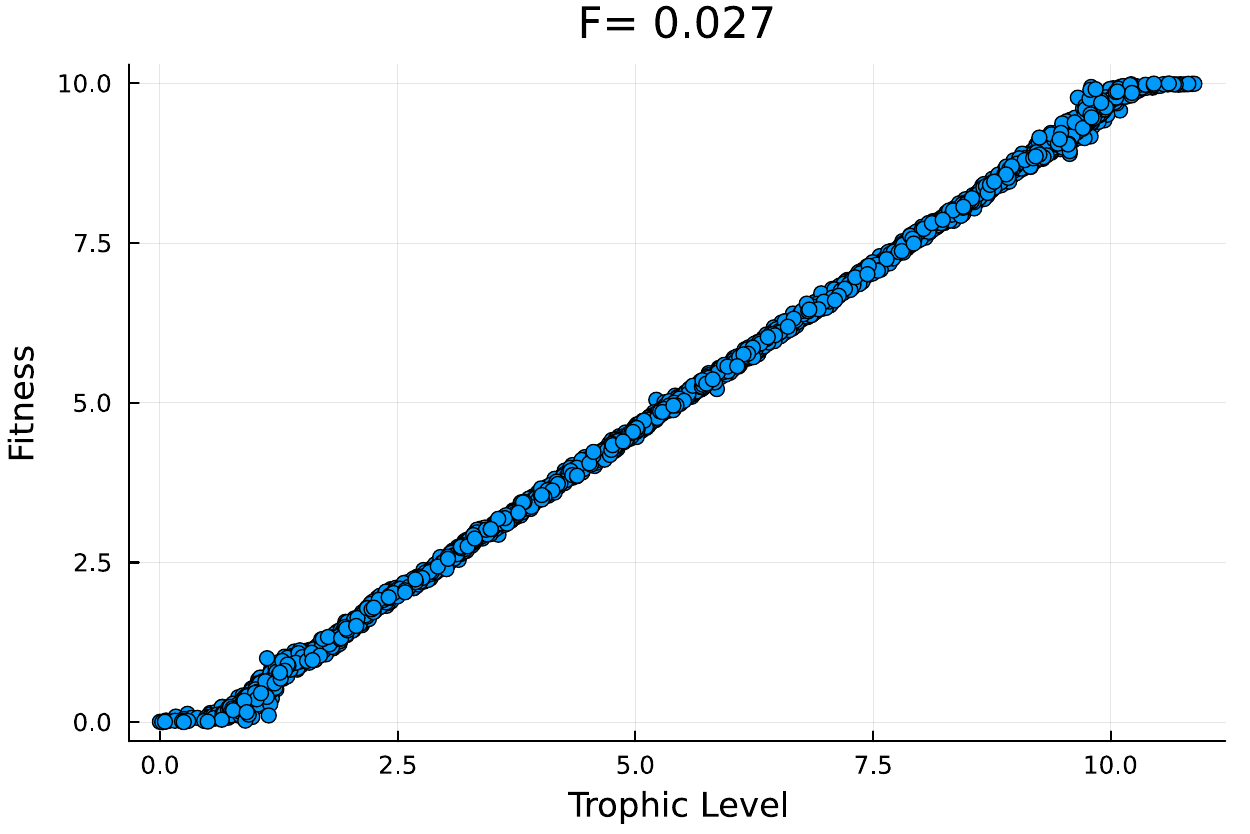}
        	\caption{ Gaussian Function. Equation \ref{eq:gaussian} $\mu_f= 1$, $\sigma_f=0.1 $ Pearson Correlation Coefficient approximately 1.0.}
        	
        \label{fig:fitness_corr_very_low}
        \end{subfigure}
        \hfill
\begin{subfigure}[t]{0.48\textwidth}       		\centering
            \includegraphics[width=\textwidth]{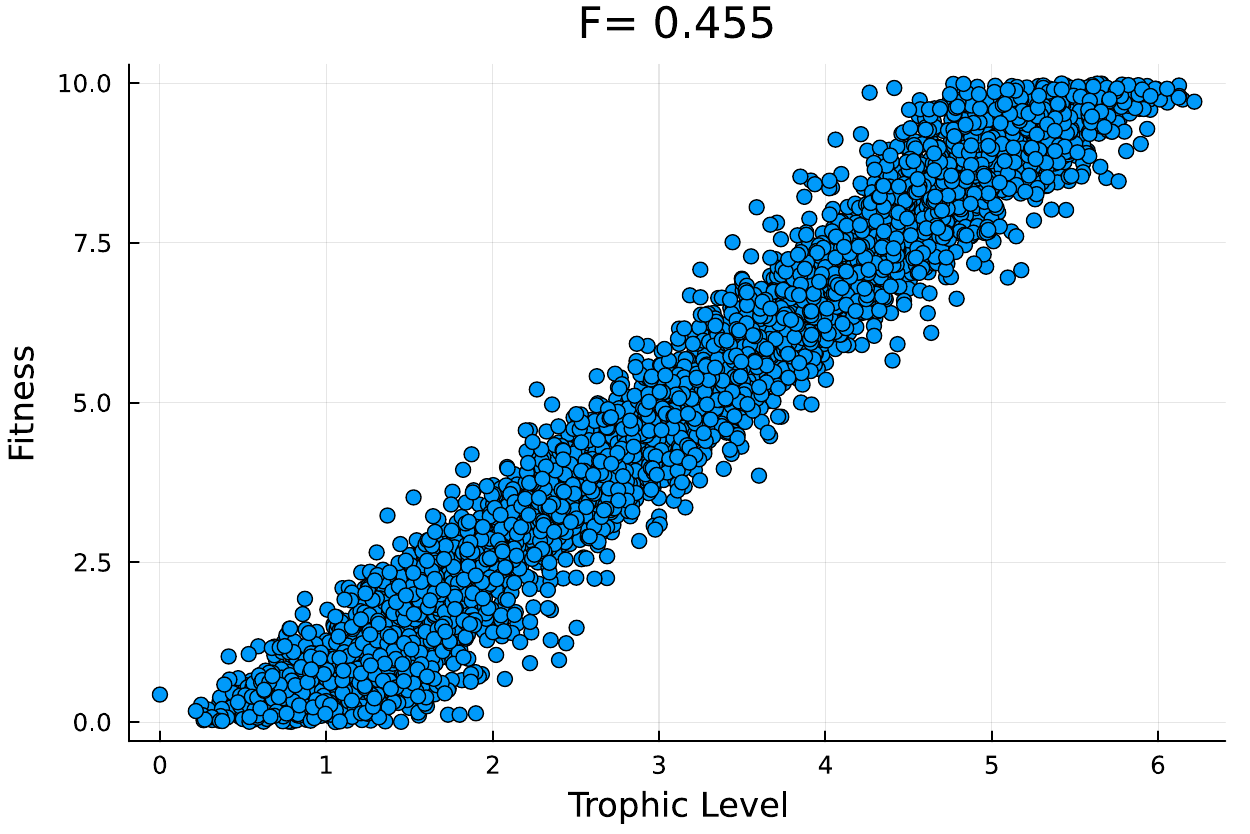}
        	\caption{ Gaussian Function. Equation \ref{eq:gaussian} $\mu_f= 1$, $\sigma_f=1 $. Pearson Correlation Coefficient approximately 0.978.}
        	
        \label{fig:fitness_corr_low}
        \end{subfigure} 
        \hfill
\begin{subfigure}[t]{0.48\textwidth}       		
\centering
            \includegraphics[width=\textwidth]{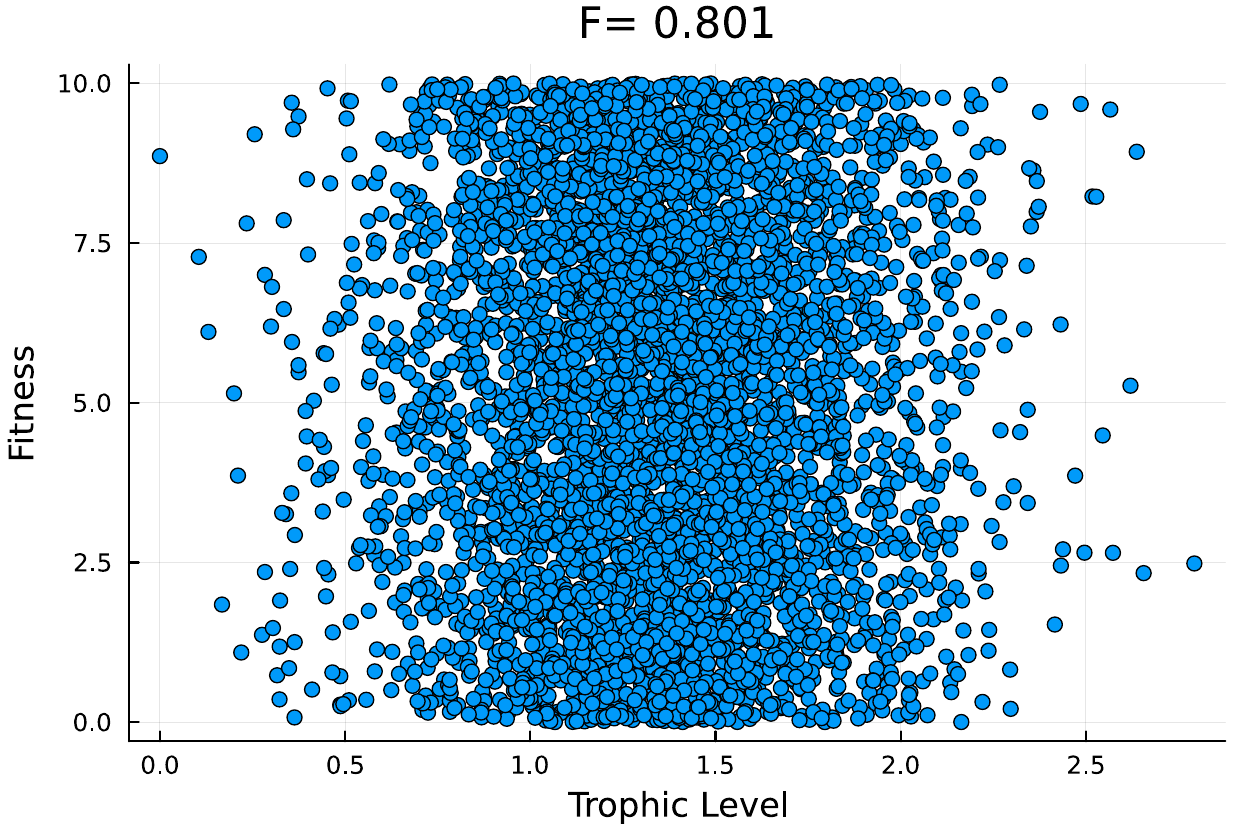}
        	\caption{ 	Gaussian Function. Equation \ref{eq:gaussian} $\mu_f= 1$, $\sigma_f= 1000$. Pearson Correlation Coefficient approximately 0.006. }
        	
        \label{fig:fitness_corr_large_T}
\end{subfigure}
\caption{N = 5000, Fitness is uniformly distributed between 0 and 10. Nodes are added with 5
new edges. Seed graph is a path of length 5.  Degree-based preferential attachment exponent set to zero.}
\label{fig:Levelandfitness}
\end{figure}

We also investigate how the trophic level of node can be used to approximate the value of the fitness of a node in networks generated with only fitness interactions with varying success depending on the trophic incoherence of the  network, figure \ref{fig:Levelandfitness}. This is a useful feature as it highlights how the structural quantity trophic level relates to the node level fitness which was involved in the network generation.

In figure \ref{fig:fitness_corr_very_low}, we see a clear correlation between fitness and trophic level for the Gaussian networks of very low $F$ showing that in this regime the level can be a good proxy for fitness. This is still maintained in figure \ref{fig:fitness_corr_low} even as $F$ reaches more intermediate values however with a broader spread of values at the same trophic level. This is due to that fact that as the network becomes more incoherent the range of trophic levels decreases and the maximum trophic level is now lower than the maximum of the fitness showing the hierarchy is less pronounced. This effect continues into the very high $F$ regime, figure \ref{fig:fitness_corr_large_T}, where there is no correlation between level and fitness and trophic level is no longer a useful parameter to predict the fitness and is likely be correlated with the degree imbalance \cite{Rodgers2023InfluenceNetworks,Peixoto2022OrderedNetworks}. The level distribution again shrinks to spanning a very small range highlighting again that the impact of hierarchical structure is not very strong in this network. As the standard deviation is very large relative to the mean of the fitness differences, the behaviour in figure \ref{fig:fitness_corr_large_T} is similar to the behaviour that would be found if the fitness function was a constant and the hierarchy was determined by the degree imbalance as in an ER random graph.

\begin{figure}[H]
   \centering
\begin{subfigure}[t]{0.48\textwidth}        		
\centering
            \includegraphics[width=\textwidth]{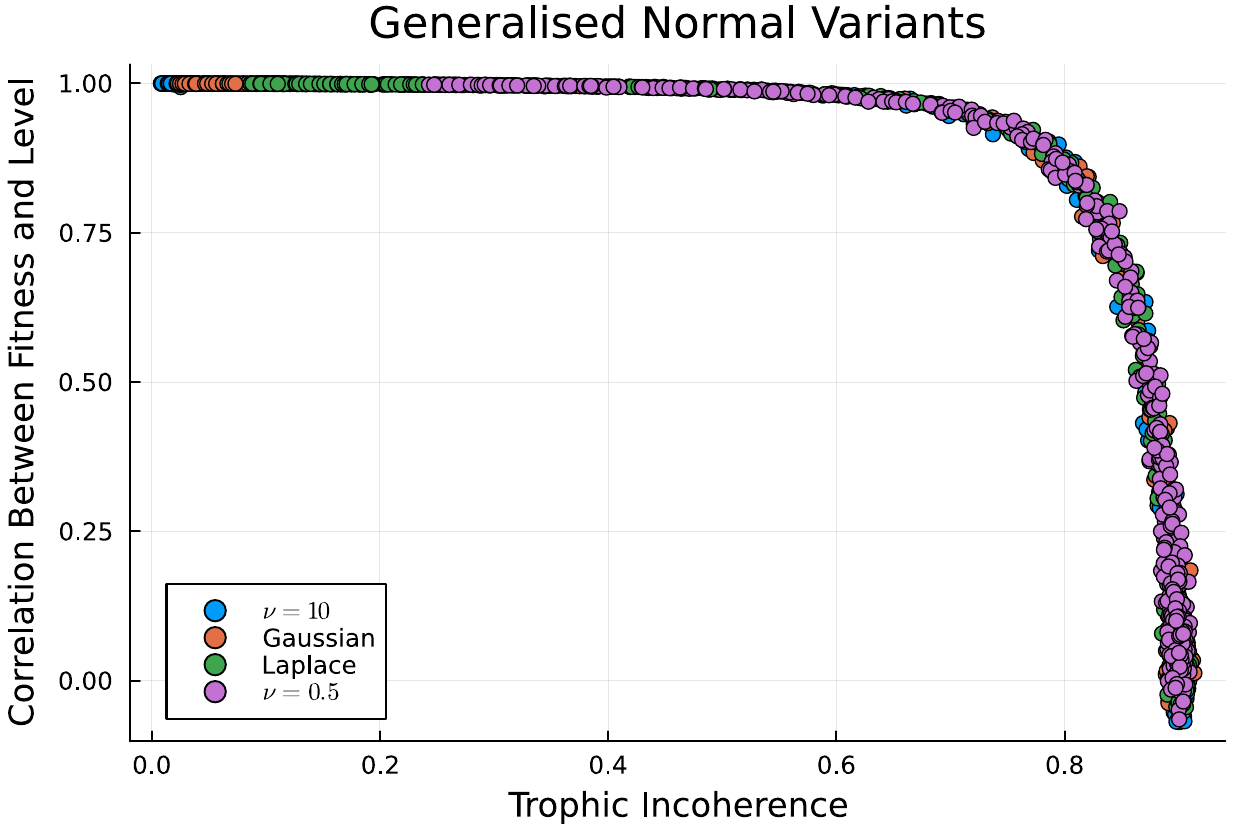}
        	\caption{Generalised Normal Variants. Equation \ref{eq:gen_normal1} $\mu_f= 1$, Parameter $b$ spaced  from $10^{-1.5}$ to $10^2$. }
        \label{fig:fitness_corr_levels}
        \end{subfigure} 
        \hfill
\begin{subfigure}[t]{0.48\textwidth}   
\centering
            \includegraphics[width=\textwidth]{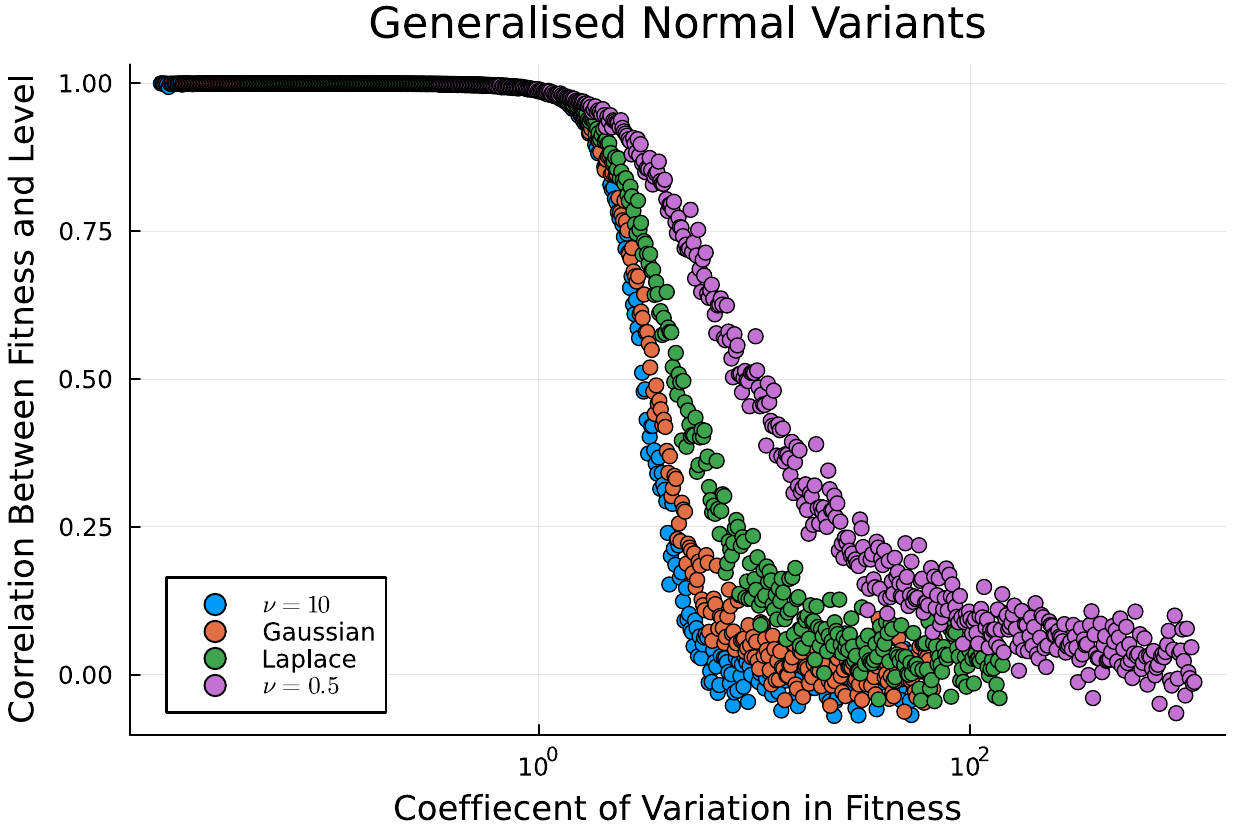}
        	\caption{Generalised Normal Variants.Equation \ref{eq:gen_normal1} $\mu_f= 1$, Parameter $b$ spaced  from $10^{-1.5}$ to $10^2$ }
        \label{fig:fitness_co_varaiation_corr}
        \end{subfigure} 
        \hfill
\begin{subfigure}[t]{0.48\textwidth}        		
\centering
            \includegraphics[width=\textwidth]{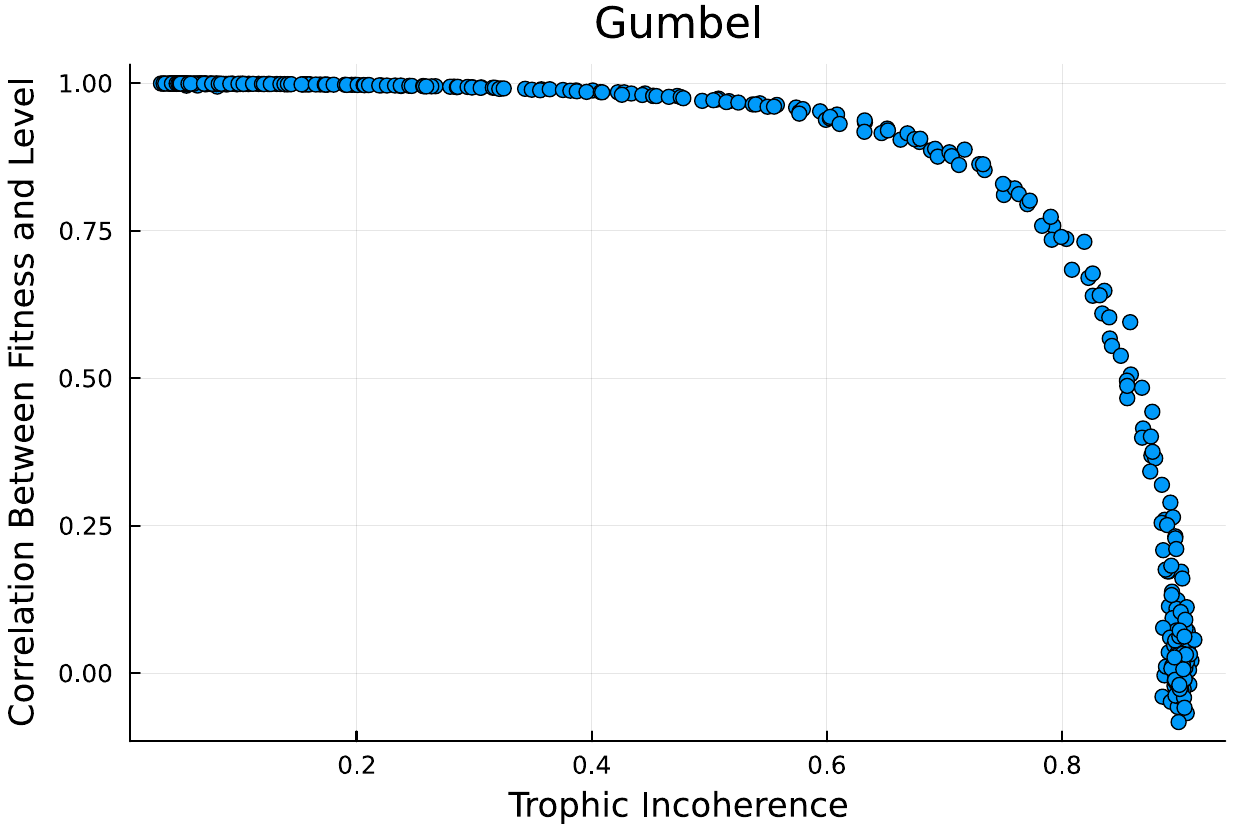}
        	\caption{Gumbel. Equation \ref{eq:gumbel} $\mu_f=1$, Parameter $b$ spaced between from $10^{-2}$ to $10^2$.}	
        \label{fig:fitness_corr_levels__gumbel}
        \end{subfigure} 
        \hfill
\begin{subfigure}[t]{0.48\textwidth}   
\centering
            \includegraphics[width=\textwidth]{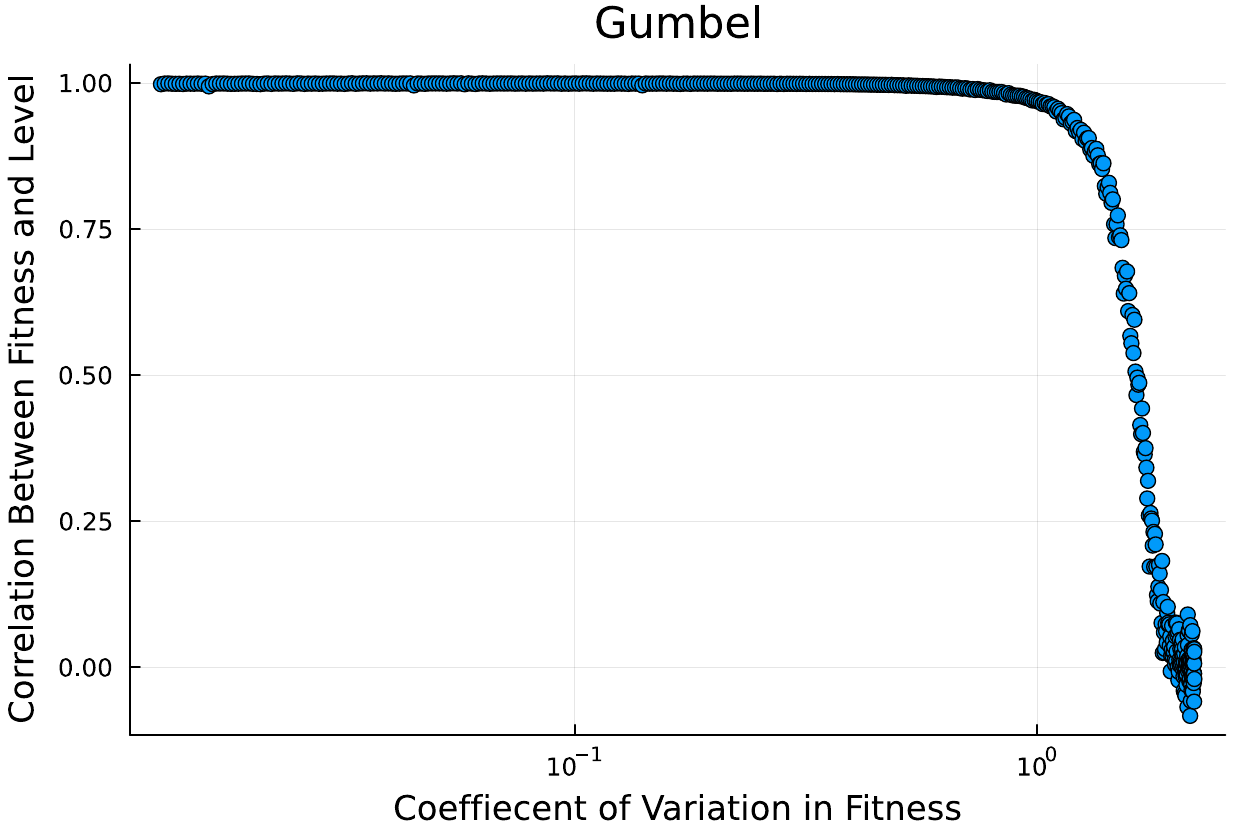}
        	\caption{Gumbel. Equation \ref{eq:gumbel} $\mu_f=1$, Parameter $b$ spaced between from $10^{-2}$ to $10^2$.}
        \label{fig:fitness_co_varaiation_corr_gumbel}
        \end{subfigure}  
        \hfill
\begin{subfigure}[t]{0.48\textwidth}        		
\centering
            \includegraphics[width=\textwidth]{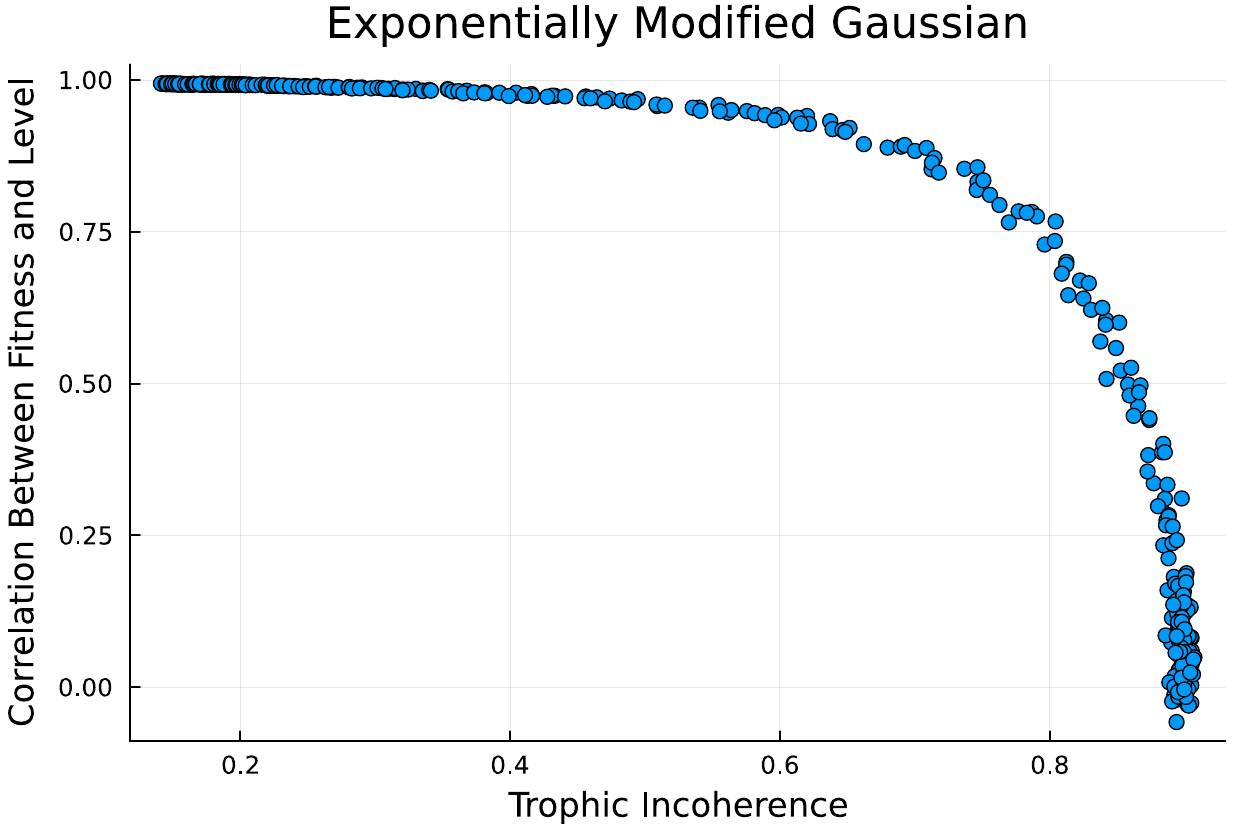}
        	\caption{Exponentially Modified Gaussian. Equation \ref{eq:exp_gaussian} $\mu_f=1$, $\lambda =1$. Parameter $\sigma_e$, spaced between $10^{-2}$ to $10^{1.5}$. }	
        \label{fig:fitness_corr_levels__exp_mod_gaussian}
        \end{subfigure} 
        \hfill
\begin{subfigure}[t]{0.48\textwidth}   
\centering
            \includegraphics[width=\textwidth]{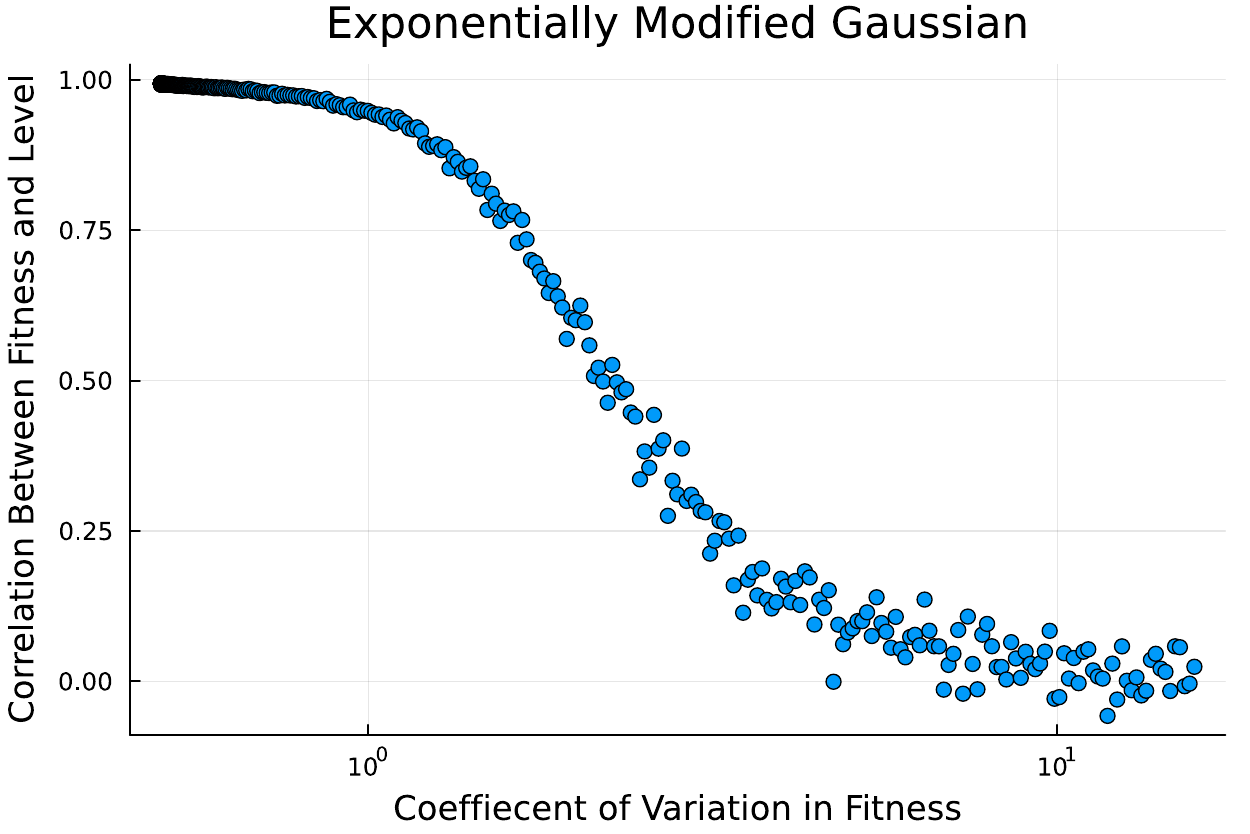}
        	\caption{Exponentially Modified Gaussian. Equation \ref{eq:exp_gaussian} $\mu_f=1$, $\lambda =1$. Parameter $\sigma_e$, spaced between $10^{-2}$ to $10^{1.5}$.}
        \label{fig:fitness_co_varaiation_corr_exp_mod_gaussian}
        \end{subfigure}     
        
\caption{Correlation between trophic levels and fitness for different fitness functions with Trophic Incoherence and Coefficient of Variation. Networks are $N=1000$ with a starting seed network which is a directed path of length 10. New nodes are added with 10 edges. 500 networks are generated for each case. Fitness is uniformly distributed between 0 and 10.  Degree-based preferential attachment exponent set to zero.}
\label{fig:corrleation_level_fitness}
\end{figure}

We investigate this phenomena in more detail by measuring the Pearson correlation coefficient between fitness and level for different fitness functions in networks of varying trophic incoherence, which we also plot in terms of the coefficient of variation (standard deviation divided by the mean) of the normalised fitness functions, in figure \ref{fig:corrleation_level_fitness}. The correlation is very good for coherent networks, which have clear hierarchical structure, but breaks down as the networks become more incoherent. A similar trend can also be seen in the coefficient of variation, figure \ref{fig:corrleation_level_fitness}.
 When this coefficient of variation is small, the standard deviation is less than the mean, there is good correlation between node fitness and trophic level. However, when the coefficient of variation is large, the mean is smaller than the standard deviation, the relationship between fitness and level breaks down. For all the functions we have roughly three regimes. When the coefficient of variation is much less than 1 then level acts as a good proxy for fitness as the correlation coefficient is near one. When the correlation coefficient approximately on the order of $1$, then we are in an intermediate regime where it is some relationship between fitness and level but the correlation is weaker and then when the coefficient of variation is much larger than 1 the correlation between level and fitness breaks down. This gives a guide as to how the underlying properties of the fitness distribution shape the usefulness of trophic analysis as a network analysis tool.

The results in figure \ref{fig:corrleation_level_fitness} extend the intuition found in \cite{DeBacco2018ANetworks} that calculated ranks in SpringRank are well correlated with the synthetic ranks. However, we extend it to edges which are created with non-Gaussian fitness difference functions, a growing network model with fixed edge number and relate the correlation to $F$ which does not depend on knowing the parameters of the generative process and can be calculated directly from network structure.

\subsection{Degree Imbalance and Trophic Level Correlations}

\begin{figure}[H]
   \centering
\begin{subfigure}[t]{0.48\textwidth}        		
\centering
            \includegraphics[width=\textwidth]{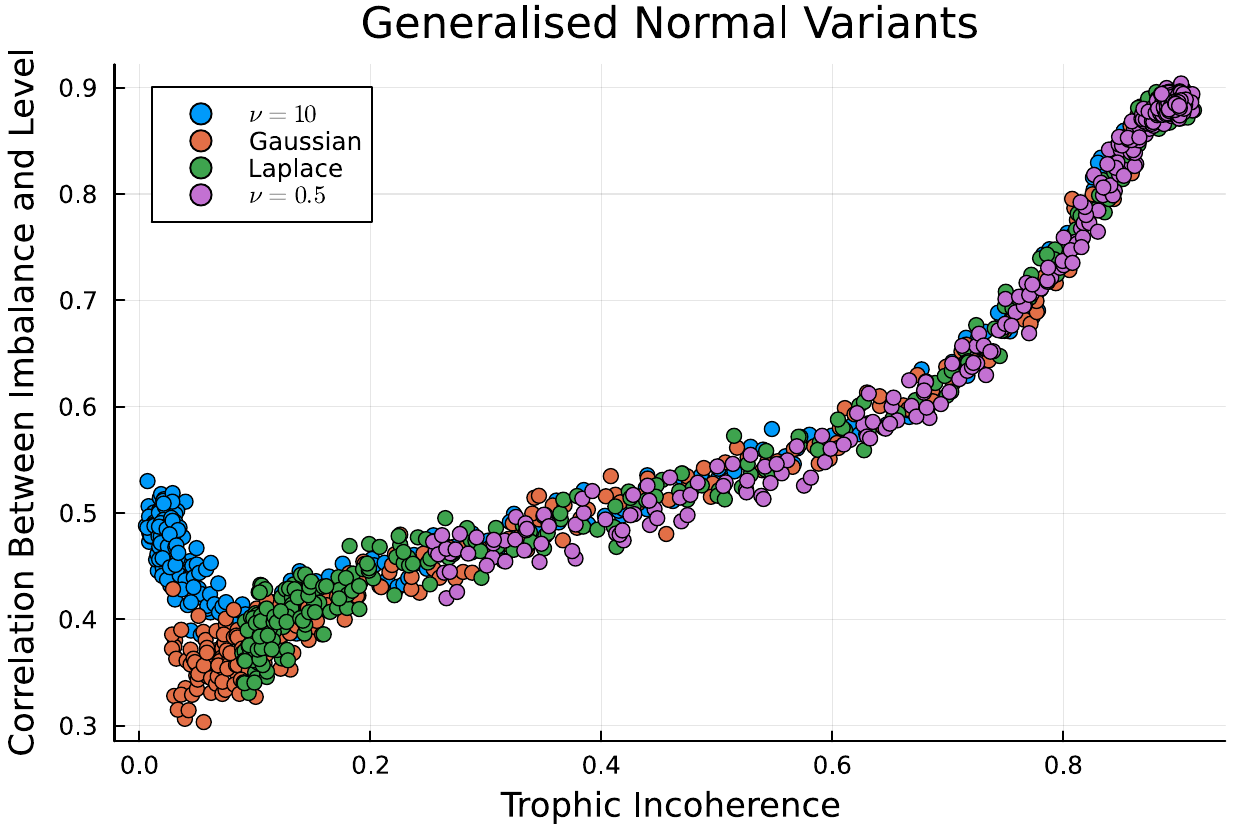}
        	\caption{Generalised Normal Variants. Equation \ref{eq:gen_normal1} $\mu_f= 1$, Parameter $b$ spaced  from $10^{-1.5}$ to $10^2$. }
        \label{fig:imb_corr_levels}
        \end{subfigure} 
        \hfill
\begin{subfigure}[t]{0.48\textwidth}   
\centering
            \includegraphics[width=\textwidth]{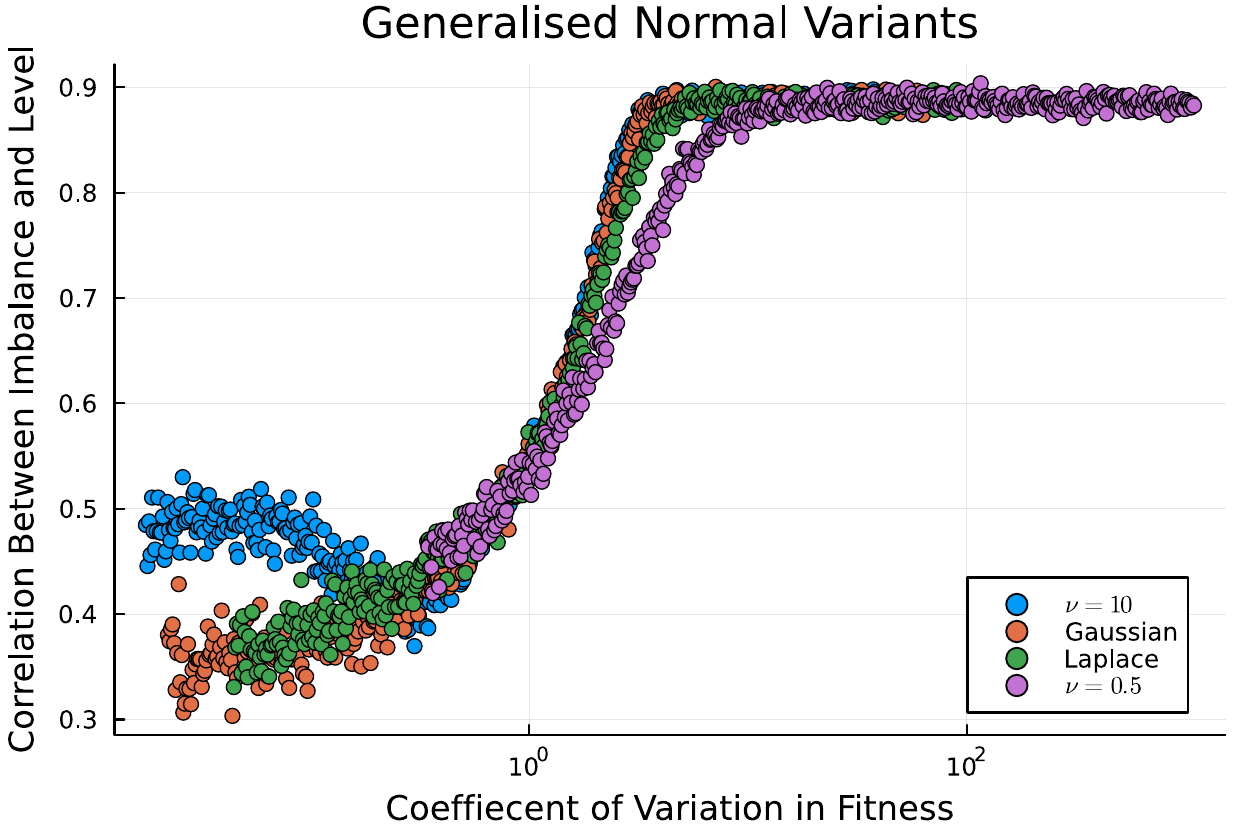}
        	\caption{Generalised Normal Variants.Equation \ref{eq:gen_normal1} $\mu_f= 1$, Parameter $b$ spaced  from $10^{-1.5}$ to $10^2$ }
        \label{fig:imb_co_varaiation_corr}
        \end{subfigure} 
        \hfill
\begin{subfigure}[t]{0.45\textwidth}        		
\centering
            \includegraphics[width=\textwidth]{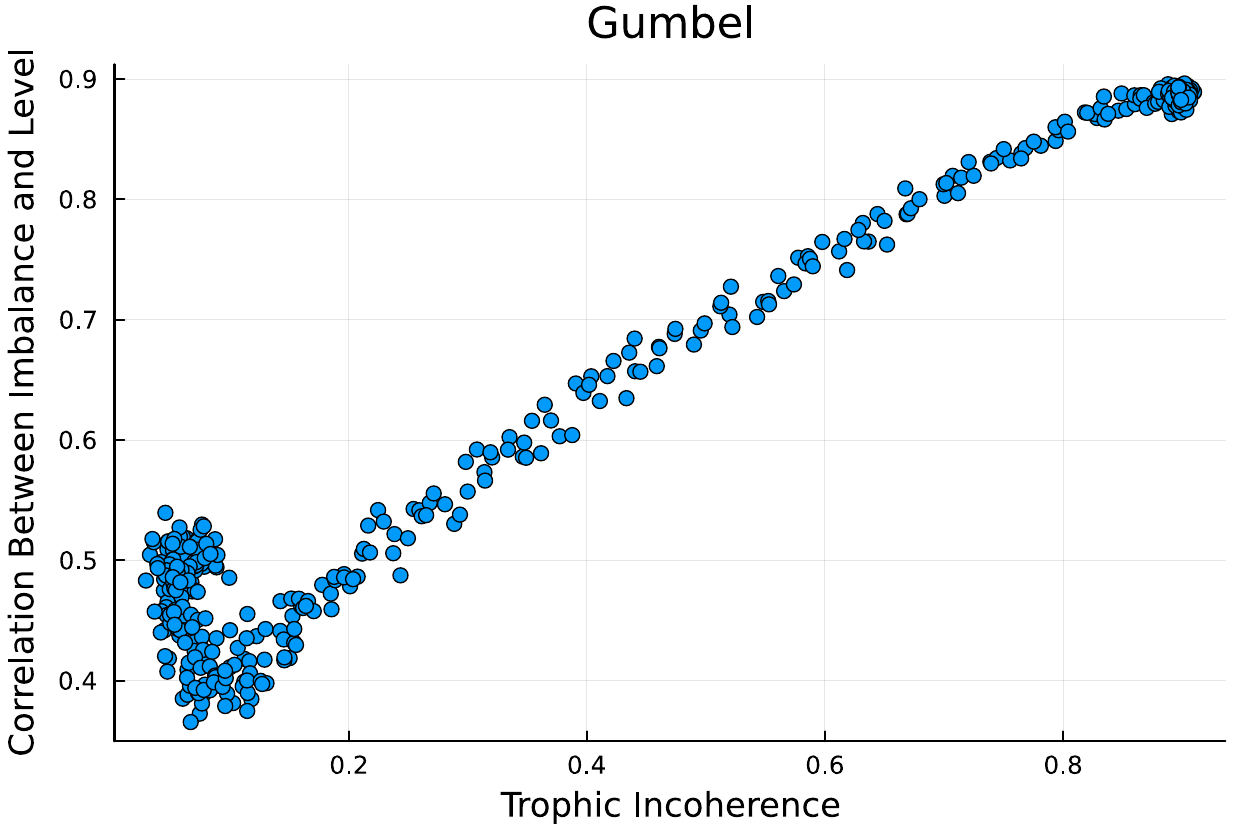}
        	\caption{Gumbel. Equation \ref{eq:gumbel} $\mu_f=1$, Parameter $b$ spaced between from $10^{-2}$ to $10^2$.}	
        \label{fig:imb_corr_levels__gumbel}
        \end{subfigure} 
        \hfill
\begin{subfigure}[t]{0.45\textwidth}   
\centering
            \includegraphics[width=\textwidth]{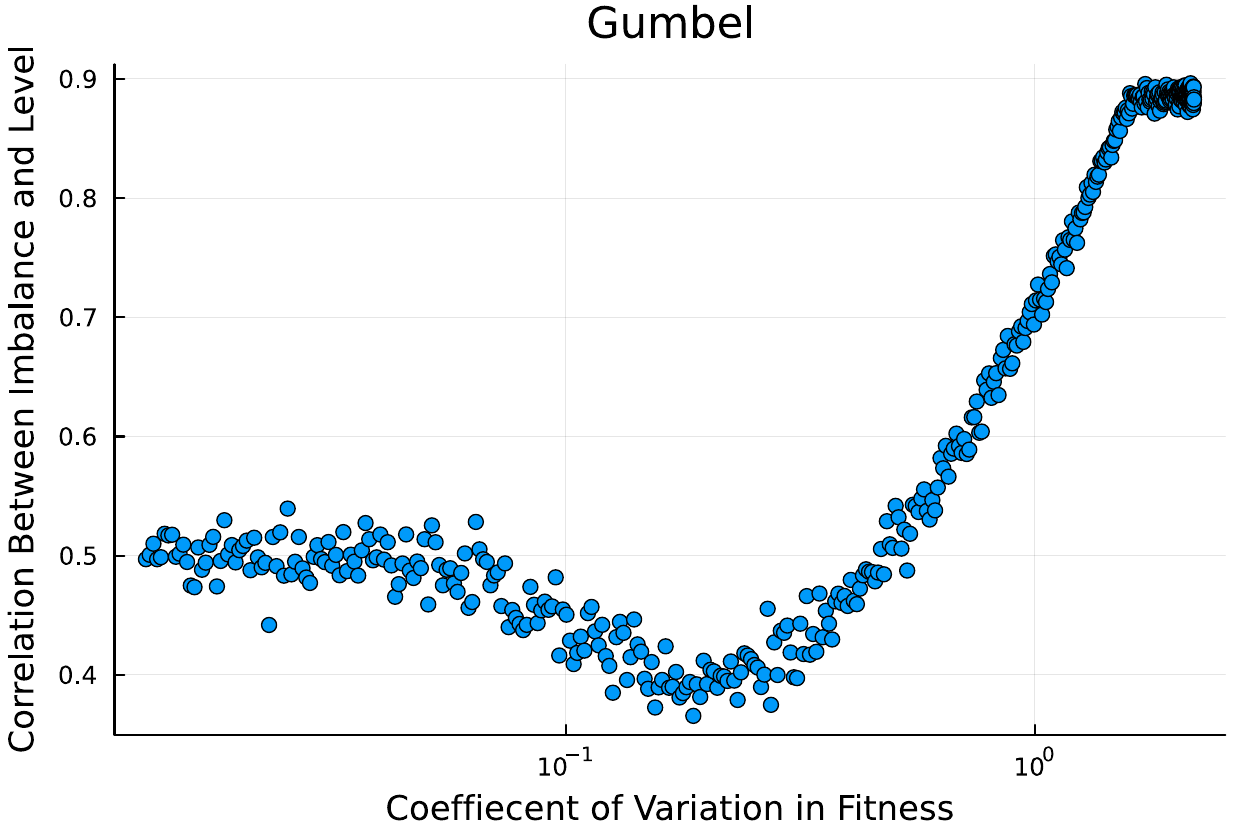}
        	\caption{Gumbel. Equation \ref{eq:gumbel} $\mu_f=1$, Parameter $b$ spaced between from $10^{-2}$ to $10^2$.}
        \label{fig:imb_co_varaiation_corr_gumbel}
        \end{subfigure}  
        \hfill
\begin{subfigure}[t]{0.45\textwidth}        		
\centering
            \includegraphics[width=\textwidth]{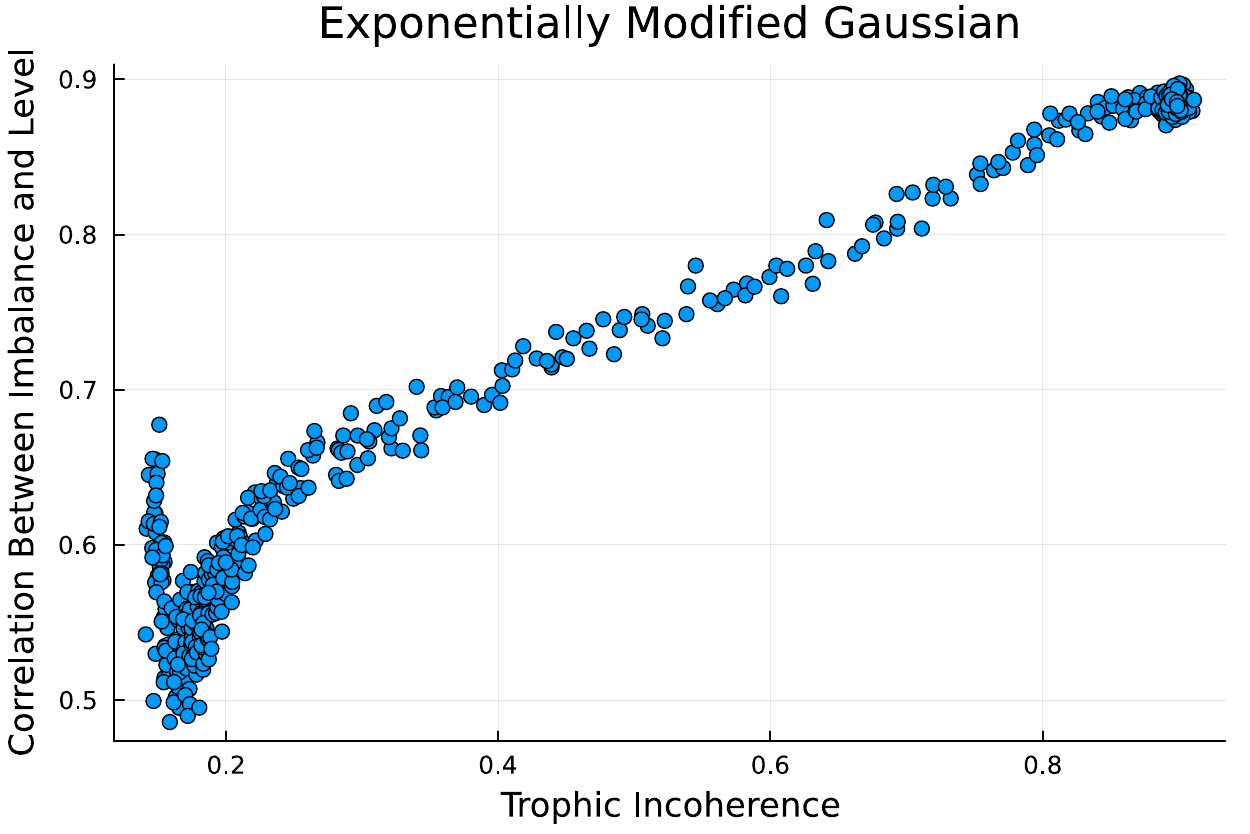}
        	\caption{Exponentially Modified Gaussian. Equation \ref{eq:exp_gaussian} $\mu_f=1$, $\lambda =1$. Parameter $\sigma_e$, spaced between $10^{-2}$ to $10^{1.5}$. }	
        \label{fig:imb_corr_levels__exp_mod_gaussian}
        \end{subfigure} 
        \hfill
\begin{subfigure}[t]{0.45\textwidth}   
\centering
            \includegraphics[width=\textwidth]{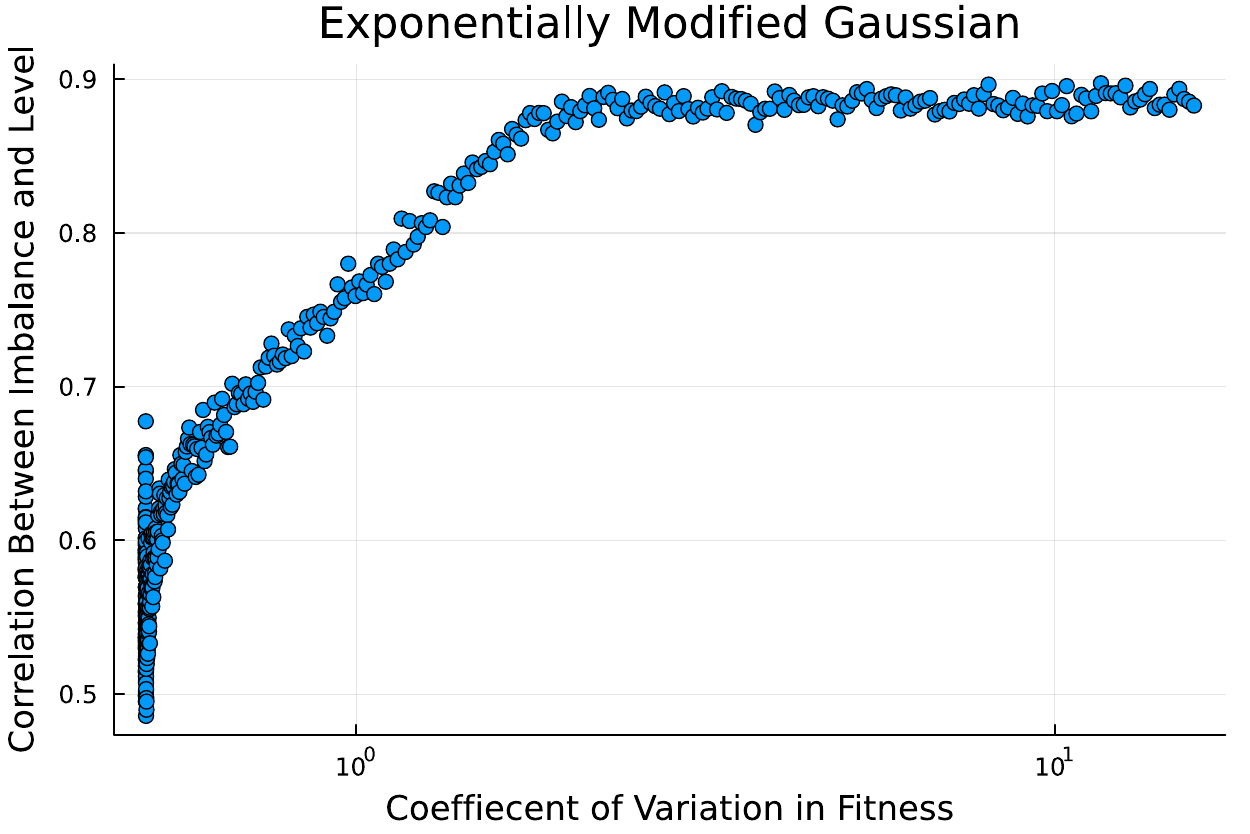}
        	\caption{Exponentially Modified Gaussian. Equation \ref{eq:exp_gaussian} $\mu_f=1$, $\lambda =1$. Parameter $\sigma_e$, spaced between $10^{-2}$ to $10^{1.5}$.}
        \label{fig:imb_co_varaiation_corr_exp_mod_gaussian}
        \end{subfigure}     
        
\caption{Correlation between trophic levels and degree imbalance (in-degree minus out-degree) for different fitness functions with Trophic Incoherence and Coefficient of Variation. Networks are $N=1000$ with a starting seed network which is a directed path of length 10. New nodes are added with 10 edges. 500 networks are generated for each case. Fitness is uniformly distributed between 0 and 10.  Degree-based preferential attachment exponent set to zero.}
\label{fig:corrleation_level_imb}
\end{figure}

We also study the variation of the correlation between trophic level and degree imbalance with trophic incoherence and coefficient of variations of the fitness functions for networks generated with only fitness interactions. This allows us to understand the regimes in which trophic analysis gives information which is separate to information gained by looking at the degree imbalance and how trophic level relates to degree imbalance. This is shown in figure \ref{fig:corrleation_level_imb}.

In figure \ref{fig:corrleation_level_imb} we see that for all fitness functions used when the incoherence or coefficient of variation is very high, a regime where there is little correlation between fitness and level, we see a strong correlation between level and degree imbalance as this is the structural feature which trophic level picks out. When the trophic incoherence and coefficient of variation are low, a regime of high level fitness correlation, then the correlation between trophic level and degree imbalance is weaker. However, they are not completely uncorrelated. This makes sense as we expect nodes which sit at the ends of the fitness distribution, and have the highest and lowest trophic levels, to have degree imbalances related to their position caused by the sharp ends of the fitness hierarchy. As a result of there being no nodes above or below respectively which are preferred by the attachment function. There is however some non-monotonic behaviour in this regime where as the hierarchy becomes less strictly enforced the correlation decreases as the sharp edge effects are less impactful. There is some variability here between function choices depending how function behaves and how strictly it enforces hierarchy when the control parameter is reduced to very small values. The results in figure \ref{fig:corrleation_level_imb} again highlight how $F$ is a useful tool to test if the hierarchy is likely to be impactful to a system and the effects that the ends of the fitness distribution can have on the behaviour of the system.

\subsection{Trophic Level Differences Distributions} \label{Sec:trophic_level_distrubtions}

We also investigate the distribution of the trophic level and fitness differences for various fitness functions in order to understand if the measured fitness differences reflect the input fitness function and if the trophic level differences are reflective of the input fitness function.   

\begin{figure}[H]
    \centering
\begin{subfigure}[t]{0.32\textwidth}        		
\centering
            \includegraphics[width=\textwidth]{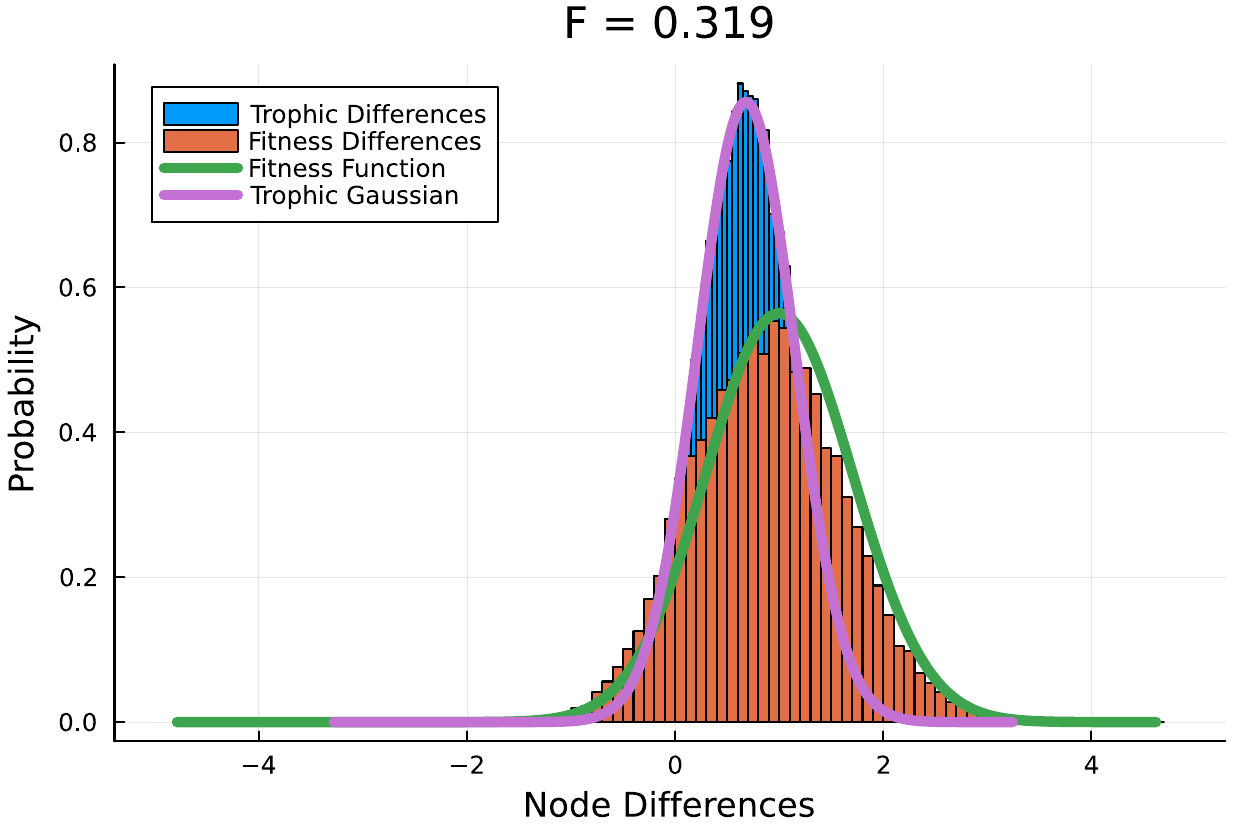}
        	\caption{Gaussian. Equation \ref{eq:gaussian} $\mu_f= 1$, $\sigma_f= \frac{1}{\sqrt{2}}$.}
        	
        \label{fig:fitness_gaussian}
        \end{subfigure} 
        \hfill
\begin{subfigure}[t]{0.32\textwidth}        		
\centering
            \includegraphics[width=\textwidth]{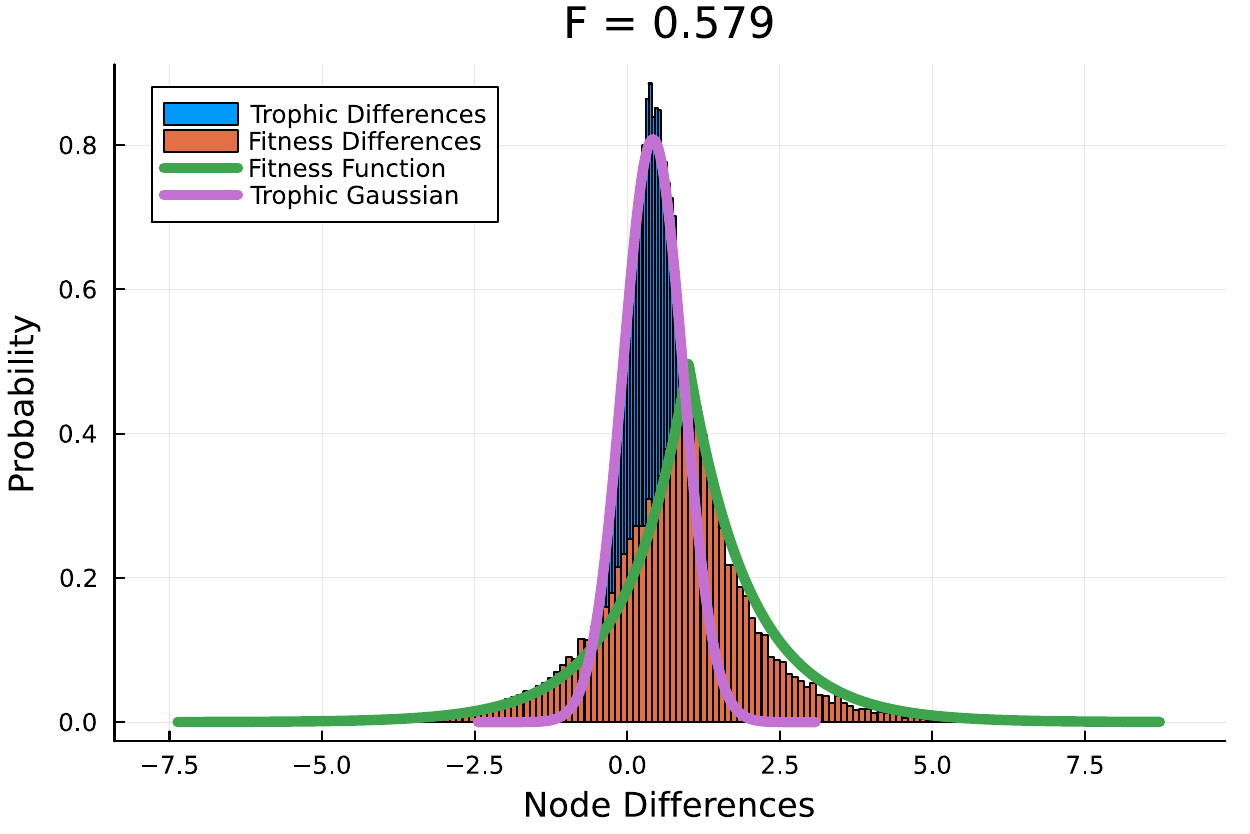}
        	\caption{Laplace. Equation \ref{eq:Laplace} $\mu_f=1$, $b=1$ }
        	
        \label{fig:fitness_Laplace}     
        \end{subfigure} 
\hfill        
\begin{subfigure}[t]{0.32\textwidth}        		
\centering
            \includegraphics[width=\textwidth]{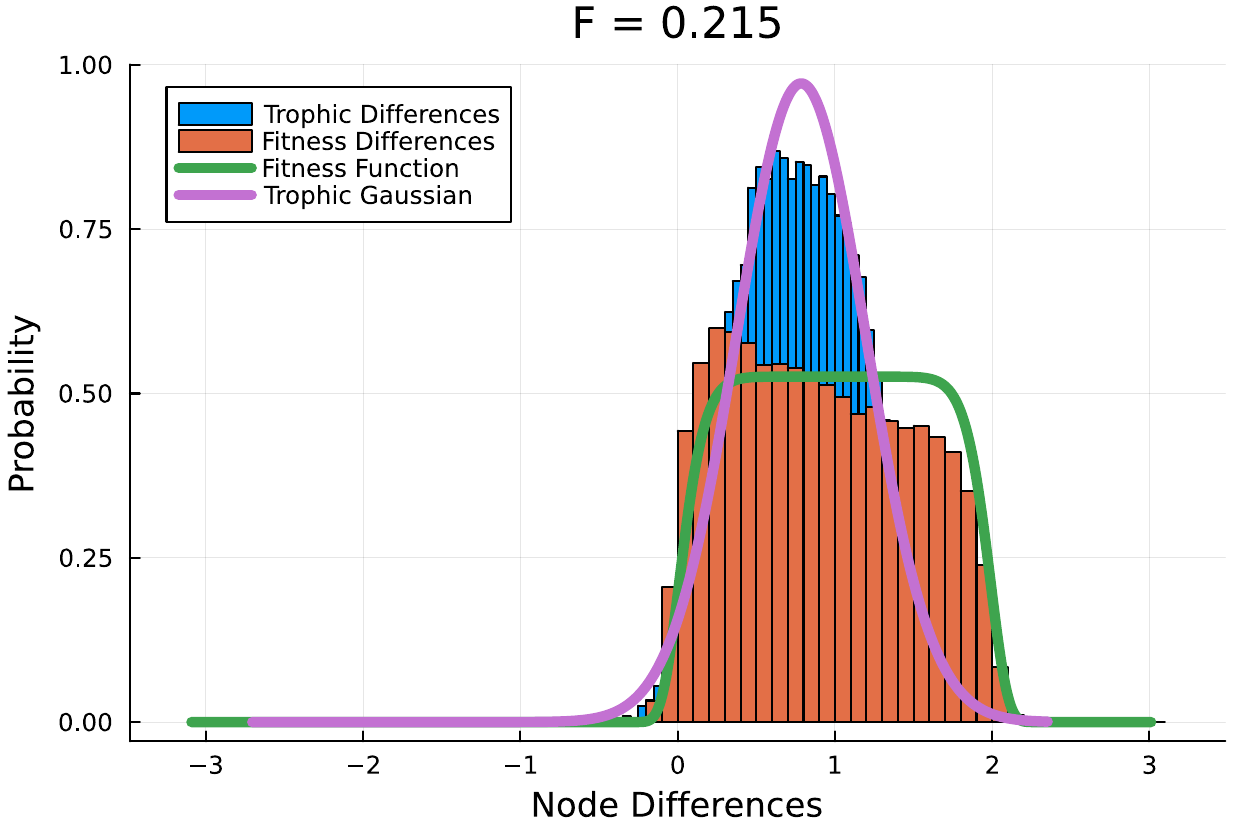}
        	\caption{Generalised Normal $\nu =10$.Equation \ref{eq:gen_normal1} $\mu_f=1$, $b=1$, $\nu =10$.}
        	
        \label{fig:fitness_gen_normal_beta}     
        \end{subfigure} 
        \hfill
\begin{subfigure}[t]{0.32\textwidth}        		
\centering
            \includegraphics[width=\textwidth]{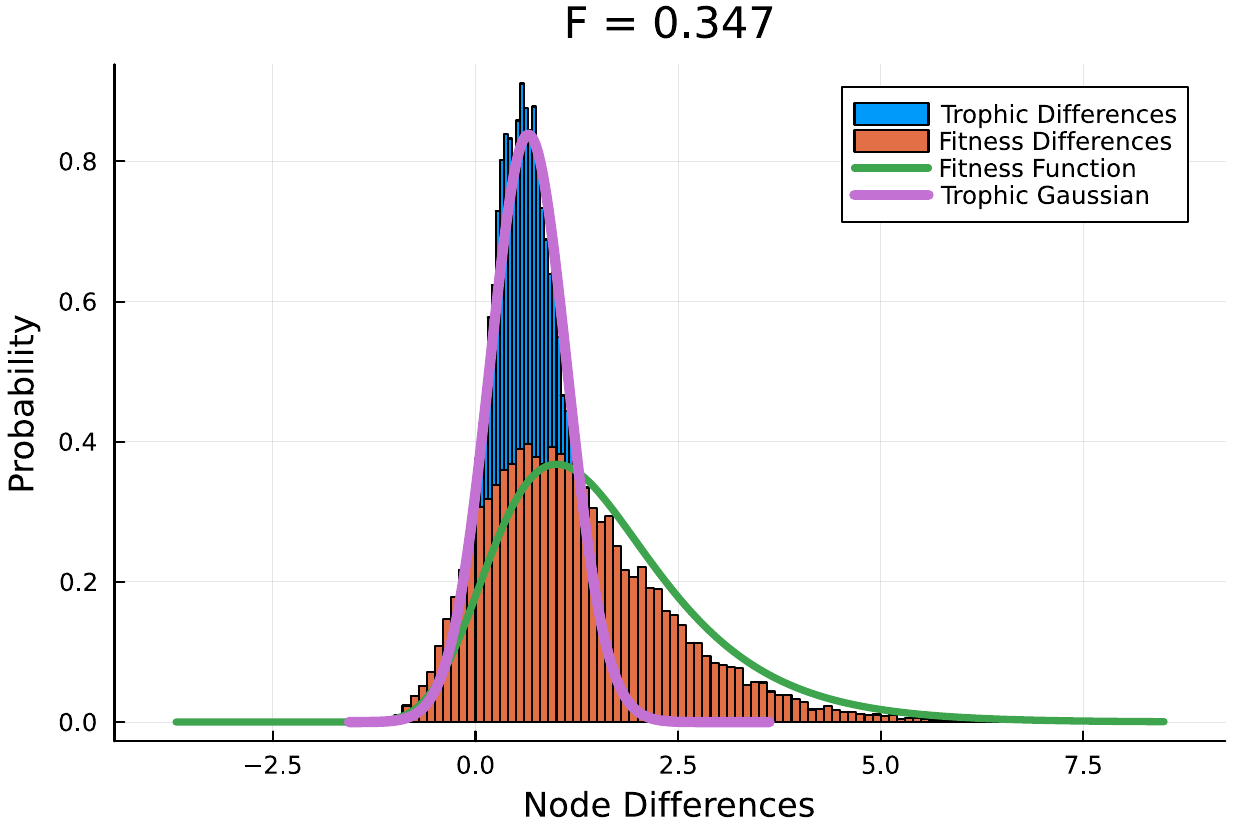}
        	\caption{Gumbel. Equation \ref{eq:gumbel} $\mu_f=1$, $b=1$.}
        	
        \label{fig:fitness_gumbel} 
        \end{subfigure} 
        \hfill
\begin{subfigure}[t]{0.32\textwidth}        		
\centering
            \includegraphics[width=\textwidth]{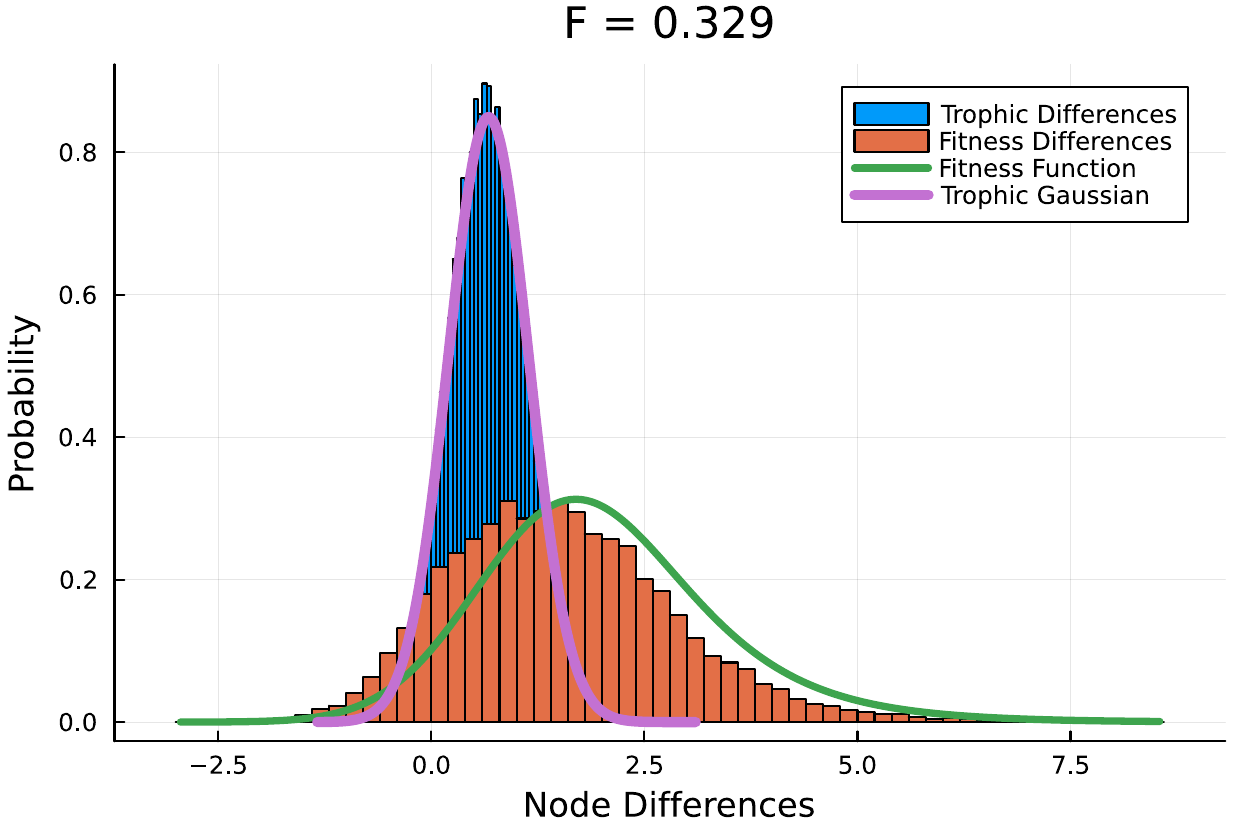}
        	\caption{Exponentially Modified Gaussian. Equation \ref{eq:exp_gaussian} $\mu_f=1$, $\sigma_e=1$, $\lambda =1$. }
        	
        \label{fig:fitness_exp_mod_gauss}
        \end{subfigure} 
        \hfill
\begin{subfigure}[t]{0.32\textwidth}        		
\centering
            \includegraphics[width=\textwidth]{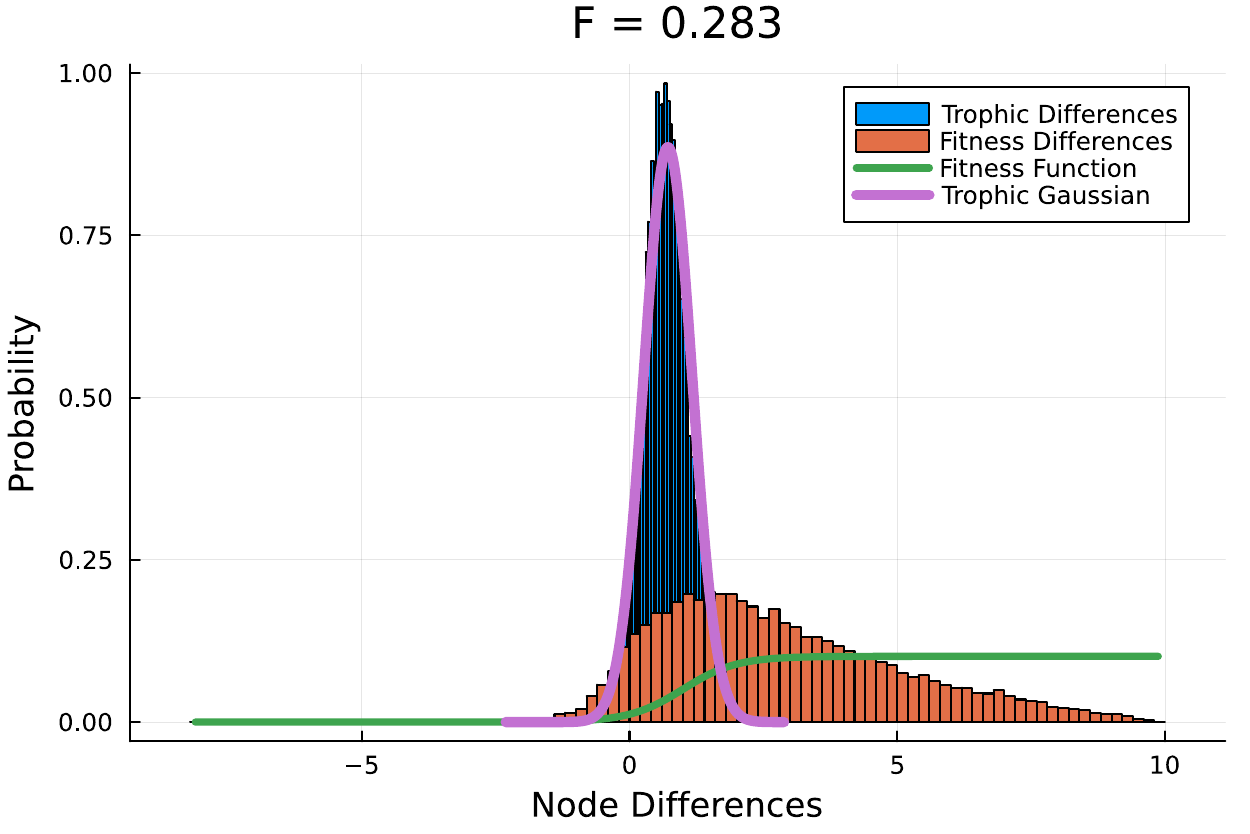}
        	\caption{Tanh. Equation \ref{eq:tanh} $a=1$, $T_1=1$.}
        	
        \label{fig:fitness_tanh}\end{subfigure} 
        \hfill
\begin{subfigure}[t]{0.32\textwidth}        		
\centering
            \includegraphics[width=\textwidth]{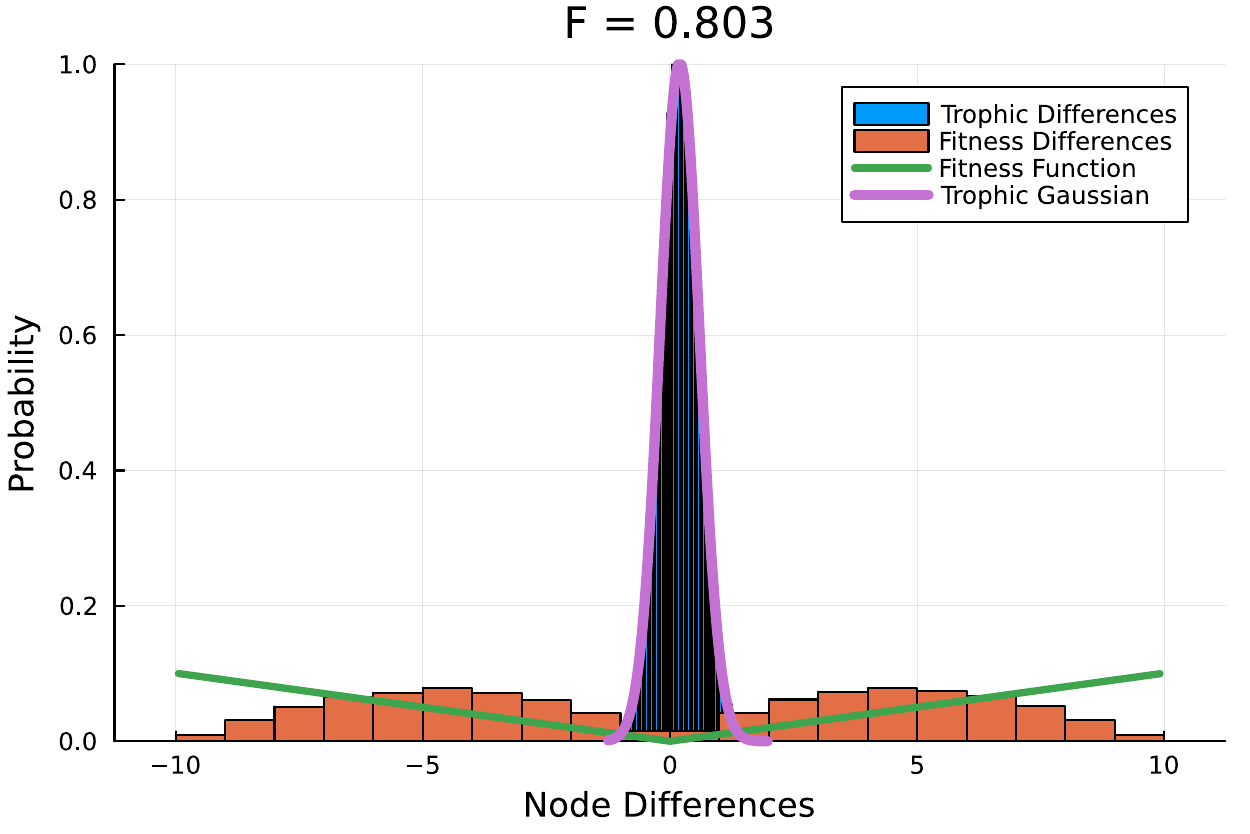}
        	\caption{Absolute Value. Equation \ref{eq:absolute} $T_2=1$.}
        	
        \label{fig:fitness_abs}\end{subfigure}
        \hfill
    \begin{subfigure}[t]{0.32\textwidth}        		
\centering
        \includegraphics[width=\textwidth]{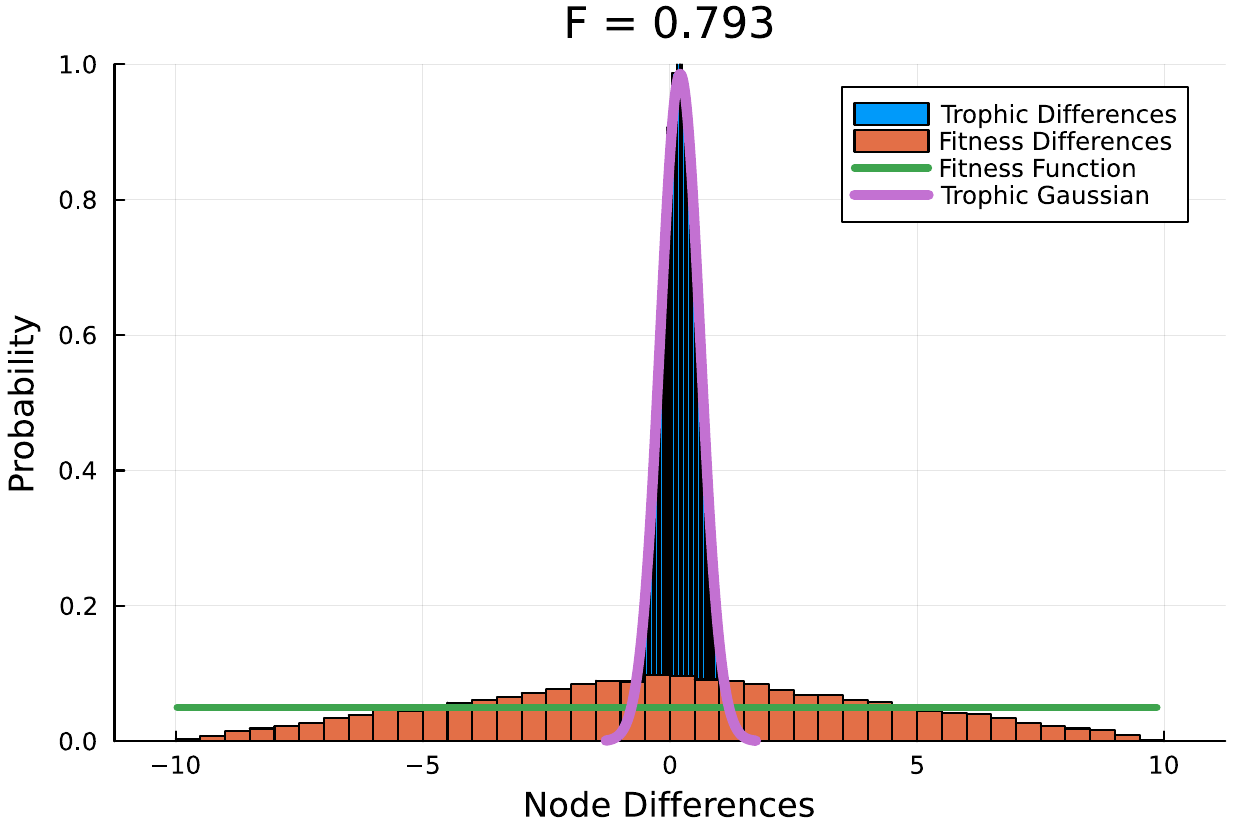}
        \caption{Uniform. Equation \ref{eq:uniform}}
        	
    \label{fig:fitness_uniform}\end{subfigure}
    \hfill
\begin{subfigure}[t]{0.32\textwidth}   
\centering
            \includegraphics[width=\textwidth]{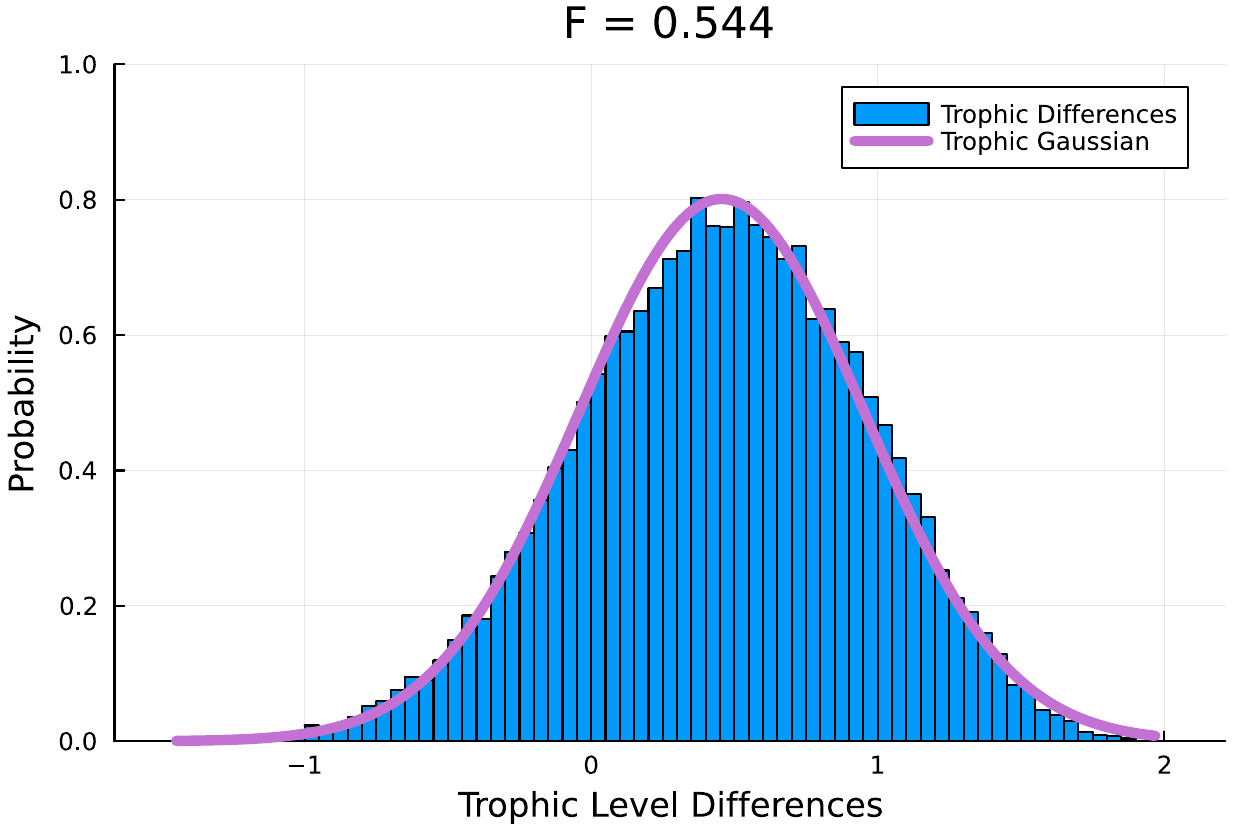}
        	\caption{ Scale Free. $\alpha =1$. Constant fitness function.}
        	
        \label{fig:fitness_scale_free}        \end{subfigure} 
        \hfill
\caption{Distributions of trophic level and fitness differences for different fitness functions, parameters and functions given in individual figure captions. Networks have $N=5000$ nodes. Fitness is uniformly distributed between 0 and 10. Nodes are added with 5 new edges. Seed graph is a path of length 5.  Degree-based preferential attachment exponent set to zero.}
\label{fig:level_distrubtions}
\end{figure}

This is shown in figure \ref{fig:level_distrubtions}. For most of the fitness distributions the measured normalised fitness distribution is very similar to the probability density function implied by the fitness function. This is useful as it shows that using this generative method preserves the structure of the fitness function. The only cases where it fails are when the fitness function is a hyperbolic tangent or uniform, figures \ref{fig:fitness_tanh} and \ref{fig:fitness_uniform}, where the very large fitness differences are rare, as it is unlikely to pair the largest and smallest fitness together when randomly pairing nodes.

When networks are generated with a variety of non-Gaussian fitness distributions the trophic level difference distributions are all well approximated by a Gaussian probability distribution with parameters which can be directly calculated from $F$, shown in figure \ref{fig:level_distrubtions}. This is done using results from \cite{MacKay2020HowNetwork} that the trophic incoherence can be written in terms of the average level difference, $\mu$, and the standard deviation of the level difference, $\sigma$, as \begin{equation}
    F = \sigma^2 + \mu^2 - 2\mu + 1. 
\end{equation} 
This is further analysed in \cite{MacKay2020HowNetwork} where it is shown that the expression for the average level difference can be written as
\begin{equation}
    \mu = 1 - F,
\end{equation}
which allows the  standard deviation to be written as \begin{equation}
    \sigma = \sqrt{F}\sqrt{1-F}.
\end{equation}

The good agreement between the level distribution and analytical Gaussian in figure \ref{fig:level_distrubtions} backs up arguments and assumptions in previous work \cite{Rodgers2023StrongNetworks,Johnson2017LooplessnessCoherence} that we expect the distribution of trophic level differences to be approximately Gaussian in real networks. However, it also demonstrates a limitation of the trophic analysis approach as information about the fitness distribution is lost in the calculation of the trophic level as all inputs are mapped to an approximate Gaussian.   

\subsubsection{Potential Explanations for Gaussian Trophic Level Difference Distributions}

 The Gaussian distribution is one which arises in many different settings and as such there are many ways for it to appear, so there are several possible explanations for the behaviour of the trophic difference distribution. One possible argument is linked to the minimisation of the squared penalty function and its Bayesian interpretation as it is well known that minimising a squared loss function can be thought of as a Gaussian likelihood and that $L_2$ regularisation can be interpreted as a zero mean Gaussian prior.

Consider a set of networks with a given set of $N$ nodes and a prescribed number of $L$ edges. Each network can be uniquely characterised by the corresponding $N \times N$ adjacency matrix $A$ of $L$ non-zero elements, $A_{ij}=1$ if and only if there is an edge from node $i$ to node $j$, otherwise $A_{ij}=0$.
Given a vector $ \Delta h $ of trophic level differences over the $L$ edges, where $\Delta h _e$ is the trophic level of the destination node of an edge, $e$, minus the trophic level of the source node of the edge.
The trophic level differences distribution yields a likelihood over possible edge 
topologies connecting the $N$ nodes, $P(A \vert \Delta h)$.
Given a prior $P(\Delta h)$ over possible trophic level difference vectors, 
if we are supplied with a particular network with edge structure $A$,
the nodes will be expected to have trophic level differences $ \Delta h$ with posterior probability,

\begin{equation}
    P(\Delta h \vert A) = \frac{P(A \vert \Delta h) P(\Delta h)}{P(A)}.
\end{equation}

Let us adopt a Gaussian likelihood model stating a preference for the trophic level difference of two nodes connected by an edge to differ by 1 (trophic level of the child larger than that of the parent), modulated by a ‘tolerance’ scale parameter $\sigma_\Delta$ (standard deviation)

\begin{equation} \label{eq:likelihood}
    P(A \vert \Delta h) \propto \prod^L_{e=1} \exp{\left[ -\frac{1}{2}\left(\frac{ \Delta h_e - 1}{\sigma_\Delta}\right)^2 \right]},
\end{equation}
where each each label $e$ represent as an edge which goes from node $i$ to $j$.

Under the assumption of flat prior over trophic level difference vectors $\Delta h $, the maximum a-posteriori estimate of the level differences, given the network A, is equivalent to minimising trophic incoherence, equation \ref{eq:F}, since the log-likelihood (up to constant terms) reads  

 \begin{eqnarray}
    \ln{P(\Delta h \vert A)} 
    &=&  -\frac{1}{2 \sigma_\Delta^2}\sum_{e=1}^L (\Delta h_e-1)^2  + \text{constant} \\
    &=& -\frac{1}{2 \sigma_\Delta^2}\sum_{ij}A_{ij}(h_{j} -h_{i}-1)^2  + \text{constant},
\end{eqnarray}
using the fact that the adjacency matrix in this case is unweighted and it can be used to indicate the existence of an edge. By this construction, we  show that the minimising of trophic incoherence can be interpreted as maximising the likelihood under the assumption of the Gaussian likelihood model for edge-based trophic differences,
equation \ref{eq:likelihood}. Under the set of assumptions made, this may explain why we observe trophic level difference distrubtions which are well approximated by Gaussian distributions in this work and in real networks \cite{Rodgers2023StrongNetworks}.

However, it is possible to come up with different generative models which lead to similar minimisation problem as a result of difference choices of likelihood functions and priors.  For example, in \cite{DeBacco2018ANetworks} it was shown that using a Poisson likelihood model, with the mean being a Gaussian function of the level differences multiplied by a sparsity control parameter, a different generative model for levels (ranks) and no fixed edge number after certain assumptions (and discarding limiting terms) maximisation of the likelihood of this model is equivalent to minimising the SpringRank Hamiltonian \cite{DeBacco2018ANetworks}. It is also argued in \cite{DeBacco2018ANetworks} that the ranks in SpringRank are distributed as a multivariate Gaussian with variable noise levels. Additionally, it can be shown that minimising trophic incoherence can be related to the likelihood of another random graph model \cite{Gong2021DirectedModels} which allows the relative importance of linear and periodic hierarchy to be studied. It was also shown in the study of node ranking in \cite{Timar2021SimpleInteractions} that the assumption that each the rank of a node represents an average performance which is normally distributed leads to the probability of a set of `results'. These are a set of edge directions determined by competition between nodes imposed on an undirected graph of interactions, and can be written as a likelihood function which is a Gaussian of the rank differences where the maximisation leads to a minimisation of the energy of a network of directed springs \cite{Timar2021SimpleInteractions}, which is similar in concept to SpringRank or Trophic Analysis.

Of course in real systems, if the differences in trophic level of nodes represents a real physical quantity they could be well approximated as a Gaussian for various reason like the central limit theorem, approximation of a binomial or the fact that the Gaussian distribution is the maximum entropy distribution with specified mean and variance. However, what we show here is that due to the form of $F$ its minimisation can be linked to a Gaussian likelihood model which may partially explain why the analytical Gaussian derived from trophic incoherence fits the level difference distrubtion so well.

Figure \ref{fig:level_distrubtions} and the link of trophic analysis to Gaussian distributions shows that trophic analysis may be less useful in situations where it is known the fitness differences are not distributed as a Gaussian. As a result,  it may be useful to cast the ranking problem in terms of a different optimisation function which corresponds to a different specifically chosen likelihood maximisation problem and create a generalised version of trophic analysis which is something that we leave for future work.

The trophic level difference distribution also has some particular properties derived in \cite{MacKay2020HowNetwork}. The mean, $\mu $, is equal to $\mu= 1-F$, which means the mean is always between 0 and 1 and the variance, $\sigma^2$ can be written as $\sigma^2 = \mu(1-\mu)$ which is very similar to the relationship between the mean and the variance of a Bernoulli distribution. This also means the variance is maximised when the mean level difference is $0.5$. This makes sense as when the network is fully coherent, $F=0$ all the level differences are $+1$ and then when the network is completely incoherent, $F=1$ every node has exactly the same trophic level so the difference across all edges is 0 and the variance is zero.

\subsection{Relating Trophic Incoherence to  Fitness Difference Distribution}

We would like to be able to relate the trophic incoherence of the network, which is a structural property, to the external parameters which make up the fitness function, which decides if nodes are connected based on an external node fitness. If the fitness distribution is ºvery tightly distributed around the mean and all the edges point upwards in fitness then it is likely that the network will be very coherent. The standard deviation and the mean of the trophic level difference distribution can both be expressed in terms of $F$ \cite{MacKay2020HowNetwork}. However, we cannot directly match each of these structural properties to the fitness difference distribution as the fitness distribution can be any scale, while the mean level difference is bound between 0 and 1 \cite{MacKay2020HowNetwork}. However, we can use the coefficient of variation, ratio of standard deviation and the mean, to provide a measure of how coherent both the distribution of level differences is and how coherent the distribution of fitness differences is. The coefficient of variation gives a measure of the size of the standard deviation relative to the mean and acts as an alternative measure of coherence: when the network has $F=0$ the coefficient of variation is zero as the mean is one and standard deviations is zero, and when $F=1$ the coefficient of variation becomes infinite as the mean level difference is zero. The key thing about the coefficient of variation is that it is a dimensionless quantity so it allows comparison between the trophic level properties and the fitness properties of real-world systems which could be off any scale and have fitness steps of any size.

By assuming the coefficient of variation of the trophic level distribution matches the coefficient of variation of the distribution of fitness differences,
\begin{equation}
    \frac{\sigma}{\mu} \approx \frac{\sigma_f}{\mu_f},
\end{equation}
we can derive an expression for the Trophic Incoherence in terms of parameters of the fitness distribution only. Where this approximation is made on the basis that even though the trophic level difference distribution is on a different scale to the fitness difference distribution the variability of the fitness difference distribution should affect the measured trophic incoherence and the coefficient of variation gives us a way to compare the distributions using a dimensionless parameter.

Using results from \cite{MacKay2020HowNetwork} that the average trophic level difference is \begin{equation}
    \mu = 1 - F
\end{equation}
and the standard deviation is given by \begin{equation}
    \sigma = \sqrt{F}\sqrt{1-F}.
\end{equation}
This leads to the expression \begin{equation}
    \sqrt{\frac{F}{1-F}} = \frac{\sigma_f}{\mu_f},
\end{equation}
which can be rearranged to give the result 
\begin{equation} \label{eq:prediction}
    F = \frac{\sigma_f^2}{\sigma^2_f + \mu^2_f}.
\end{equation}
This can also can also be expressed directly in terms of the coefficient of variation as \begin{equation}
    F = \frac{(\frac{\sigma_f}{\mu_f})^2}{(\frac{\sigma_f}{\mu_f})^2 + 1 }.
\end{equation} 

\begin{figure}[H]
    \centering
\begin{subfigure}[t]{0.45\textwidth}        		
\centering
            \includegraphics[width=\textwidth]{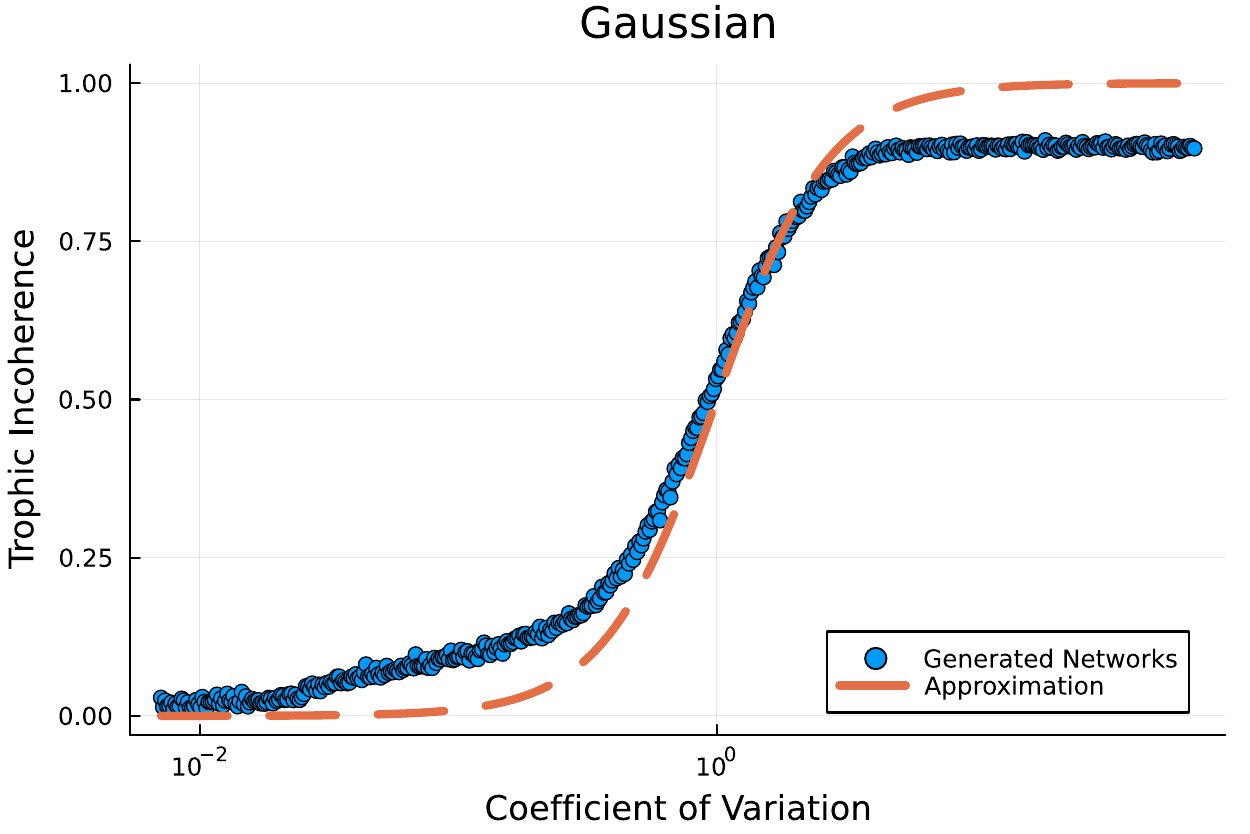}
        	\caption{ Gaussian. Equation \ref{eq:gaussian} $\mu_f= 1$, $\sigma_f$ spaced $\frac{10^{-2}}{\sqrt{2}}$ and $\frac{10^{2}}{\sqrt{2}}$}
        	
        \label{fig:gaussian_predict}
        \end{subfigure}
        \hfill
        \begin{subfigure}[t]{0.45\textwidth}        		
\centering
            \includegraphics[width=\textwidth]{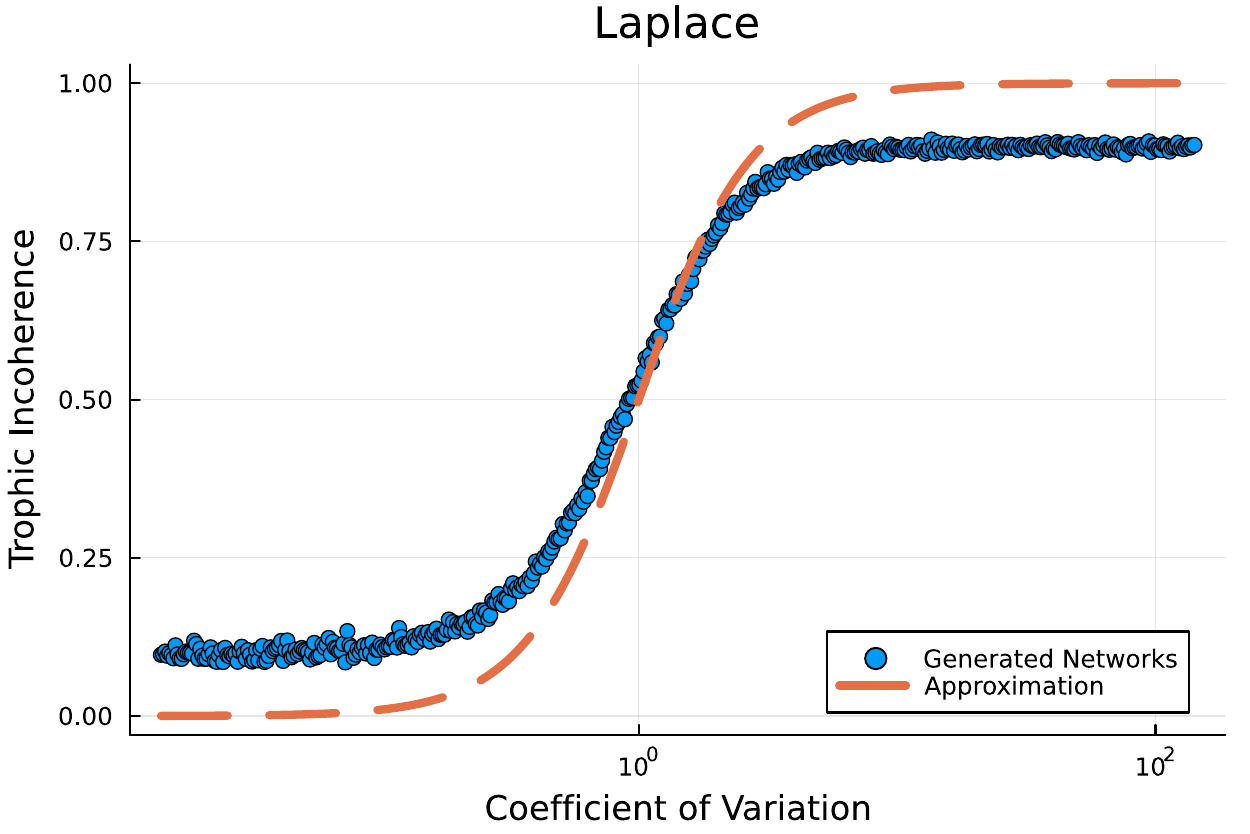}
        	\caption{Laplace. Equation \ref{eq:Laplace} $\mu_f=1$. Parameter $b$ spaced between from $10^{-2}$ to $10^2$.}
        	
        \label{fig:laplace_predcit}    
        \end{subfigure} 
        \hfill
\begin{subfigure}[t]{0.45\textwidth}        		
\centering
            \includegraphics[width=\textwidth]{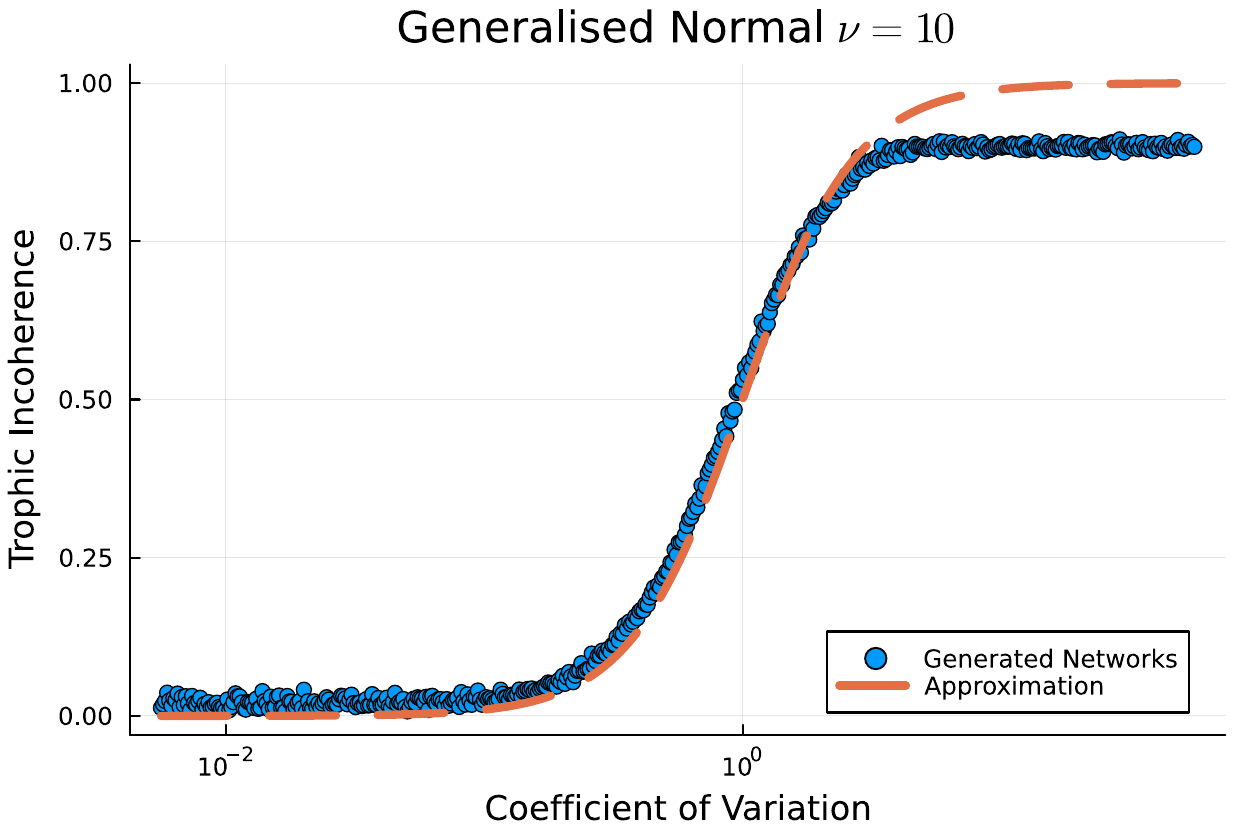}
        	\caption{Generalised Normal $\nu =10$. Equation \ref{eq:gen_normal1} $\mu_f=1$, $\nu =10$. Parameter $b$ spaced between from $10^{-2}$ to $10^2$. }
        	
        \label{fig:generlaised normal_10_predcict}   
        \end{subfigure}
        \hfill
\begin{subfigure}[t]{0.45\textwidth}        		
\centering
            \includegraphics[width=\textwidth]{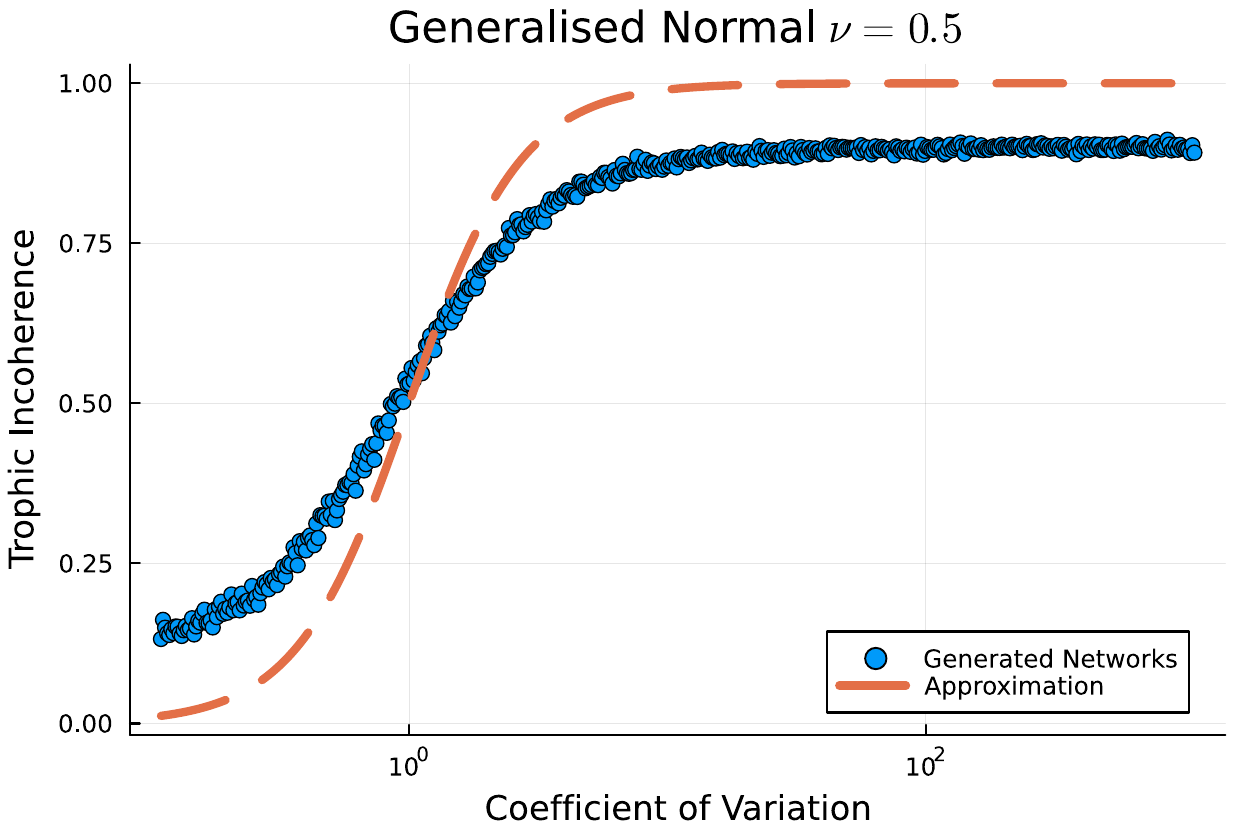}
        	\caption{Generalised Normal $\nu =0.5$.Equation \ref{eq:gen_normal1} $\mu_f=1$, $\nu =0.5$. Parameter $b$ spaced between from $10^{-2}$ to $10^2$. }
        	
        \label{fig:genelatsied_normal_0.5_predict}   
        \end{subfigure}
        \hfill
\begin{subfigure}[t]{0.45\textwidth}        		
\centering
            \includegraphics[width=\textwidth]{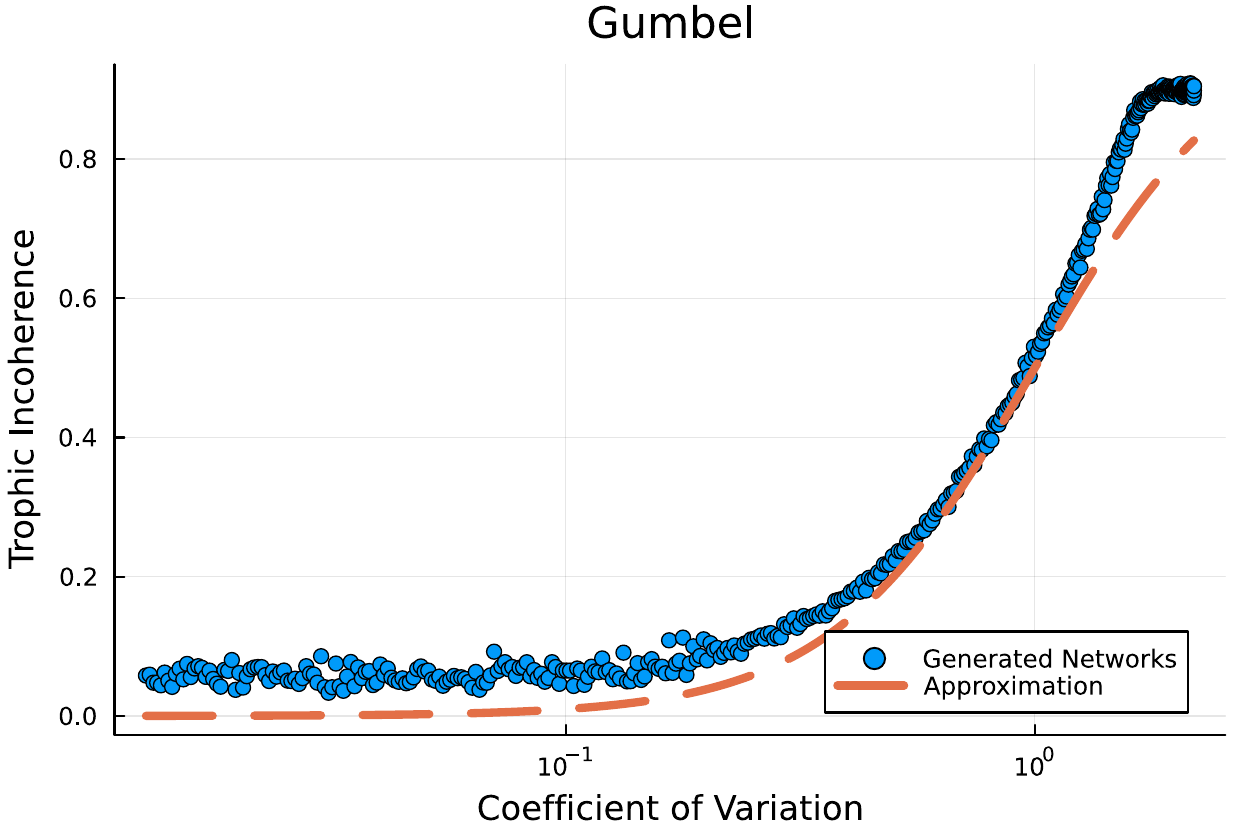}
        	\caption{Gumbel. Equation \ref{eq:gumbel} $\mu_f=1$, Parameter $b$ spaced between from $10^{-2}$ to $10^2$.}
        	
        \label{fig:gumbel_predict}   
        \end{subfigure} 
        \hfill
\begin{subfigure}[t]{0.45\textwidth}        		
\centering
            \includegraphics[width=\textwidth]{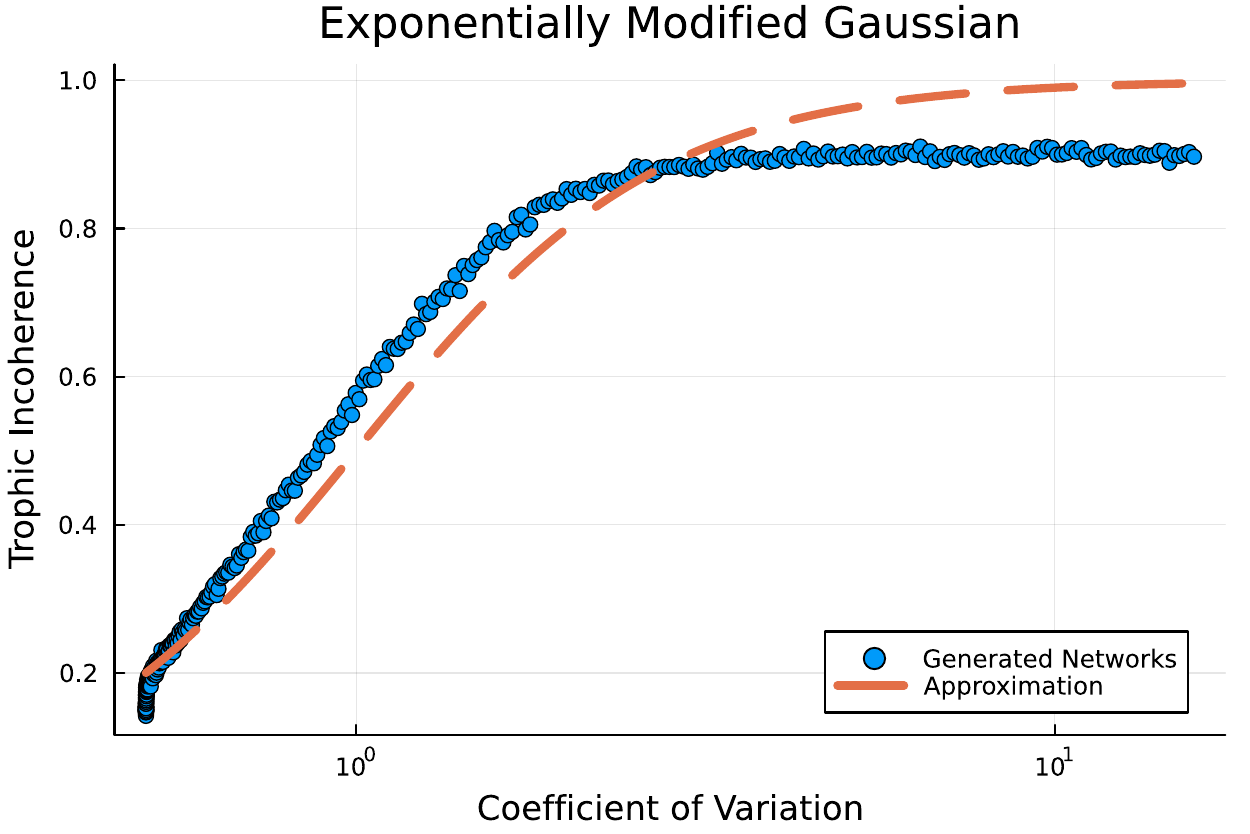}
        	\caption{Exponentially Modified Gaussian. Equation \ref{eq:exp_gaussian} $\mu_f=1$, $\lambda =1$. Parameter $\sigma_e$ spaced between $10^{-2}$ to $10^{1.5}$.}        	
        \label{fig:exp_mod_gauss_predict}   
        \end{subfigure} 

\caption{ Trophic incoherence and analytical approximation of trophic incoherence based on the mean and standard deviation of the fitness difference via equation \ref{eq:prediction}. Each network has $N=1000$ nodes. 10 edges are added with every new node with equal probability of being an in or out edge of the new node. Seed graph is a path of length 10. Fitness uniformly distributed between 0 and 10. Parameters and functions given in each sub-figure caption with the spacing of all varied parameters being logarithmic.  Degree-based preferential attachment exponent set to zero.}
\label{fig:prediction_of_F_with_coeff_var}
\end{figure}

This is a useful prediction as it allows the incoherence to be expressed as function of the fitness distribution which is an external property imposed on the network and not a structural property like incoherence.  The accuracy of this prediction, equation \ref{eq:prediction}, is shown in figure \ref{fig:prediction_of_F_with_coeff_var}. This prediction works reasonably well, even for networks which are generated with a non-Gaussian fitness functions, which shows how trophic incoherence depends on how the tightly the network follows the defined fitness hierarchy. It fails slightly at very low incoherence as the specific functions and finite networks may not generate networks of extremely low coherence as predicted by the approximation and similarly at high incoherence, the generative model does not produce the balanced networks which correspond to maximally incoherent networks, instead producing variants of random graphs where the trophic incoherence plateaus below $F=1$.

The relationship between trophic incoherence and the coefficient of variation follows a sigmoid shaped curve when plotted on a  logarithmic axis and can be thought of as three distinct regions. Firstly, the sigmoid shape of the curves when plotting on a logarithmic axis, figure \ref{fig:prediction_of_F_with_coeff_var} can be explained by equation \ref{eq:prediction} which can be reparameterised such that $(\frac{\sigma_f}{\mu_f})= 10^x $. This leads to \begin{equation}
    F = \frac{1}{1 + 10^{-2x}},
\end{equation}
which is a sigmoid curve in log coefficient of variation in fitness. When the coefficient of variation of the network is much less than 1 we are in the region where we predict a very coherent network with clear hierarchy, when the coefficient of variation is much larger than 1 we expect the network to be very incoherent and there to be little hierarchy determined by fitness; and when the coefficient of variation is around the order of $1$, we are in an intermediate regime where moderate incoherence is expected and there is some hierarchy and ordering but not as strict as in a very coherent network. 

Moreover, the approximation also works for the more complicated functions like the Gumbel, figure \ref{fig:gumbel_predict}, and Exponentially Modified Gaussian \ref{fig:exp_mod_gauss_predict} where the coefficient of variation tends to a constant as the standard deviation grows large in the case of Gumbel and to constant as the standard deviation becomes very small in the case of the exponentially modified Gaussian. This figure demonstrate how measuring the trophic incoherence can give you information about the parameters used to generate the network and information about the fitness distribution, even if the actual distribution information is lost when the trophic levels are calculated.

In this work, we have varied the coefficient of variation by modifying the standard deviation but an alternative approach could modify the mean and fix the standard deviation instead. However, care would need to taken with zero mean and it may be necessary to work with the inverse of the coefficient of variation. This could link to other work on Trophic Analysis where the trophic incoherence has been shown to equal to one minus the mean trophic level difference \cite{MacKay2020HowNetwork}.

\section{Interplay of Fitness Interactions and Degree-based Preferential Attachment}

The final way we can explain the ubiquity of coherent networks in nature as well as understand the relationship between trophic level and fitness is to look at the interplay of degree-based preferential attachment and connection based on fitness hierarchy. This is done by using an addition probability which is formed by a Gaussian fitness interaction, of which we vary the standard deviation, that we multiply by degree-based preferential attachment, where we vary the attachment exponent.

\begin{figure}[H]
    \centering
\begin{subfigure}[t]{0.48\textwidth}        		
\centering
            \includegraphics[width=\textwidth]{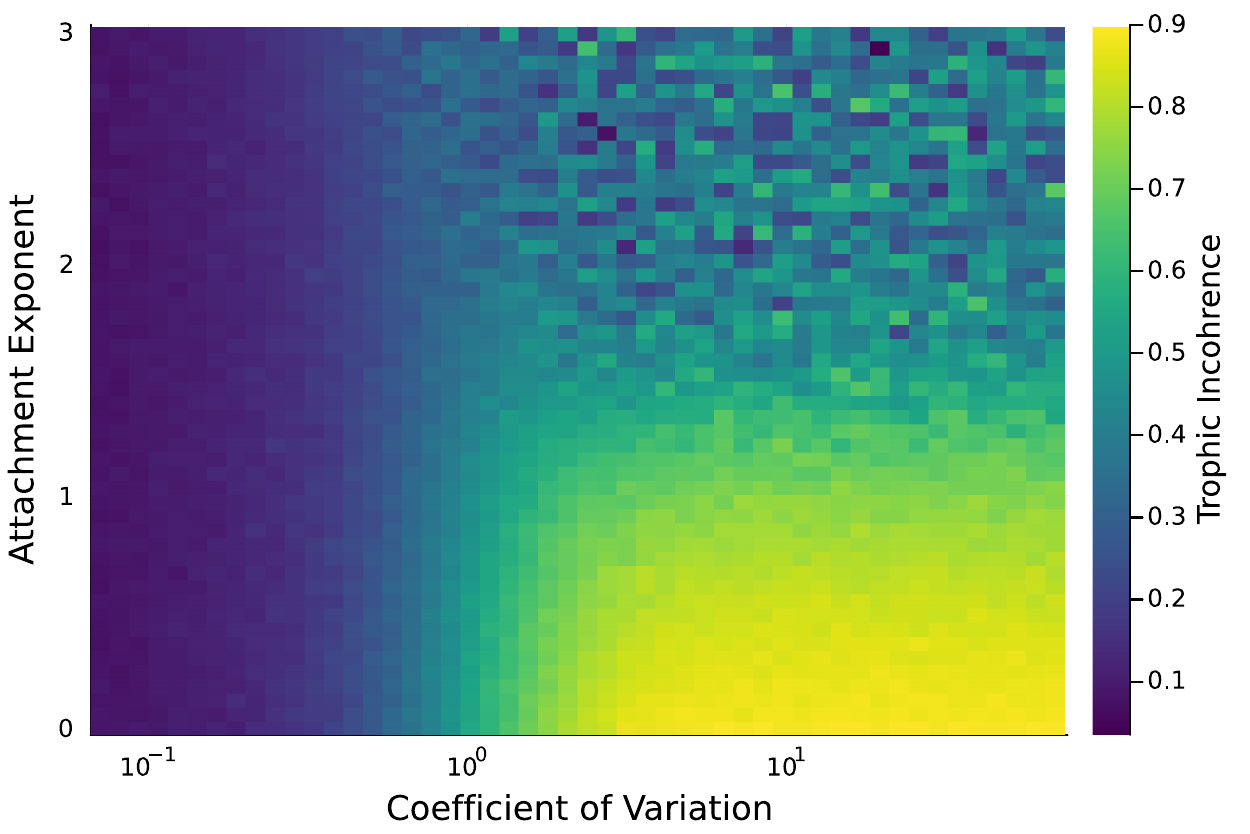}
        	\caption{Trophic Incoherence}
        \label{fig:combination_F}\end{subfigure}
        \hfill
\begin{subfigure}[t]{0.48\textwidth}   
\centering
            \includegraphics[width=\textwidth]{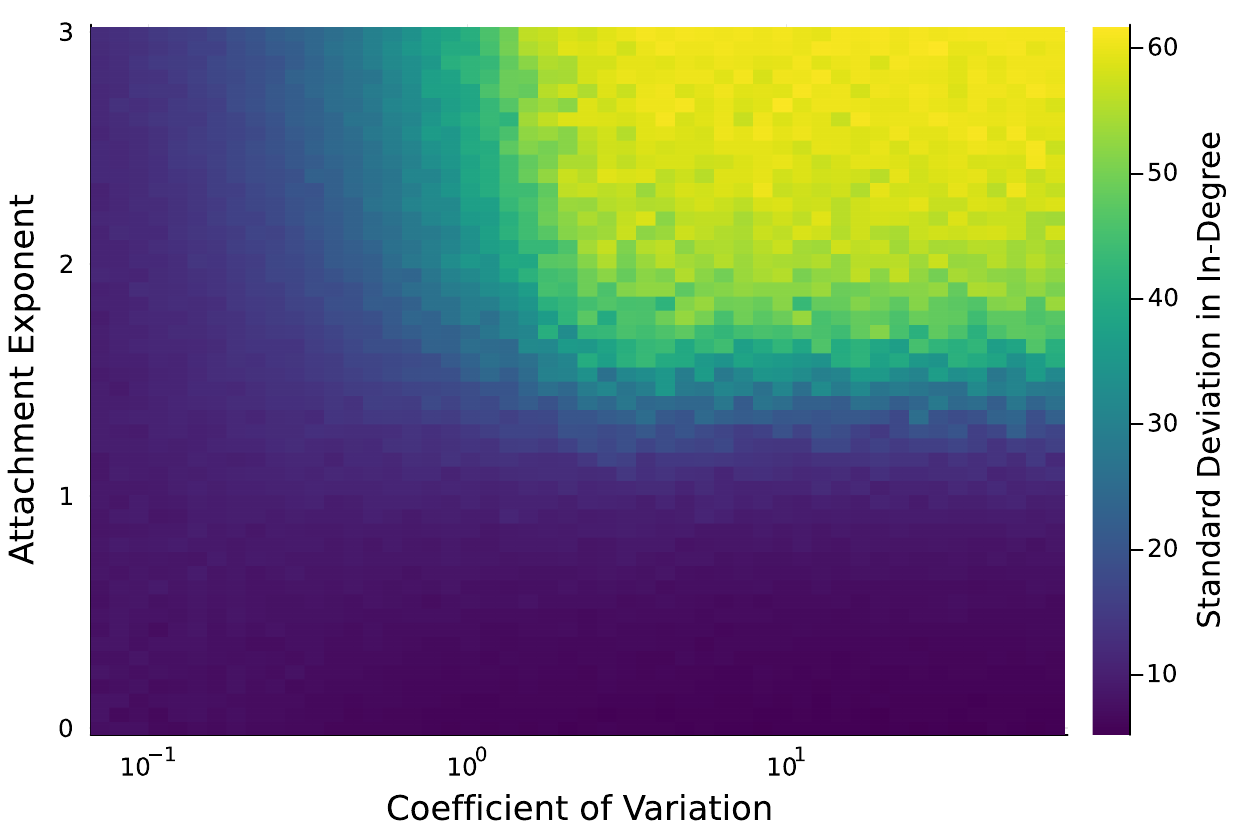}
        	\caption{Standard Deviation of In-Degree}
        	
        \label{fig:combination_degree}
        \end{subfigure} 
\begin{subfigure}[t]{0.48\textwidth}   
\centering
            \includegraphics[width=\textwidth]{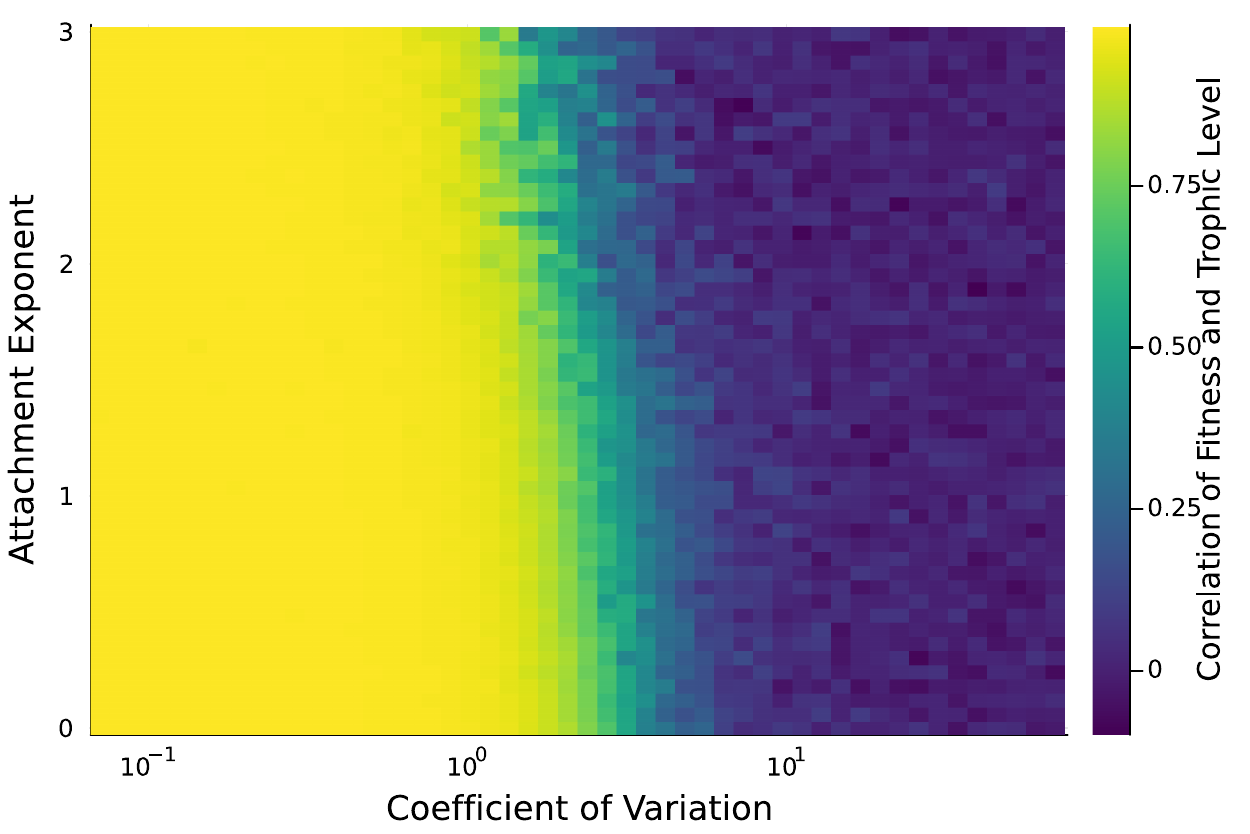}
         \caption{Correlation of Fitness and level}
         \label{fig:correlations}
\end{subfigure}
\hfill
\begin{subfigure}[t]{0.48\textwidth} 
\centering
           \includegraphics[width=\textwidth]{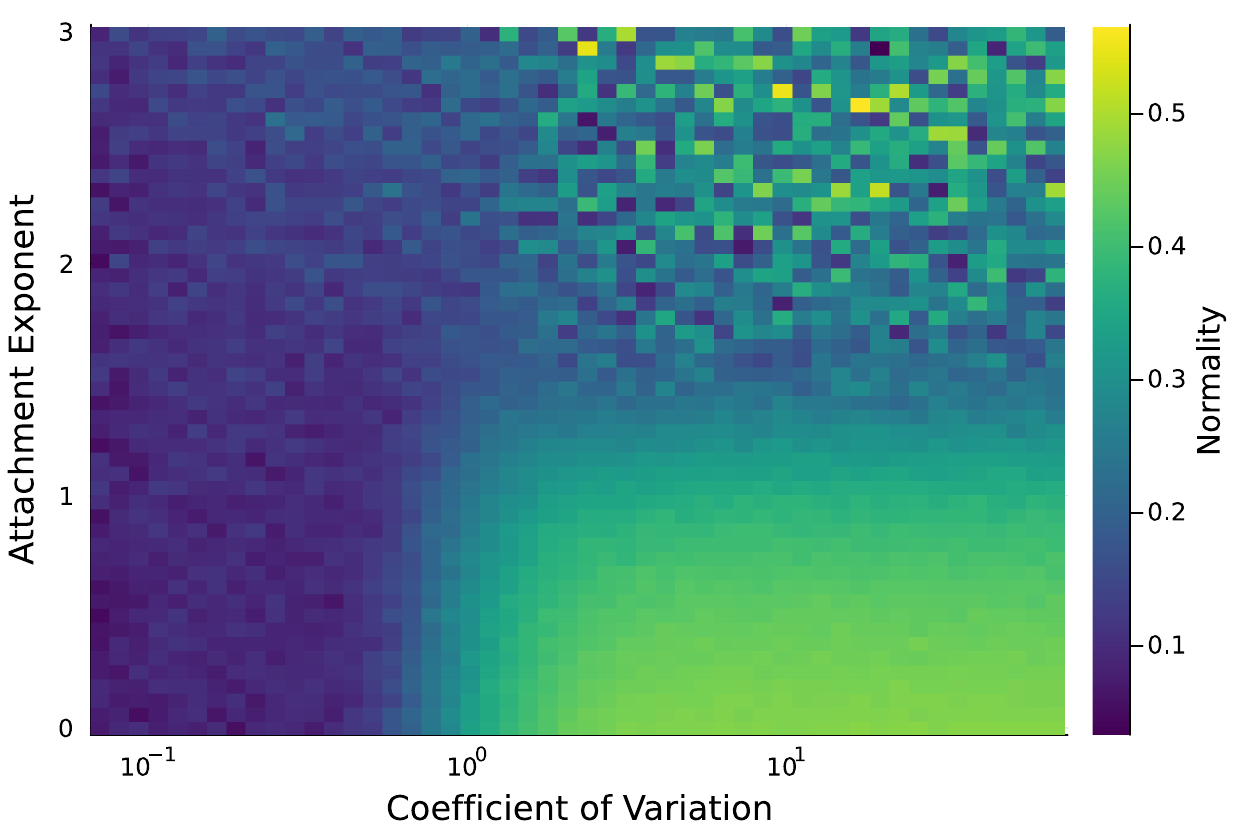}
       	\caption{Matrix Normality measured using convention in \cite{MacKay2020HowNetwork}, equation \ref{eq:normality}.}
       \label{fig:normality}
\end{subfigure}
\caption{Impact of varying coefficient of variation of a Gaussian fitness function and degree-based preferential attachment exponent on network properties. Each networks has  1000 nodes with 10 edges added with each new node with equal probability connecting in or out of the newly added node. The seed graph is a path of length 10. Fitness is uniformly distributed between zero and 10. We use a 50 by 50 grid of points with the preferential attachment evenly spaced between 0 and 3 and the fitness standard deviations logarithmically spaced between $\frac{10^{-1}}{\sqrt{2}}$ and $\frac{10^{2}}{\sqrt{2}}$. The mean fitness difference, $\mu_f =1$, for all points.}
\label{fig:combination_affects}
\end{figure}

We plot the changes in trophic incoherence and other network properties as networks are generated with varying exponent, $\alpha$, and coefficient of variation of the Gaussian fitness function in figure \ref{fig:combination_affects}.  In figure \ref{fig:combination_F}, we see that there are three distinct regimes. When the coefficient of variation of fitness is small the networks are very coherent as the hierarchy is strongly enforced by the fitness function. When the coefficient of variation of the fitness function is high we have two behaviours. A smooth regime, where the incoherence increases as $\alpha$ gets closer to 0 and a regime where $\alpha$ is greater than 1 and the preferential attachment reaches the super-linear regime. In this region, the networks become more coherent but the behaviour is much more variable as the networks are dominated by a few super-hubs. This can be understood by looking at figure \ref{fig:combination_degree} which shows the standard deviation of the in-degree distribution for the same sample of networks. The standard deviation in the degree distribution grows large when $\alpha>1$ and the coefficient of variation of the fitness function is large. This is due to hubs with very large degree being formed by the strong degree-based preferential attachment and this not being restricted by a fitness function which limits the range of nodes which can be connected to. In figure \ref{fig:correlations}, we show how interplay of the degree-based preferential attachment and fitness functions affect the correlation coefficient of the trophic level and the fitness. The results show a similar trend to when the preferential attachments is fixed, figure \ref{fig:corrleation_level_fitness}. With the correlation being very strong when the coefficient of variation is low and network is coherent and the correlation being weak when the coefficient of variation is high and the network is incoherent. This shows that even with additional degree-based preferential attachment contribution that the trophic level can be taken as a good proxy for the network fitness, when the coefficient of variation of the fitness function is low. Figure \ref{fig:correlations} also demonstrates an interesting feature that the line separating the two regimes of high and low correlation between fitness and level is diagonal, as when the preferential attachment is weaker the correlation of fitness and level is maintained for fitness functions which are less strict, while when the preferential attachment exponent is higher the correlation is more quickly destroyed. A similar trend can also be seen if we take one of the quantities which is related to trophic incoherence such as matrix normality \cite{MacKay2020HowNetwork,Nartallo-Kaluarachchi2024BrokenNetworks} which is measured following the convention of \cite{MacKay2020HowNetwork}, where normality, $y$, is measured by \begin{equation} \label{eq:normality}
    y =\frac{\sum_j\lvert \lambda_j\rvert^2}{\lvert\lvert A \rvert\rvert^2_F}.
\end{equation} 
Here, $\lambda_j$ are the eigenvalues of the adjacency matrix and $\lvert\lvert A \rvert\rvert_F$ is the Frobenius norm of the matrix $A$ defined as $\lvert\lvert A \rvert\rvert_F = \sqrt{\sum_{ij} \lvert A_{ij} \rvert^2}$. The full justification and explanation of this measure are given in \cite{MacKay2020HowNetwork}.

Figure \ref{fig:normality} shows a similar trend in normality as \ref{fig:combination_F} does for trophic incoherence. This fits with the close relationship between non-normality and trophic coherence described previously \cite{Johnson2020DigraphsSystems,MacKay2020HowNetwork}.

The fact that a combination of these effects can lead to many different values of trophic coherence suggests and explanation for why we observe non-trivial trophic coherence in many real-world systems. It can also give us information about how the networks are formed as the prevalence of large degree imbalance and its association with the trophic level can be an indication as to whether the trophic coherence derives from an underlying fitness interaction or strong  degree-based preferential attachment. Figure \ref{fig:correlations} also shows how trophic analysis can be useful as a tool to predict fitness variables as there is a good correlation between fitness and level in the low coefficient of variation regime even when the  degree-based preferential attachment is strong.

\subsection{Example Application to Historical Network of Ragusan Nobility}

\begin{figure}[H]
      		\centering
            \includegraphics[width=0.8\linewidth]{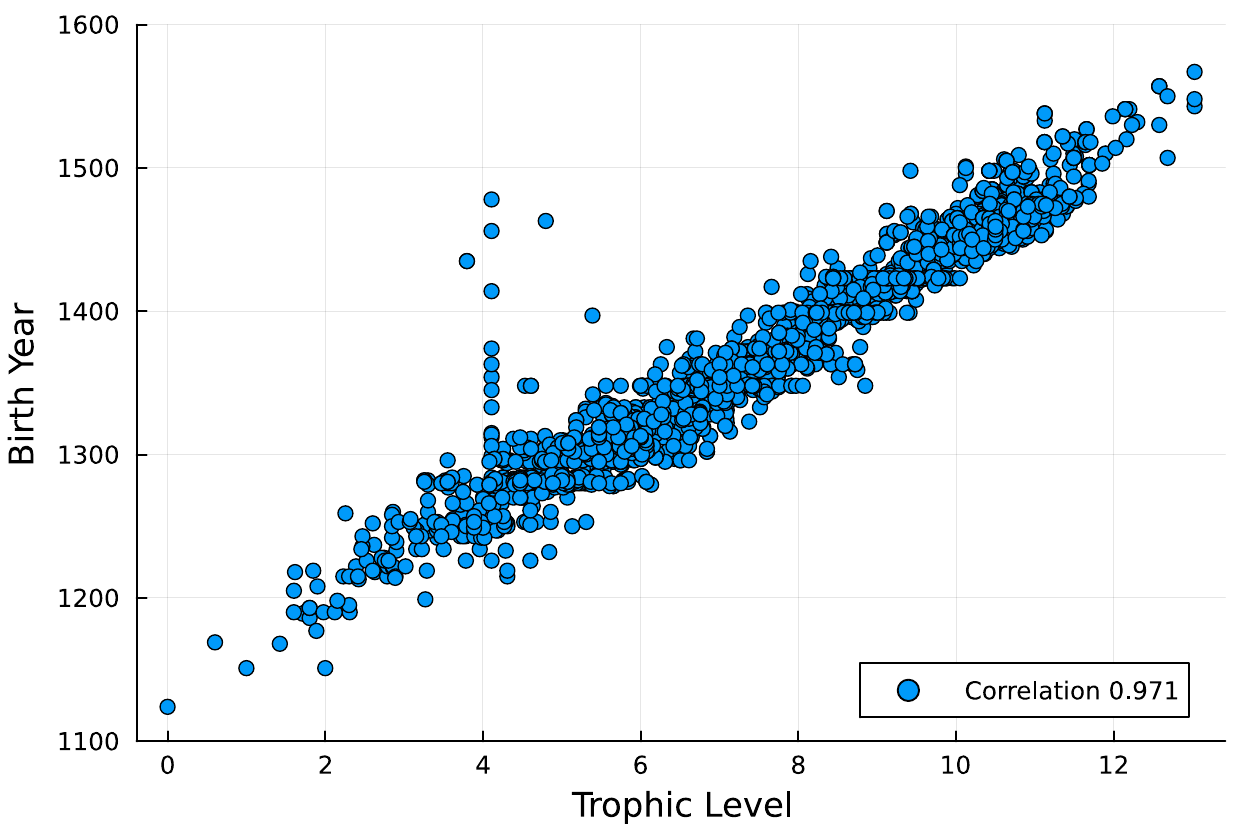}
        	\caption{  Known Birth Dates (3065 individuals) of Members of the Ragusan Nobility, \cite{Ragusan_Data_Pajek}, against Trophic Level with Pearson correlation coefficient given in the legend.}
        	
        \label{fig:ragusan_birth_level}
        \end{figure}

To highlight how our methods can be used to study real systems, we demonstrate with an example of a real network. The family tree of the nobility from the Republic of Ragusa from the 12th to the 16th century, data from \cite{Ragusan_Data_Pajek} and converted using \cite{Gedcom_parser}, with metadata which roughly corresponds to node fitness which we compare to the quantities found via trophic analysis. The Republic of Ragusa was a merchant republic located in Dubrovnik, Croatia and ruled by a number of noble merchant families. We construct a network containing 5,999 people and draw 9315  edges from parent to child and making no distinction between males and females in the network. We compute the trophic levels of the nodes in this network and compare with the birth dates of the nodes for which this is known (3065 individuals). This is shown in figure \ref{fig:ragusan_birth_level}. Here we find a good correlation between the year in which someone was born and their trophic level. This is to be expected as the year you are born can only be a within a range after your patents birth, hence we expect trophic analysis of the network to detect this structure. The size of each trophic level roughly corresponds to generation of people which shows that trophic analysis is detecting some real features of the network. There are some issues in the correlation around the mid 14th century however there are several factors which could explain this. The structure of the nobility was changing between the 12th and 14th centuries and was established by stature in 1332 \cite{Ragusan_Data_Pajek} with no new families accepted after this time. So the changing legal basis of who is included and recorded in the nobility is likely to explain the lack of correlation at this time. This is also the time of the `Black Death' plague in Europe so we expect a loss of population to occur at that time.

\begin{figure}[H]
    \centering
\begin{subfigure}[t]{0.48\textwidth}        		
\centering
            \includegraphics[width=\textwidth]{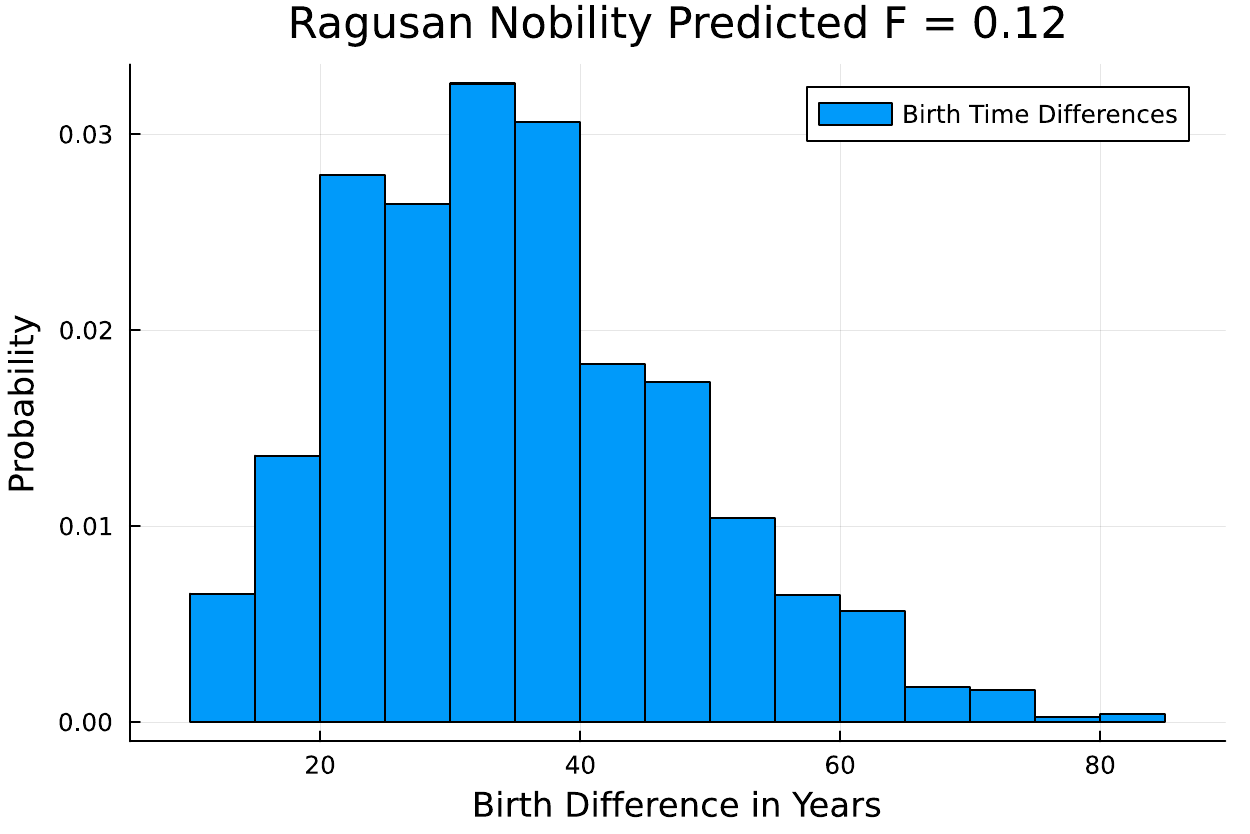}
        	\caption{Birth Year difference between parent and child in Ragusan Nobility Network \cite{Ragusan_Data_Pajek}. Across 2442 edges where this is known for both nodes.}
        	
        \label{fig:ragusan_birth_time_differences}        \end{subfigure}  
\begin{subfigure}[t]{0.48\textwidth}   
\centering
            \includegraphics[width=\textwidth]{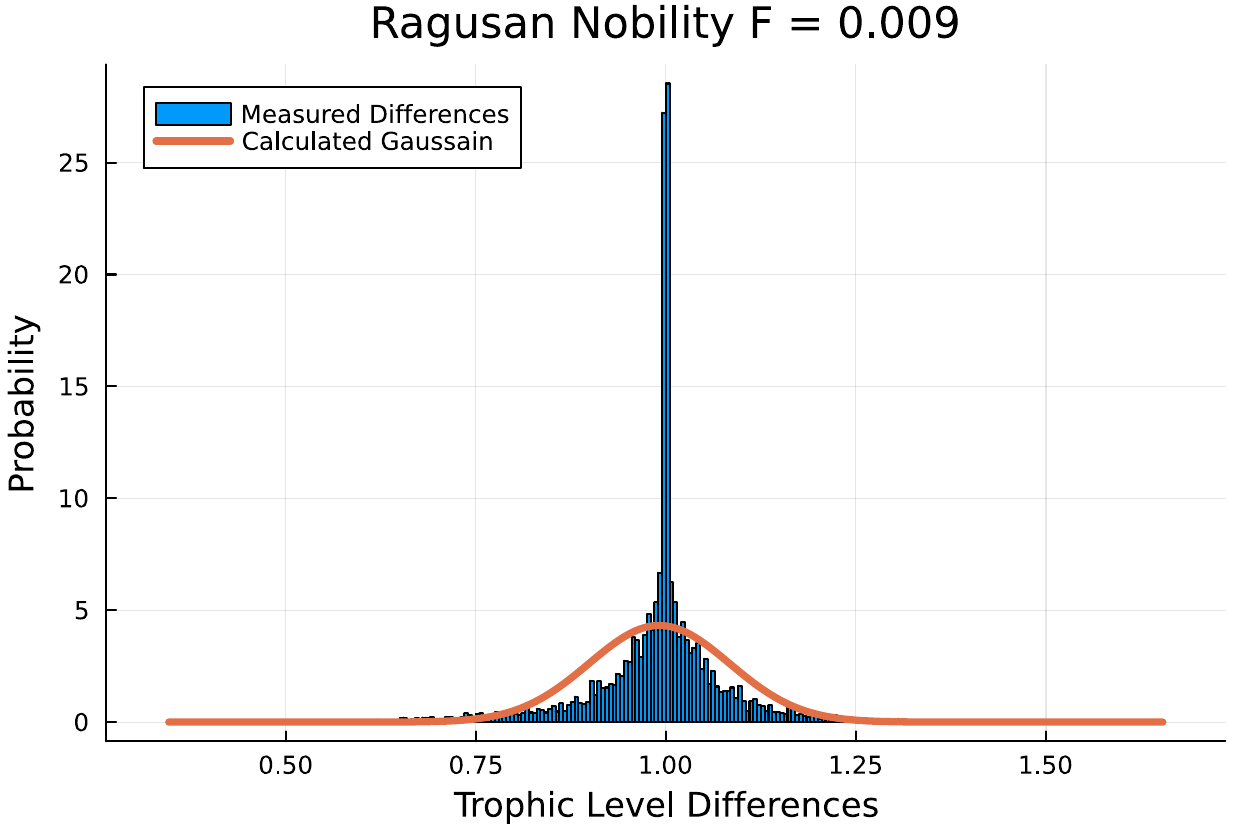}
        	\caption{Trophic level differences between nodes in Ragusan Nobility Network \cite{Ragusan_Data_Pajek} across the whole network.}
        	
        \label{fig:Ragusan_level_differences}        
        \end{subfigure}  
\caption{Comparison of trophic level differences and birth year differences in real historical network \cite{Ragusan_Data_Pajek}}
\label{fig:nobility_differences}
\end{figure}

We also note that for this network the data may be ``missing not at random". This means that the fact the birth dates for certain nodes is missing may be correlated with the time period the node is from, events that were ongoing at that time or another property of the node. In figure \ref{fig:ragusan_birth_time_differences}, we present the differences in birth time for the parent-child pairs where we have data for both parties. This constrains how much we can say definitively as this is a small subset of the network data. We prune the data-set of obvious errors, making sure that the differences are positive to preserve causality, that you are born after your parents, as well as removing any differences outside what is biologically possible to exist (birth time differences smaller than 10 and greater than 120). This data is also a mixture of the mother-child and father-child differences which may be different for biological reasons and due to the patriarchal society from which we draw the data. Based on this difference distribution, we predict a Trophic incoherence of $F=0.12$ which makes sense. We predict a very coherent network as edges are constrained to only go forward in time in this network. This is more than the measured trophic incoherence of $F=0.009$. This is not surprising as there is a large variation in birth difference between parent and child as well as the fact that the full network is sparse, the number of children and parents you can have is limited, and acyclic so it is not surprising that $F$ is very low. The two values do not agree very closely but do imply the same kind of regime where the network is very hierarchical and ordered which is what we expect when analysing this network. Additionally, we do not know the birth year difference across all nodes in the network as this data is missing for some pairs so the trophic incoherence is being measured over a much larger samples so this may account for some of the differences as the sample size could affect the ratio of mean and standard deviation for the birth year differences.

Additionally, Trophic level has been used to analyse another historical network of Islamic scholars and infer missing data. \cite{Shuaib2022TrophicInformation}. Our work explains why this was possible and justifies the assumption that trophic level could be used as a good approximation for the relevant ``fitness'' parameter in that network. We also analyse the difference in death dates between individuals and the relationship between degree imbalance and birth year in the network in appendix \ref{sec:Death_differences}, where we show how both of these are not useful factors for predicting the network structure and how understanding the trophic structure shows that certain factors do not contribute to the formation of a network. The study of this historical network is not meant to unveil new information as the process of how family trees form is well understood. However, we demonstrate how the technique of trophic analysis can be linked to node properties, used to infer missing data and analyse network structure. We hope that techniques in this work can be extend to study real-world systems which are not as well understood.

\section{Discussion}

In this work, a uniform fitness distribution was used as we wished to create the simplest possible structures where the number of nodes are equally distributed across the fitness spectrum. However, this need not be the case and many fitness distributions could be used. If the fitness is thought to represent a property of a real system it may be more realistic if this is a Pareto or Gaussian distribution for example. Modifying the fitness distribution may impact the degree distribution and arrangement of levels which could be the subject of future work. Changing the range of the fitness distribution and the size of the preferred fitness difference will affect the maximum number of levels with the fitness density in a particular regime translating into the number of nodes of a particular trophic level. The number of levels and the fitness distance between them could reflect an underlying feature of a real system, such as the `` depth of compettion'' of a game or range of social status of individuals \cite{Jerdee2023LuckHierarchies}. The effects associated with the sharp end of the uniform distribution could also be mitigated by having a distribution which decays of towards the maximum and minimum fitness; although, in some circumstances, the  fitness edge affects may be a realistic behaviour which is important to the dynamics of the system. For example, consider a network of sports teams where edges represent player transfers between teams. The top teams clearly have different dynamics as there are no ``better'' teams for their players to move to compared to a team which buys players from lower leagues but also sells to better teams. A similar effect can be imagined in food-webs or production networks, where the nodes which consume the raw material or energy input to the system behave differently to nodes in the bulk of the system. Related to this, it is known that the presence of source/sink nodes \cite{OBrien2021HierarchicalNetworks} can be related to non-normality and that nodes at the ends of the trophic hierarchy are important for influencing the network \cite{Rodgers2023InfluenceNetworks}. It is also possible to have a fitness which is multi-dimensional, where different functions act on different fitness aspects. This could allow a mixing of heterophilic and homophilic behaviour where a different interaction could be present in each fitness. Our model provides a natural way to move between linear hierarchy and homophillic behaviour with a single fitness as the difference parameter can be decreased to zero. Community structure could also be studied through fitness functions and distributions as densely connected groups with few connections between groups could be modelled. And in some systems there can be an interaction of the hierarchical structure of the network with organisation of the network into communities \cite{Iacovissi2022TheNetworks,Peixoto2022OrderedNetworks}.

Another possible extension of this work would be to use a fitness space which is not linear. For example, the node fitness could be an angle on a circle and used to study periodic structures \cite{Gong2021DirectedModels}. It may also be possible to link network hierarchy and fitness, linear or periodic, to the field of network geometry which has recently been extend to the directed case \cite{Allard2023GeometricNetworks} where the topological features found in these models, such as reciprocity and the prevalence of different types of directed cycle, may be analysed through various tools associated with network hierarchy \cite{Gong2021DirectedModels, Peixoto2022OrderedNetworks, DeBacco2018ANetworks,MacKay2020HowNetwork, Nartallo-Kaluarachchi2024BrokenNetworks}.

In this work, we present a model to explain how a simple fitness-based mechanism can lead to a hierarchical network structure. However, it is also possible to view the emergence of directed hierarchy as the result of dynamics taking place which alter the network structure \cite{Kawakatsu2021EmergenceDynamics,Zamani2017GlassyOrganizations, Nepusz2013HierarchicalIndividuals}. Faculty hiring dynamics were studied in \cite{Lee2021TheNetworks} where hierarchy steepness, fraction of edges pointing up in rank, and other network measures were analysed as the institution hiring preference functions were varied. This has also being studied in time varying endorsement networks where the probability to endorse depends on the preference for endorsing up the hierarchy and preference for endorsing nodes close in the hierarchy \cite{Kawakatsu2021EmergenceDynamics}. Hierarchy linked dynamics can also take place in animal social interactions \cite{Ward2022NetworkInteractions} with position in the hierarchy affecting the behaviour of individuals. These works link to the kind of dynamics which could be built into extensions of our growing fitness model where hierarchy can be mixed with other interactions. As the network evolves in time, it may be possible to analyse the time evolution of the network hierarchy either using standard trophic level or the recent reformulation of SpringRank for ranking in temporal networks \cite{DellaVecchia2023ANetworks}. Inspiration for creating a temporal network could also be taken from the approach of \cite{Hartle2021DynamicModels} where the network structure updates according to interaction between dynamic hidden variables assigned to each node. In the directed hierarchical case, a model could be constructed where the fitness of a node varies over time and edges are broken and formed as it moves through the hierarchy.  

This work is also closely related to the study of non-normality \cite{Duan2022NetworkSystems,Asllani2018StructureNetworks, Nartallo-Kaluarachchi2024BrokenNetworks} in networks and provides a generative model to create networks with varying degrees of normality while controlling how the levels are formed by varying the fitness distribution. It also provides a justification for the ubiquity of non-normal networks by creating a model for them which does not only depend on ordering by node arrival time and reciprocal edges \cite{OBrien2021HierarchicalNetworks,Sornette2023Non-normalBubbles, Nartallo-Kaluarachchi2024BrokenNetworks}. The fact that it was observed in a real network in \cite{Sornette2023Non-normalBubbles} that the number of reciprocal edges varies with level is something that could be built into the fitness function we use, showing the flexibility of the model. The model we present also comes with the advantage that the number of edges and nodes is fixed and that the model always generates connected networks which may not be the case in some static fitness models. Due to the similarities between trophic analysis and the study of non-normality it is hoped that trophic analysis and models of the type presented in this work could be a useful tool to augment areas where non-normality is being applied such as has been done in \cite{Nartallo-Kaluarachchi2024BrokenNetworks}.

One of the key results of this work is that it provides some explanation of why coherent networks are common in nature as they can arise by multiple mechanisms which act simultaneously. We have shown that coherence can be induced by the fraction of edges which go in and out of newly introduced nodes, by the strong degree imbalance induced by degree-based preferential attachment or by the node ordering induced by fitness interactions. These simple effects could combine in many real-world systems to create the varied spectrum of trophic incoherence observed in real networks \cite{Johnson2017LooplessnessCoherence,MacKay2020HowNetwork}. The general model we provide captures both the hierarchy induced by time ordering as in citation networks and the niche-based hierarchy studied in ecology.

This work also provides some insight into the utility and limitations of trophic analysis as a tool. We have shown that when incoherence is low trophic analysis can be taken as good proxy for node level fitness and can be useful to infer some properties about the nodes. However, when a network is very incoherent this is not the case and trophic level gives you no information about the node fitness and simply correlates with degree imbalance. We also show how trophic analysis does not fully capture the underlying fitness distribution as all the trophic level differences are approximately mapped to Gaussian even when the underlying fitness distribution is not a Gaussian. We do however show that trophic incoherence can be used as a good proxy for estimating the scale of the coefficient of variation of the underlying fitness distribution, if the network is built using a fitness hierarchy. Linking trophic analysis to node fitness and the properties of the fitness distribution also allows the fitness properties to be linked to the network properties which have been shown to be related to trophic structure \cite{MacKay2020HowNetwork,Johnson2017LooplessnessCoherence,Johnson2014TrophicStability,Rodgers2023InfluenceNetworks,Rodgers2023StrongNetworks}. 
Another limitation of this work is that we assume that each edge represents one type of interaction whereas in real systems the ranking of elements may depend on the interplay between different types of edge interactions \cite{Newman2022RankingComparisons}. It is also true in some systems that being at the top of the ranking may not represent some intrinsic skill or ability but could just be related to luck \cite{DeDomenico2024ImitationDynamics}. For such systems, we could use a variant of our model where some edges can be added without respect for fitness or hierarchy representing luck or intrinsic variability in the system taking inspiration from the work of \cite{Jerdee2023LuckHierarchies}, where luck-based variability was incorporated into their model of rankings based on pairwise comparisons.

In this work, we have also shown the interplay between trophic analysis and degree imbalance. In some circumstances, this is useful as it has been shown that there is a relationship between non-normality and degree imbalance and that both are prevalent in nature \cite{Duan2022NetworkSystems}. However, there may be situations where this degree imbalance just arises from randomness like in an ER random graph and then trophic level simply correlates with this feature. This may be useful in some cases if the degree imbalance is an important feature in your dynamics however this is something to be aware of when using methods like Trophic Analysis or SpringRank \cite{DeBacco2018ANetworks}. Alternative ranking methods exist which mitigate this affect based on the stochastic block model \cite{Peixoto2022OrderedNetworks}. This allows the statistical significance of the observed structure to be understood. The approach of \cite{Peixoto2022OrderedNetworks} is very useful for understanding hierarchy and we recommend that it is used when trying to understand the significance of ranks in an application. In Trophic Analysis, the incoherence, $F$, can be used as a guide to understand the importance of hierarchy to the system and we would argue that ranks should always be given with a parameter like $F$ or another quantification of the importance of hierarchy like the normality or significance of the ranks \cite{Peixoto2022OrderedNetworks}. A stochastic block model framework has also been used to study the ability of ``agony'' to detect planted ranks \cite{Letizia2018ResolutionNetworks}. $F$ and other parameters which involves an average over the network structure do however have some limitations. For example, if we used a fitness function which creates a very strict hierarchy in one fitness region and no significant hierarchy in the other we lose the information about the fact that the network has two distinct behaviours as $F$ takes a value based on the average across both. This means that we also need to look at the distribution of levels to fully understand the hierarchy in a network and the incoherence has an implicit assumption of the hierarchical behaviour being homogeneous across the system. Additionally, source/sink nodes which always have non-zero degree imbalance play a unique role in the network \cite{OBrien2021HierarchicalNetworks} with their number potentially affecting the dynamics.

Our results may also be affected by varying the sparsity of the networks used. If a networks is extremely dense, with the precise threshold depending on the fitness function and expected number of levels implied by the fitness function, it may be impossible to build very coherent networks as there are too many edges to have only edges which go up by exactly one trophic level. Additionally, depending on the generative process, compared to dense networks, very sparse networks may have fewer cycles and more nodes of zero in or out-degree. Hence these are generally more coherent, as there is more likely to exist an arrangement such that the trophic level differences are close to one across all edges.

This work could also prompt further examination of the relationship between trophic analysis and models of ecological network formation, as has been studied using a previous definition of trophic level which required basal nodes and a network formation model which was not based on fitness interactions \cite{Dominguez-Garcia2016IntervalityNetworks}. Our work explains features like how the hierarchical arrangement and level structure change as the fitness mean and standard deviation are varied \cite{Loeuille2005EvolutionaryWebs}. Our model could also be extended in a similar way to in the probabilistic niche model \cite{Williams2010TheWeb,Jacobs2015UntanglingModels}: we could assign the preferred fitness difference and fitness function standard deviation to each node individually to account for the fact that in some systems, such as food webs, different nodes may be able to interact over different fitness distances with different standard deviations, which may change the structure of the network formed, depending on the distributions of parameters chosen.

Trophic coherence has also been found to affect opinion dynamics in social network models \cite{pilgrim2020organisational}. These results could be linked to social network formation via the work of \cite{Ball2013FriendshipStatus}, where it was shown that there is a hierarchical structure present in some social networks, due to a phenomenon that edges going from lower to higher status individuals are less likely to be reciprocated. This concept could be incorporated into a variant of our model with a reciprocity parameter and analysed via Trophic Analysis.

\section{Conclusion}

In conclusion, we have presented a model of growing directed networks based on the principle of assigning nodes fitness and the network being built from a mixture of degree-based preferential attachment and fitness interactions.
We have shown how these simple effects can produce a wide range of values of trophic incoherence which may explain the ubiquity of trophic coherence in real-world systems. We have shown how the properties of the networks, as measured with trophic analysis, can be related to the properties of the fitness function used in the generative model. In addition, we have shown how preferential attachment interacts with the trophic structure of the network and how fitness hierarchy affects the degree distribution in networks generated with our model. We have provided 
a variety of possible extensions and use cases of this model and other models of this type, in the hope that it can generate ideas for more work on the interplay between network hierarchy and both the structure and dynamics of directed networks.

\section*{Acknowledgements}

All graph manipulation and construction was carried out using the software package Graphs.jl \cite{Graphs2021}. The authors would like to thank the Centre for Doctoral Training in Topological Design and Engineering and Physical Sciences Research Council (EPSRC) for funding this research, Grant EP/S02297X/1. S.J. also acknowledges the support from the Alan Turing Institute under EPSRC Grant EP/N510129/1.

\section*{Data Availability Statement}

The data and code used as part of this work can be found at \url{https://github.com/nrodgers1/directed_networks_with_fitness_and_hierarchy}. Original network data can be found at \cite{Ragusan_Data_Pajek}.

\section*{Conflict of Interest Statement}

The authors declare no conflicts of interest.

\appendix

\section{Additional Analysis of Historical Network Data} \label{sec:Death_differences}

We also extract from the historical network, \cite{Ragusan_Data_Pajek}, the death year of the individuals and the difference in death year between individuals and their parents. The death years were found to be related to the trophic level however the difference distribution predicts a different kind of network structure than the one we observe. Showing that not all network metadata can be used to understand the hierarchical nature of network formation.

\begin{figure}[H]
    \centering
\begin{subfigure}[t]{0.48\textwidth}        		
\centering
            \includegraphics[width=\textwidth]{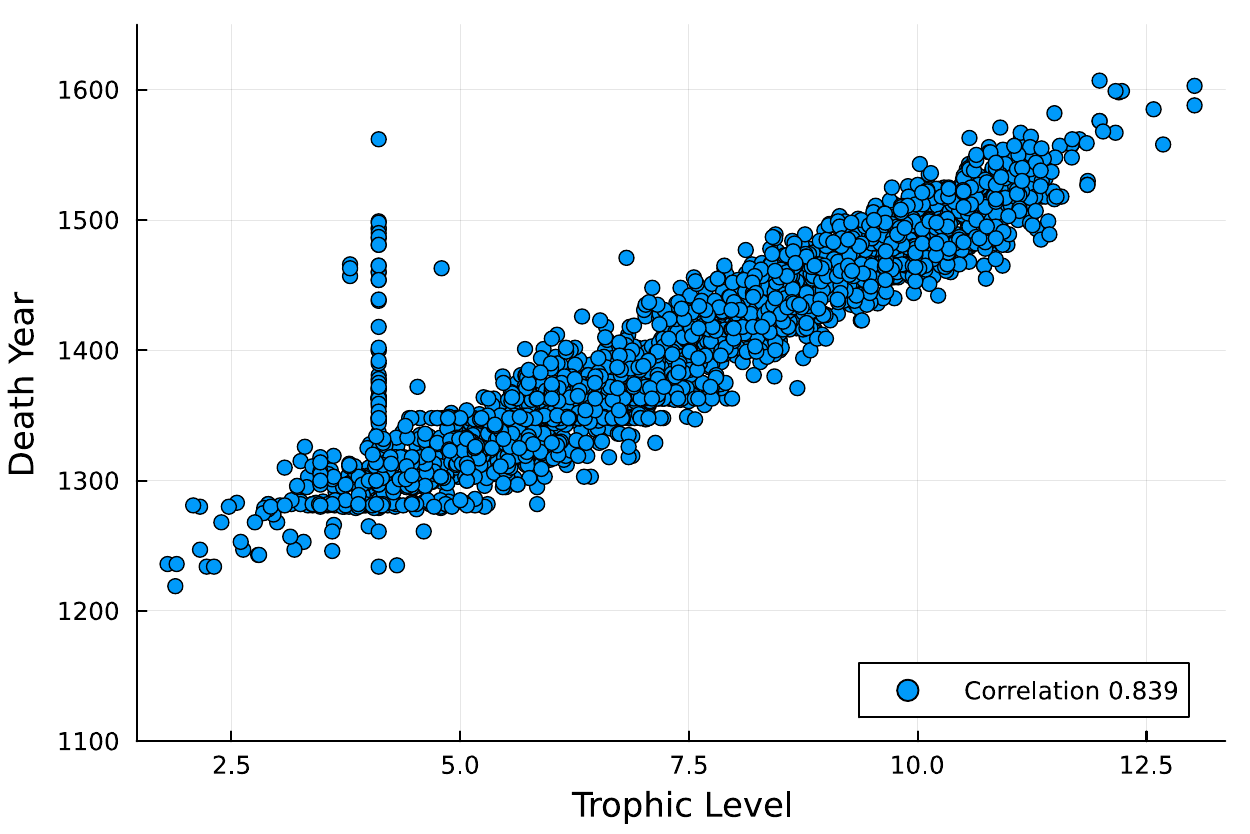}
        	\caption{Death Year against Trophic Level of nodes in Ragusan Nobility Network  \cite{Ragusan_Data_Pajek}	for 3043 nodes where this is known.}
        	
        \label{fig:ragusan_death_time_level}        \end{subfigure}  
\begin{subfigure}[t]{0.48\textwidth}   
\centering
            \includegraphics[width=\textwidth]{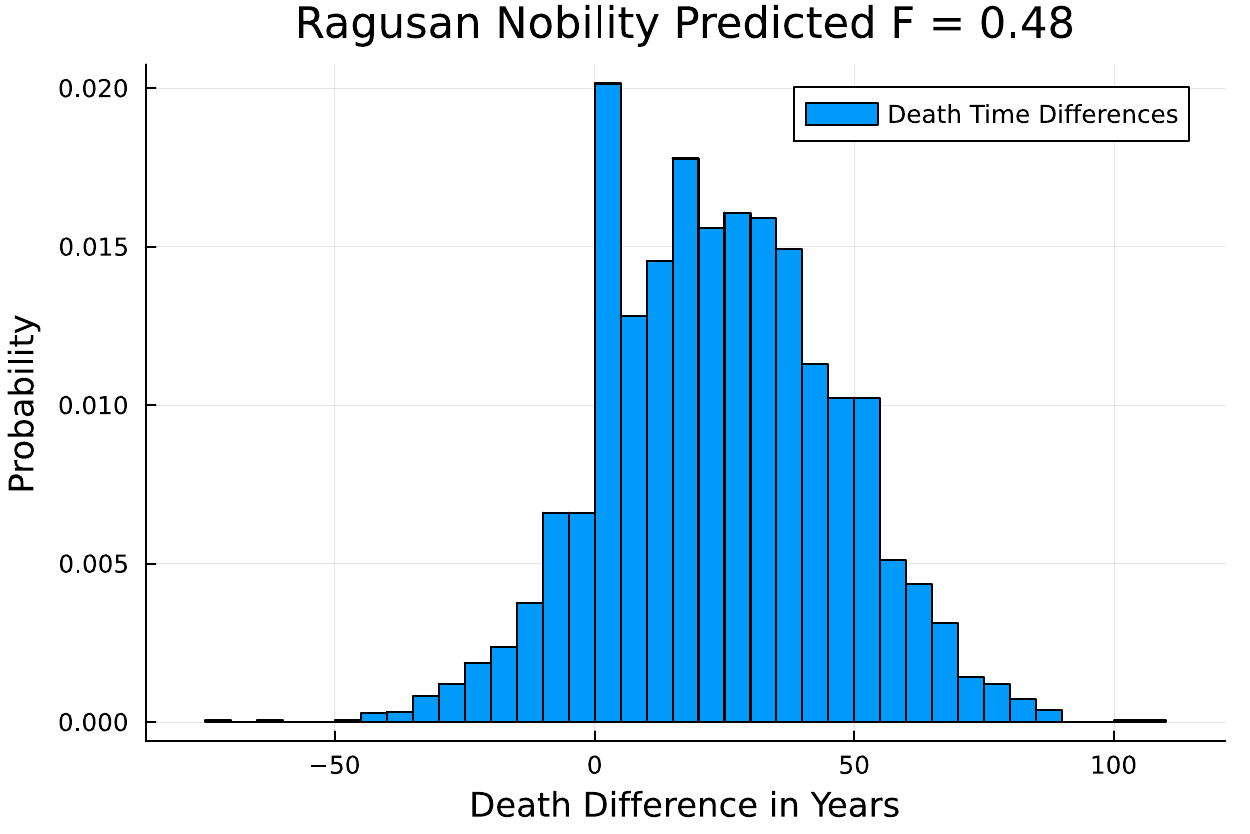}
        	\caption{Death Time differences between nodes in Ragusan Nobility Network \cite{Ragusan_Data_Pajek} across the 3633 edges where the year is known for both nodes.}
        	
        \label{fig:Ragusan_death_differences}        
        \end{subfigure}  
\caption{Comparison of trophic level differences and birth year differences in real historical network \cite{Ragusan_Data_Pajek}}
\label{fig:nobility_differences_death}
\end{figure}

In figure \ref{fig:nobility_differences_death} we analyse the relationship between the trophic parameters and the other data we have from this network which is the date of death of some of the individuals (3043). There is a strong correlation between the year an individual dies, figure \ref{fig:ragusan_death_time_level}, which makes sense as we expect individuals who die later in time to be further along the family tree. However, the correlation is not as strong as in the case of birth times which makes sense when we consider figure \ref{fig:Ragusan_death_differences}. We see that the difference in death year (year of child death minus year of parent death) can be negative which is not the case of the birth difference. Hence based on death differences you would not predict an acyclic and very coherent network as seen in the real data so this rules this out as a generative mechanism for the network. Also, it explains why the correlation between death times and trophic level is worse than the birth-level correlation as there is much more variability in the difference in death times as it can be negative and there is a large spike at zero which is not the case in the birth year difference. We assume this is the affect of disease and high childbirth mortality in the historical era in which this data is taken from. This highlights how trophic analysis can be used to understand the parameters that contribute to network formation in a simple real-world example but we hope this can be extended to more complex real world systems where the answers are less obvious.    

\begin{figure}[H]
      		\centering
            \includegraphics[width=0.8\linewidth]{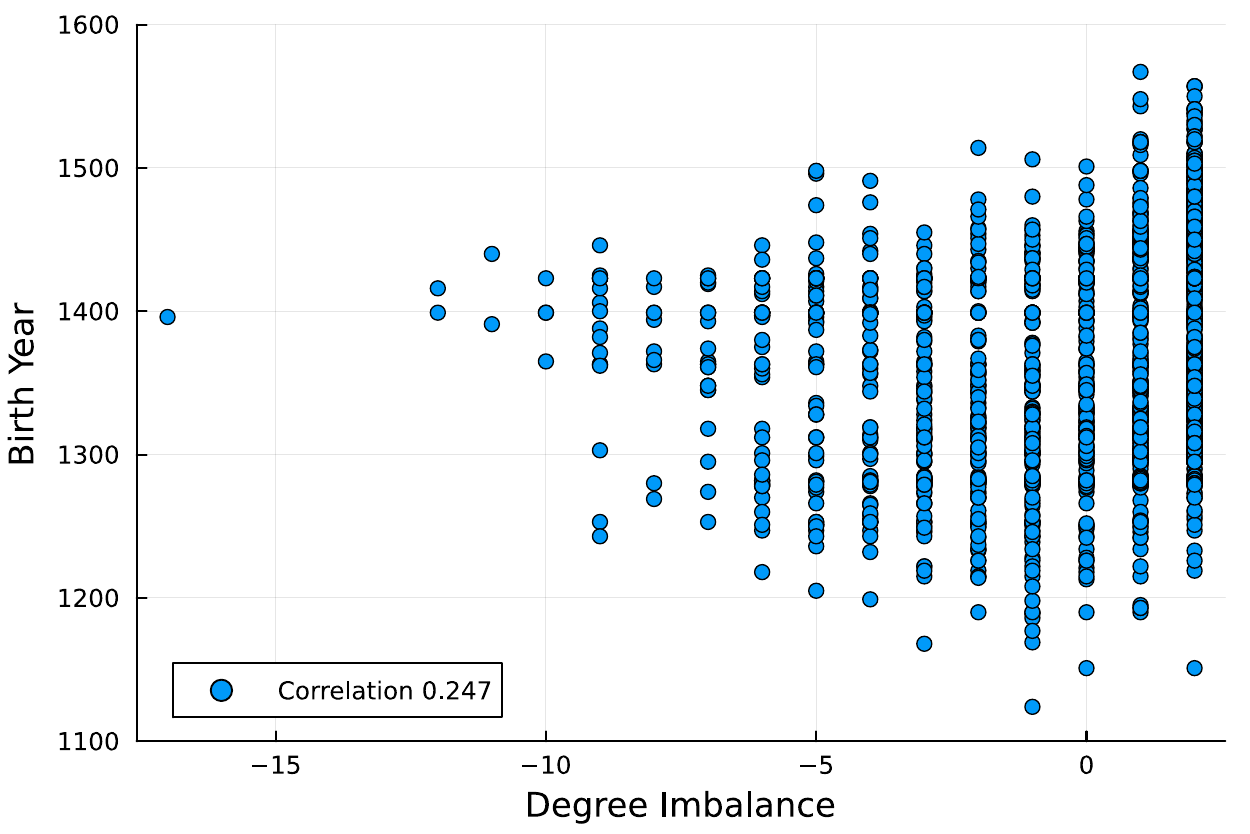}
        	\caption{  Known Birth Date (3065 individuals)  of Members of the Ragusan Nobility, \cite{Ragusan_Data_Pajek}, against Degree Imbalance (in-degree minus out-degree) with Pearson correlation coefficient given in the legend.}
        	
        \label{fig:ragusan_imbalance_level}
        \end{figure}

It can also be demonstrated, figure \ref{fig:ragusan_imbalance_level}, that degree imbalance is not a useful metric to predict birth year as you would expect in the low incoherence regime as there is little correlation between the two quantities unlike was observed with trophic level. We also see that as the network is a network of parent-child relationships the maximum degree imbalance is two from having two parents and no children and the minimum is constrained by the number of children someone can have minus two for their parents.  

\bibliographystyle{unsrt}
\bibliography{references,ref1, extra_ref}

\end{document}